\documentclass[]{mn2e}
\usepackage{deluxetable}
\usepackage{graphicx}
\usepackage{lscape}

\newcommand{\mgii}{\mbox{Mg\,{\sc ii }}}
\newcommand{\hi}{\mbox{H\,{\sc i }}}

\title[DLA, subDLA, and LLS galaxies]{A Groundbased Imaging Study of Galaxies Causing DLA, subDLA, and
LLS Absorption in Quasar Spectra\thanks{Based on data obtained from the Sloan Digital Sky
Survey,  the WIYN telescope, the MDM Observatory 2.4 m Hiltner
telescope, the KPNO  2 m telescope, and the NASA IRTF 3 m
telescope. The WIYN Observatory is a  joint facility of the University
of Wisconsin-Madison, Indiana University, Yale University, and the
National Optical Astronomy Observatories. The Infrared Telescope
Facility is operated by the University of Hawaii under a cooperative
agreement with the National Aeronoautics and Space Administration. The
Hiltner 2.4 m Telescope on Kitt Peak is operated by MDM Observatory,
which at the time was a joint facility of University of Michigan, Dartmouth
College, Ohio State University, and Columbia University.}}
\author[S. M. Rao et al.]
{\parbox[t]{\textwidth}{\raggedright Sandhya M. Rao$^{1}$\thanks{Visiting Astronomer,
Kitt Peak National Observatory, National Optical Astronomy
Observatory, which is operated by the Association of Universities for
Research in Astronomy, Inc. (AURA), under cooperative agreement with
the National Science Foundation.  Visiting Astronomer at the Infrared
Telescope Facility, which is operated by the University of Hawaii
under Cooperative Agreement number NCC 5-538 with the National
Aeronautics and Space Administration, Science Mission Directorate,
Planetary Astronomy Program.}\thanks{E-mail: srao@pitt.edu}, 
Mich\`ele Belfort-Mihalyi$^{1}$\footnotemark[2], 
David A. Turnshek$^{1}$\footnotemark[2],
Eric M. Monier$^2$\footnotemark[2],
Daniel B. Nestor$^3$\footnotemark[2],
and
Anna Quider$^4$\footnotemark[2]}
\vspace*{6pt}\\
$^{1}$Department of Physics and Astronomy,  University of Pittsburgh,
Pittsburgh, PA 15260, USA\\
$^{2}$Department of Physics, The College at Brockport, State
University of New York, Brockport, NY 14420, USA\\
$^{3}$Department of Physics and Astronomy, University of California, Los Angeles, CA 90095, USA\\
$^{4}$Institute of Astronomy, University of Cambridge, Madingley Road,  Cambridge, CB3 0HA
}

\begin{document}

\date{}

\pagerange{\pageref{firstpage}--\pageref{lastpage}} \pubyear{2011}

\maketitle

\label{firstpage}

\begin{abstract}
We present results from a search for galaxies that give rise to damped
Lyman alpha (DLA),  subDLA, and Lyman limit system (LLS) absorption at
redshifts $0.1 \la z \la 1$ in the spectra of background quasars. The
sample was formed from a larger sample of strong \mgii
absorbers ($W_0^{\lambda2796} \ge 0.3$ \AA) whose \hi column densities
were determined by measuring the Ly$\alpha$ line in HST UV
spectra.  Photometric redshifts, galaxy colours, and proximity to the
quasar sightline, in decreasing order of importance, were used to
identify galaxies responsible for the absorption. Our sample includes
80 absorption systems for which the absorbing galaxies have been
identified,  of which 54 are presented here for the first time. In 
some cases a reasonable identification for the absorbing galaxy
could not be made. 

The main results of this study are: (i) the surface density of galaxies falls off exponentially 
with increasing impact parameter, $b$, from the quasar sightline relative 
to a constant background of galaxies, with an {\it e}-folding length of $\approx 46$
kpc.  Galaxies with $b \ga 100$ kpc calculated at the
absorption redshift are statistically consistent with being unrelated
to the absorption system, and are either background or foreground
galaxies. (ii) $\log N_{HI}$ is inversely correlated with $b$ at
the 3.0$\sigma$ level of significance.  DLA galaxies
are found systematically closer to the quasar sightline, by a factor
of two, than are galaxies which give rise to subDLAs or LLSs.
 The median impact parameter is 17.4 kpc for the DLA
galaxy sample, 33.3 kpc for the subDLA sample, and 36.4 kpc for the
LLS sample.   We also find that the decline in 
$\log N_{HI}$ with $b$ can be roughly described by an
exponential with an {\it e}-folding length of 12 kpc that occurs at
$\log N_{HI} = 20.0$. (iii) Absorber galaxy luminosity relative to $L^*$, 
$L/L^*$, is not significantly correlated
with $W_0^{\lambda2796}$, $\log N_{HI}$, or $b$. (iv) DLA, subDLA, 
and LLS galaxies comprise a mix of spectral
types, but are inferred to be predominantly late type galaxies based
on their spectral energy distributions. (v) The properties of low-redshift DLAs and
subDLAs are very different in comparison to the properties of gas-rich
galaxies at the present epoch. A significantly
higher fraction of low-redshift absorbers have large $b$ values, and a
significantly higher fraction of the large $b$ value galaxies have
luminosities $L<L^*$. The implications of these results are discussed.

\end{abstract}

\begin{keywords}
quasars: absorption lines -- galaxies: ISM -- galaxies: statistics
\end{keywords}

\section{Introduction}

The recognition that damped Ly$\alpha$ absorption-line systems (DLAs)
seen in quasar spectra arise in neutral-gas-rich foreground galaxies
(Wolfe et al. 1986) motivated new methods for high-redshift galaxy
studies $\approx25$ years ago. These high HI column density systems,
with $N_{HI} \ge 2\times 10^{20}$ atoms cm$^{-2}$, trace the bulk of
the observed neutral gas in the Universe, and they are, therefore,
powerful probes of galaxy formation and evolution back to the
redshifts of the most distant quasars. Larger datasets and deeper
surveys (e.g. Prochaska, Herbert-Fort, \& Wolfe 2005; Rao, Turnshek,
\& Nestor 2006, henceforth, RTN06; Noterdaeme et al. 2009) have
improved our knowledge of the neutral gas content and distribution at
all observable redshifts, including the present epoch (Ryan-Weber et
al. 2003; Zwaan et  al. 2005). No other technique has revealed
comparable reservoirs of neutral gas beyond the local Universe. See
Rao (2005) and Wolfe et al. (2005) for some past reviews. Few DLAs
were known at low redshift prior to the turn of the century because
the Ly$\alpha$ line falls in the UV for $z<1.65$. Thus, in the absence
of large UV spectroscopic surveys, this meant that studies of neutral
gas in what corresponds to the most recent $\approx70$\% of the age of
the Universe\footnote{The ``737'' cosmology is used throughout:
($\Omega_\Lambda, \Omega_m, h) = (0.7,0.3, 0.7)$}  were
problematic. But now MgII-based UV spectroscopic surveys (with
HST-FOS, HST-STIS, HST-ACS Grism, and GALEX Grism) are identifying
significant numbers of low-redshift DLAs and subDLAs (Rao \& Turnshek
2000; RTN06; Monier et al. 2009a; Turnshek et al. in prep). The
\mgii-based surveys for DLAs,  which are designed to be unbiased
(RTN2006), can be used to infer the incidence and cosmic neutral gas
mass density at $z<1.65$ (e.g. RTN06). Also, while subDLA absorbers,
those with $10^{19} \le N_{HI} < 2\times10^{20}$ atoms cm$^{-2}$, do
not contribute much to the cosmic neutral gas mass density (P\'eroux
et al. 2005), they are often found to have higher metallicities than
DLAs (Kulkarni et al. 2007 and references therein). Lyman Limit System
(LLS) absorbers are simply  those with $ N_{HI} \ga
3\times10^{17}$ atoms cm$^{-2}$. Both subDLAs and LLSs generally
exhibit \mgii absorption, and all strong \mgii systems,
those with $W_0^{\lambda2796} \ge 0.3$ \AA, are Lyman limit systems
(e.g., Churchill et al. 2000). However,
unbiased \mgii-based surveys for subDLAs and LLS have  never been implemented.

With the identification of DLAs (and subDLAs and LLSs), follow-up work
involving the study of their host galaxies, environments,
neutral-gas-phase metallicities, kinematics, 21 cm spin temperatures
(when possible), ionization conditions, and numerical and
semi-analytic modeling has kept many astronomers busy for
decades.\footnote{A recent search of the SAO/NASA Astrophysics Data
System database identifies $>2000$ papers.} Despite this, a consensus
on certain aspects of DLAs is still lacking. Our previous studies
indicated that DLA galaxies are of mixed morphology and that the
highest $N_{HI}$ systems have the smallest impact parameters, but are
hosted by low luminosity ($L<0.2L^*$) galaxies (Rao et al. 2003;
Turnshek et al. 2001; see also Chun et al. 2006). On the other hand,
while Chen \& Lanzetta (2003) and Chen, Kennicutt, \& Rauch (2005)
conclude that DLA galaxies span a  mix of morphological types, they
also propose that a large contribution from dwarf galaxies is not
required to explain the properties of DLAs. In addition, Zwaan et
al. (2005) suggest that the local galaxy population can completely
explain the properties of known low-redshift DLA galaxies.  Studies of
\mgii galaxies, of which DLAs form a subset, have  also revealed
a mix of morphological types (e.g. Churchill, Kacprzak,  and Steidel
2005; Kacprzak et al. 2007), although most appear to be
spirals and  the majority exhibit minor perturbations (as seen in HST
images).  From stacked images of over 2800 SDSS quasar sightlines
containing MgII absorption,  Zibetti et al. (2007) derive an Sbc-type
average colour and $0.5L^*$ average  luminosity for the absorbing
galaxies. Images of quasar sightlines with ``ultra-strong''
\mgii systems\footnote{Those systems with \mgii rest
equivalent width $W_0^{\lambda2796} \ge 3$ \AA.} point to outflows
from bright  ($L>L^*$) starbursting galaxies as the cause of the
kinematically-complex absorption (Nestor et al. 2007; 2010). A large
fraction of these are known to be DLAs (RTN06). However, the
ultra-strong \mgii regime is not  addressed in this paper.

Semi-analytic and numerical models, some of which are based on results
from high-resolution spectroscopic data of DLA metal absorption lines,
have resulted in a variety of often competing scenarios for DLAs:
large rapidly rotating protogalactic disks (Prochaska \& Wolfe 1997,
1998; Wolfe \& Prochaska 1998), merging protogalactic clumps in a
hierarchical merging scenario (Haehnelt et al. 1998), low surface
brightness galaxies (Jimenez et al. 1999), dwarf galaxies (Okoshi \&
Nagashima 2005), compact, faint galaxies with impact parameters
smaller than 5 kpc at $z\sim 3$ (Nagamine et al. 2007), and the outer
regions of high-$z$ Lyman break galaxies (M{\o}ller et al. 2002; Wolfe
et al. 2003). Recent high-quality \hi 21 cm data of local galaxies
indicate that DLA gas velocity widths are more consistent with tidal
gas related to galaxy interactions or superwinds rather than galaxy
disks (Zwaan et al. 2008). In addition, some recent compelling
cosmological simulations relevant to interpreting the nature of
\mgii-selected galaxies in general, and DLA galaxies in
particular, have been presented by Kacprzak et al. (2010).  

However, there is a dearth of identified DLA (and subDLA) galaxies, 
and this has undoubtedly motivated the various interpretive
scenarios. Therefore, larger samples of neutral-gas-selected galaxies
are required to investigate the possibilities, which in turn will help
constrain models of  galaxy evolution and better establish the galaxy
population that harbors  the bulk of the neutral gas in the
Universe. Traditional galaxy surveys trace galaxies by virtue of their
luminous emission. Beyond the local Universe far less is known about
neutral-gas-selected galaxies and  their relationship to
luminosity-selected galaxies.

As an extension of our earlier work (Rao et al. 2003), we have
undertaken a large multi-colour optical/IR imaging programme of quasar
fields containing \mgii absorbers with measured $N_{HI}$. The
absorbers were selected from Table 1 of RTN06, and were required to
have absorption redshifts $z_{abs} \la 1$ to optimize the
possibilities for detection and characterization of galaxies in these
fields.  Data and results from eight DLAs in six quasar fields were
presented in Turnshek et al. (2001), Rao et al. (2003), and Turnshek
et al. (2004). Other observations from the literature were also
considered. In total, 27 DLA, 30 subDLA, and 23 LLS galaxies  (i.e. 80 absorbers in
total) have been identified. In this paper, we present and analyse the entire dataset. These
observations are described in \S2. The identification of galaxies is
described in \S3 through specific examples. The entire dataset can be
accessed on line at http://enki.phyast.pitt.edu/Imaging.php. A
summary of the data and statistical inferences are presented in
\S4. Conclusions and discussion  are presented in \S5 and 6,
respectively. Among other findings this work demonstrates that 
the neutral hydrogen column density, $N_{HI}$, is strongly correlated
with impact parameter, $b$, in the sense that DLA galaxies are
systematically closer to the quasar sightline, by a factor of two,
than are galaxies which give rise to subDLAs and LLSs. We also find
that the properties of low-redshift ($0.1 \la z \la 1$)
DLAs and subDLAs are very different in comparison to the properties of
\hi-rich galaxies at the present epoch. A significantly higher
fraction of low-redshift absorbers have large $b$, and a significantly
higher fraction of the large $b$ galaxies have luminosities $L<L^*$.

All magnitudes reported in this paper are in the AB system, and all
distance related quantities are calculated using the ``737'' cosmology
with ($\Omega_\Lambda, \Omega_m, h$) = (0.7,0.3,0.7).

\section{Observations}

The imaging data were obtained between December 1998 and June 2005
through  community-access time at national facilities as well as
through Ohio State University's share of time at the MDM
Observatory. The various telescopes and detectors that were employed,
as well as the varying observing conditions that prevailed over the
better part of a decade of observations, resulted in an unavoidably
inhomogeneous dataset. Nevertheless, since most of the data are well
calibrated and reach fainter magnitudes than large groundbased surveys
such as the Sloan Digital Sky Survey, this observing programme has
yielded the most useful and comprehensive set of images of DLA,
subDLA, and LLS absorption-line-producing galaxies that has thus far
been obtained.

The optical images were obtained at Kitt Peak National Observatory in
Arizona. The telescopes and corresponding detectors used were 1) the
KPNO 2.1 m  with the T2KA or T2KB CCDs covering a 10.4\arcmin\
$\times$ 10.4\arcmin\ field-of-view at a scale of
0.305$\arcsec$/pixel, 2) the MDM Observatory 2.4 m Hiltner with the
1024 $\times$ 1024  Templeton CCD covering a 4.72\arcmin\ $\times$
4.72\arcmin\ field-of-view  at 0.275 $\arcsec$/pixel, and 3) the 3.5 m
WIYN with the Tip-Tilt Module (WTTM) covering a  3.84\arcmin\ $\times$
4.69\arcmin\ field-of-view at 0.1125$\arcsec$/pixel.  The
near-infrared images were obtained on Mauna Kea,  Hawaii, with the
NASA IRTF 3.0 m telescope using NSFCAM which has a 76.8$\arcsec$
$\times$ 76.8$\arcsec$ field-of-view at  0.30 $\arcsec$/pixel. The
detector was the 256$\times$256 InSb  array. A few images were
obtained with the SpeX infrared slit-viewer/guider covering a
60$\arcsec$ $\times$60$\arcsec$ field-of-view at 0.12 $\arcsec$/pixel
with  the Raytheon 512 $\times$ 512 InSb array. The optical images
taken with the KPNO 2.1 m or WIYN telescopes were obtained using the
Johnson-Cousins $U,B,R,I$ or $u', g', r', i'$  KPNO SLOAN  filters,
and those observed at MDM were observed with the  MDM Gunn-Thuan $u,
g, r, i$ filters.   Henceforth, we ignore the differences between the
SDSS and Gunn-Thuan filter sets since the transformation between them
is small.  Near-infrared images  were taken using the standard Mauna
Kea Observatory $J, H, K$  filter set.

Optical data were taken in groups of 3 or 4 offset exposures ranging
from 900 to 1800 seconds per exposure, and standard data reduction
procedures were followed. The infrared observations were carried out
using a series of either 30 or 60 dithered short exposures ranging
from 2 to 20 seconds per exposures. The individual exposure times were
chosen to prevent the quasar point spread function from
saturating. Flat fielding was done using sky frames constructed from
the dithered object frames.  Landolt standards were used to calibrate
the Johnson-Cousins observations (Landolt 1992). Photometric
calibration of fields that overlapped with  SDSS images was performed
by comparing our instrumental magnitudes with the SDSS DR4 photometry
of point sources.  The photometric zeropoint solution with
corresponding errors for each frame were determined by a least squares
fit to the SDSS magnitudes and our instrumental magnitudes.
Generally, 10 or more isolated, unsaturated stars that were common to
our  images and the SDSS fields were used in the calibration.  UKIRT
faint photometric standards (Hawarden et al. 2001)  were used to
calibrate the near-infrared  observations.

Table 1 lists the fields for which we obtained imaging data. The first
six columns give details about the absorption-line system from
RTN06. Column 1 gives the  quasar name, column 2, the quasar
magnitude, column 3, the quasar  emission redshift, column 4 gives the
redshift of the \mgii absorption-line system,  column 5  the rest
equivalent width of the stronger member ($\lambda$2796)  of the \mgii
doublet, and column 6 gives the \hi column density as measured from
the {\it HST} UV spectrum. Column 7 lists the  optical and infrared
filters through which images of each quasar field were obtained.  Here
we summarize a few salient features of the dataset.

\begin{table*}
\caption{The Imaging Sample}
\begin{tabular}{lcccccl}
\hline \hline Quasar & Mag.\tablenotemark{a} & $z_{\rm em}$ & MgII
$z_{\rm abs}$ & MgII $W^{\lambda2796}_{0}$ (\AA) & $\log
N_{HI}$(cm$^{-2}$) & Filters \\ \hline 0021+0043 & 17.7 & 1.245 &
0.5203 &
0.533$\pm$0.036	&	$19.54^{+0.02}_{-0.03}$	&	JHK	\\
\nodata & \nodata & \nodata & 0.9420 &
1.777$\pm$0.035	&	$19.38^{+0.10}_{-0.15}$	&	\nodata	\\
0041$-$266 & 17.8 & 3.053
&	0.8626	&	0.67$\pm$0.06	&	$<$18.00
&	R	\\ 0058+019 & 17.2 & 1.959 & 0.6127 &
1.666$\pm$0.003\tablenotemark{b}&$20.04^{+0.10}_{-0.09}$&UIK	\\
0107$-$0019 & 18.3 & 0.738 & 0.5260 &
0.784$\pm$0.080	&	$18.48^{+0.30}_{-0.63}$	&	JHK	\\
0116$-$0043 & 18.7 & 1.282 & 0.9127 &
1.379$\pm$0.096	&	$19.95^{+0.05}_{-0.11}$	&	JHK	\\
0117+213 & 16.1 & 1.491 & 0.5764 &
0.91$\pm$0.04	&	$19.15^{+0.06}_{-0.07}$	&	UBRIJHK	\\
0123$-$0058 & 18.6 & 1.551
&	0.8686	&	0.757$\pm$0.098	&	$<$18.62
&	JHK	\\ 0138$-$0005 & 18.7 & 1.340 & 0.7821 &
1.208$\pm$0.096	&	$19.81^{+0.06}_{-0.11}$	&	JHK	\\
0139$-$0023 & 19.0 & 1.384 & 0.6828 &
1.243$\pm$0.102	&	$20.60^{+0.05}_{-0.12}$	&	JHK	\\
0141+339 & 17.6 & 1.450 & 0.4709 &
0.78$\pm$0.07	&	$18.88^{+0.08}_{-0.10}$	&	g'r'i'JK\\
0152+0023 & 17.7 & 0.589 & 0.4818 &
1.340$\pm$0.057	&	$19.78^{+0.07}_{-0.08}$	&	H	\\
0153+0009 & 17.8 & 0.837 & 0.7714 &
2.960$\pm$0.051	&	$19.70^{+0.08}_{-0.10}$	&	JHK	\\
0253+0107 & 18.8 & 1.035 & 0.6317 &
2.571$\pm$0.166	&	$20.78^{+0.12}_{-0.08}$	&	g'r'i'JHK\\
0254$-$334 &	16.0	&	1.849	&	0.2125	&	2.23
&	$19.41^{+0.09}_{-0.14}$	&	BRIJK	\\ 0256+0110 & 18.8 &
1.349 & 0.7254 &
3.104$\pm$0.115	&	$20.70^{+0.11}_{-0.22}$	&	g'HK	\\
0420$-$014 & 17.0 & 0.915 & 0.6331 &
0.75$\pm$0.02\tablenotemark{b}	&$18.54^{+0.07}_{-0.10}$&BRIJHK	\\
0454+039 & 16.5 & 1.343 & 0.8596 &
1.45$\pm$0.01	&	$20.67^{+0.03}_{-0.03}$	&	JHK	\\
0710+119 & 16.6 & 0.768
&	0.4629	&	0.62$\pm$0.06	&	$<$18.30 &	g'r'
\\ 0735+178 & 14.9 & \nodata
&	0.4240	&	1.32$\pm$0.03	&	$<$19.00
&	UBRIJHK	\\ 0843+136 & 17.8 & 1.877 & 0.6064 &
0.938$\pm$0.035\tablenotemark{c}&$19.56^{+0.11}_{-0.14}$&u'g'r'i'JHK\\
0953$-$0038 & 18.4 & 1.383 & 0.6381 &
1.668$\pm$0.080	&	$19.90^{+0.07}_{-0.09}$	&	urJHK	\\
0957+003 & 17.6 & 0.907 & 0.6720 &
1.936$\pm$0.118\tablenotemark{c}&$19.59^{+0.03}_{-0.03}$&UBRIJHK\\
1009$-$0026 & 17.4 & 1.244 & 0.8426 &
0.713$\pm$0.038	&	$20.20^{+0.05}_{-0.06}$	&	JHK	\\
\nodata & \nodata & \nodata & 0.8866 &
1.900$\pm$0.039	&	$19.48^{+0.01}_{-0.08}$	&	\nodata	\\
1009+0036 & 19.0 & 1.699 & 0.9714 &
1.093$\pm$0.111	&	$20.00^{+0.11}_{-0.05}$	&	JHK	\\
1019+309 & 17.5 & 1.319 & 0.3461 &
0.70$\pm$0.05	&	$18.18^{+0.08}_{-0.10}$	&	K	\\
1028$-$0100 & 18.2 & 1.531 & 0.6322 &
1.579$\pm$0.087	&	$19.95^{+0.05}_{-0.08}$	&	JHK	\\
\nodata & \nodata & \nodata & 0.7087 &
1.210$\pm$0.066	&	$20.04^{+0.07}_{-0.04}$	&	\nodata	\\
1047$-$0047 & 18.4 & 0.740 & 0.5727 &
1.063$\pm$0.117	&	$19.36^{+0.17}_{-0.19}$	&	JHK	\\
1048+0032 & 18.6 & 1.649 & 0.7203 &
1.878$\pm$0.063	&	$18.78^{+0.18}_{-0.48}$	&	u'g'r'i'JHK\\
1107+0048 & 17.5 & 1.392 & 0.7404 &
2.952$\pm$0.025	&	$21.00^{+0.02}_{-0.05}$	&	ugriJHK	\\
1109+0051 & 18.7 & 0.957 & 0.4181 &
1.361$\pm$0.105	&	$19.08^{+0.22}_{-0.38}$	&	g'r'i'JHK\\
\nodata & \nodata & \nodata & 0.5520 &
1.417$\pm$0.085	&	$19.60^{+0.10}_{-0.12}$	&	\nodata	\\
1209+107 & 17.8 & 2.193 & 0.3930 &
1.00$\pm$0.07	&	$19.46^{+0.08}_{-0.08}$	&	u'g'r'i'JHK\\
\nodata & \nodata & \nodata & 0.6295 &
2.619$\pm$0.083\tablenotemark{c}&$20.30^{+0.18}_{-0.30}$&\nodata\\
1225+0035 & 18.9 & 1.226 & 0.7730 &
1.744$\pm$0.138	&	$21.38^{+0.11}_{-0.12}$	&	JHK	\\
1226+105 & 18.5 & 2.305 & 0.9376 &
1.646$\pm$0.110\tablenotemark{c}&$19.41^{+0.12}_{-0.18}$&UBRJHK	\\
1323$-$0021 & 18.2 & 1.390 & 0.7160 &
2.229$\pm$0.071	&	$20.54^{+0.15}_{-0.15}$	&	ugriJHK	\\
1342$-$0035 & 18.2 & 0.787 & 0.5380 &
2.256$\pm$0.068	&	$19.78^{+0.12}_{-0.14}$	&	ugriJHK	\\
1345$-$0023 & 17.6 & 1.095 & 0.6057 &
1.177$\pm$0.049	&	$18.85^{+0.15}_{-0.24}$	&	u'g'r'i'JHK\\
1354+258 & 18.0 & 2.006 & 0.8585 &
1.176$\pm$0.076\tablenotemark{c}&$18.57^{+0.07}_{-0.08}$&BRIJK	\\
\nodata & \nodata & \nodata & 0.8856 &
0.489$\pm$0.069\tablenotemark{c}&$18.76^{+0.10}_{-0.13}$&\nodata\\
1419$-$0036 & 18.3 & 0.969 & 0.6238 &
0.597$\pm$0.069	&	$19.04^{+0.07}_{-0.14}$	&	HK	\\
\nodata & \nodata & \nodata & 0.8206 &
1.145$\pm$0.057	&	$18.78^{+0.26}_{-0.23}$	&	\nodata	\\
1426+0051 & 18.8 & 1.333 & 0.7352 &
0.857$\pm$0.080	&	$18.85^{+0.03}_{-0.03}$	&	u'g'r'i'JHK\\
\nodata & \nodata & \nodata & 0.8424 &
2.618$\pm$0.125	&	$19.65^{+0.09}_{-0.07}$	&	\nodata	\\
1431$-$0050 & 18.1 & 1.190 & 0.6085 &
1.886$\pm$0.076	&	$19.18^{+0.30}_{-0.27}$	&	ugriJHK	\\
\nodata & \nodata & \nodata & 0.6868 &
0.613$\pm$0.066	&	$18.40^{+0.06}_{-0.08}$	&	\nodata	\\
1436$-$0051 & 18.5 & 1.275 & 0.7377 &
1.142$\pm$0.084	&	$20.08^{+0.10}_{-0.12}$	&	u'g'r'i'JHK\\
\nodata & \nodata & \nodata
&	0.9281	&	1.174$\pm$0.065	&	$<$18.82
&	\nodata	\\ 1437+624 & 19.0 & 1.090
&	0.8723	&	0.71$\pm$0.09	&	$<$18.00
&	K	\\ 1521$-$0009 & 19.0 & 1.318 & 0.9590 &
1.848$\pm$0.096	&	$19.40^{+0.08}_{-0.14}$	&	u'r'g'i'J\\
1525+0026 & 17.0 & 0.801 & 0.5674 &
1.852$\pm$0.035	&	$19.78^{+0.07}_{-0.08}$	&	BRIJHK	\\
1622+239 & 17.5 & 0.927 & 0.6561 &
1.471$\pm$0.050\tablenotemark{c}&$20.36^{+0.07}_{-0.08}$	&K\\
\nodata & \nodata & \nodata & 0.8913 &
1.622$\pm$0.042\tablenotemark{c}&$19.23^{+0.02}_{-0.03}$&\nodata\\
\hline
\end{tabular}
\end{table*}

\begin{table*}
	\addtocounter{table}{-1} \small
\caption{Continued}
\begin{tabular}{lcccccl}
\hline \hline Quasar & Mag.\tablenotemark{a} & $z_{\rm em}$ & MgII
$z_{\rm abs}$ & MgII $W^{\lambda2796}_{0}$ (\AA) & $\log
N_{HI}$(cm$^{-2}$) & Filters \\ \hline

1704+608 & 15.3 & 0.371 & 0.2220 &
0.562$\pm$0.013\tablenotemark{b}&$18.23^{+0.05}_{-0.05}$&IJK	\\
1714+5757 & 18.6 & 1.252 & 0.7481 &
1.099$\pm$0.084	&	$19.23^{+0.17}_{-0.33}$	&	ugriJHK	\\
1715+5747 & 18.3 & 0.697 & 0.5579 &
1.001$\pm$0.067	&	$19.18^{+0.15}_{-0.18}$	&	u'g'r'i'JHK\\
1716+5654 & 19.0 & 0.937 & 0.5301 &
1.822$\pm$0.130	&	$19.98^{+0.20}_{-0.28}$	&	ugri'JH	\\
1722+5442 & 18.8 & 1.215 & 0.6338 &
1.535$\pm$0.098	&	$19.00^{+0.30}_{-0.22}$	&	u'g'r'i' \\
1727+5302 & 18.3 & 1.444 & 0.9448 &
2.832$\pm$0.070	&	$21.16^{+0.04}_{-0.05}$	&	u'g'r'i'JHK\\
\nodata & \nodata & \nodata & 1.0312 &
0.922$\pm$0.057	&	$21.41^{+0.03}_{-0.03}$	&	\nodata	\\
1729+5758 & 17.5 & 1.342 & 0.5541 &
1.836$\pm$0.046	&	$18.60^{+0.18}_{-0.43}$	&	u'g'r'i'JHK\\
1733+5533 & 18.0 & 1.072 & 0.9981 &
2.173$\pm$0.069	&	$20.70^{+0.04}_{-0.03}$	&	u'g'r'i'HK\\
1857+566 &	17.3	&	1.578	&	0.7151	&	0.65
&	$18.56^{+0.05}_{-0.06}$	&	UBRIJHK	\\ 2149+212
&	19.0	&	1.538	&	0.9114	&	0.72
&	$20.70^{+0.08}_{-0.10}$	&	UBRIJK	\\ \nodata
&	\nodata	&	\nodata	&	1.0023	&	2.46
&	$19.30^{+0.02}_{-0.05}$	&	\nodata	\\ 2212$-$299 & 17.4 &
2.706 & 0.6329 &
1.15$\pm$0.02\tablenotemark{b}&	$19.75^{+0.03}_{-0.03}$	&BRJK	\\
2223$-$052 & 18.4 & 1.404 & 0.8472 &
0.586$\pm$0.012\tablenotemark{b}&$18.48^{+0.41}_{-0.88}$&BI	\\
2328+0022 & 17.9 & 1.308 & 0.6519 &
1.896$\pm$0.077	&	$20.32^{+0.06}_{-0.07}$	&	g'r'i'JHK\\
2334+0052 & 18.2 & 1.040 & 0.4713 &
1.226$\pm$0.107	&	$20.65^{+0.12}_{-0.18}$	&	g'r'i'JHK\\
2353$-$0028 & 17.9 & 0.765 & 0.6044 &
1.601$\pm$0.082	&	$21.54^{+0.15}_{-0.15}$	&	JHK	\\
\hline
\vspace{-0.5cm}
\end{tabular}
\tablenotetext{a}{Here we provide $m_V$ for quasars that pre-date SDSS
(these can be identified by their 3-digit, 1950,  Dec designation),
and $(m_g+m_r)/2$ (approximately $m_V$) for SDSS quasars, which can be
identified by their 4-digit,  2000, Dec designation. For consistency
with our earlier work as well as for historical reasons, 1950 quasar
names have  not been altered to reflect 2000 co-ordinates. }
\tablenotetext{b}{Measurements of $W^{\lambda2796}_{0}$ have been
changed from RTN06 values to reflect the more  recent measurements of
Mathes et al. (in preparation).}  \tablenotetext{c}{Measurements of
$W^{\lambda2796}_{0}$ have been changed from RTN06 values to reflect
the more  recent measurements of Quider et al. (2011).}
\end{table*}

\begin{figure*}
\begin{minipage}[c]{0.95\textwidth}
\includegraphics[angle=0,width={0.9\textwidth}]{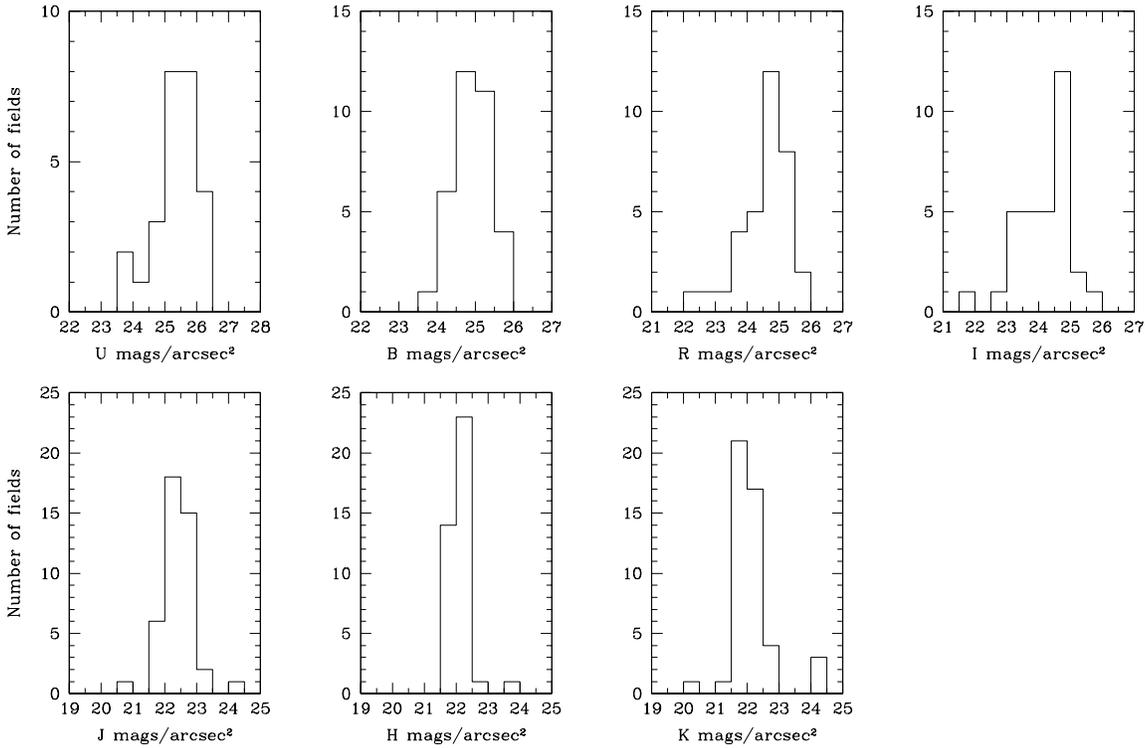}
\caption{Distribution of 3$\sigma$ surface brightness limits reached
in the final reduced images for each of the seven filters.}
\label{allSBhist}
\end{minipage}
\end{figure*}

We were able to obtain a complete optical and infrared  ({\it UBRIJHK}
or {\it ugriJHK}) dataset for 18 of the 60 fields.  The largest number
of fields, 53, were observed in {\it K}, while the fewest, 26, were
observed in {\it U} (or {\it u}).  Figure \ref{allSBhist} gives the
distribution of 3$\sigma$  surface brightness limits reached in each
filter.  The infrared data show the least spread since almost all were
obtained with the IRTF NSFCAM, with the exception of two observations
with the (low-sensitivity) IRTF SpeX guide camera.

\begin{figure*}
\includegraphics[angle=0,width={0.9\columnwidth}]{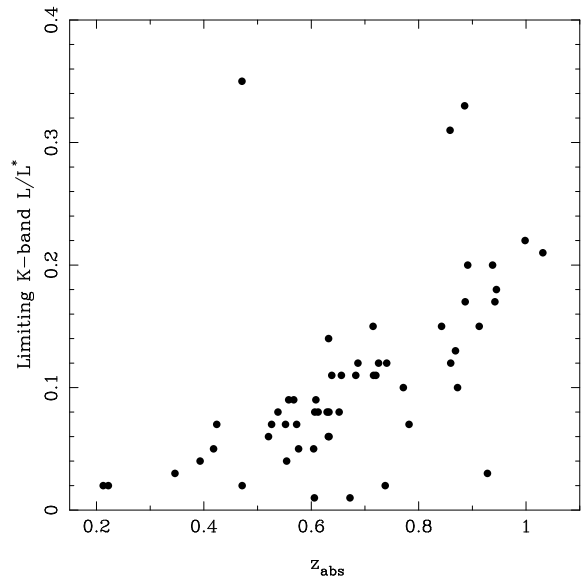}
\caption{{\it K}-band luminosity of a fiducial 10-kpc sized galaxy
that would be detectable at the redshift of the absorber as a function
of redshift for our dataset.  The luminosity is expressed in terms of
$L_K^*$ and is estimated from the 3$\sigma$ limiting surface
brightness  achieved for each field. K-corrections appropriate for an
Sb type galaxy have been applied.}
\label{LKvsz}
\end{figure*}

Since the {\it K}-band data are the most extensive as well as uniform,
we illustrate a few properties of the dataset using the {\it K}-band
data  sample.   Figure \ref{LKvsz} is a plot of the {\it K}-band
galaxy luminosity limit  (in terms of $L_K^*$, where $M_K^* = -22.86$
for $z<1$, Cirasuolo et al. 2006) as a function of redshift. Here, the
$3\sigma$ limiting {\it K}-band surface brightness was used to
estimate the luminosity of a fiducial 10 kpc-sized galaxy that can be
detected at the redshift of the  absorber. K-corrections appropriate
for an Sb type  galaxy have been applied to all datapoints.

\begin{figure*}
\includegraphics[angle=0,width={0.9\textwidth}]{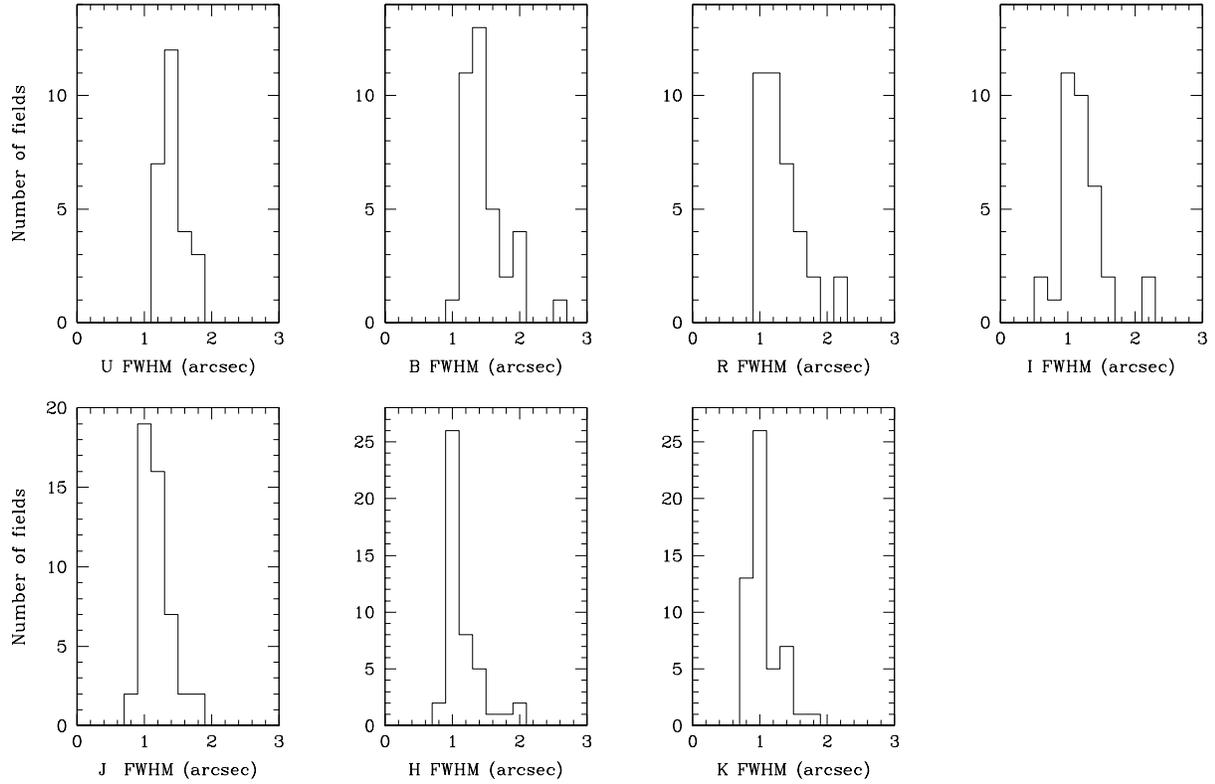}
\caption{Distribution of seeing values obtained for observations in
each filter. FWHM of  point sources in the final reduced images are
reported.}
\label{allseehist}
\end{figure*}

\begin{figure*}
\includegraphics[angle=0,width={0.45\textwidth}]{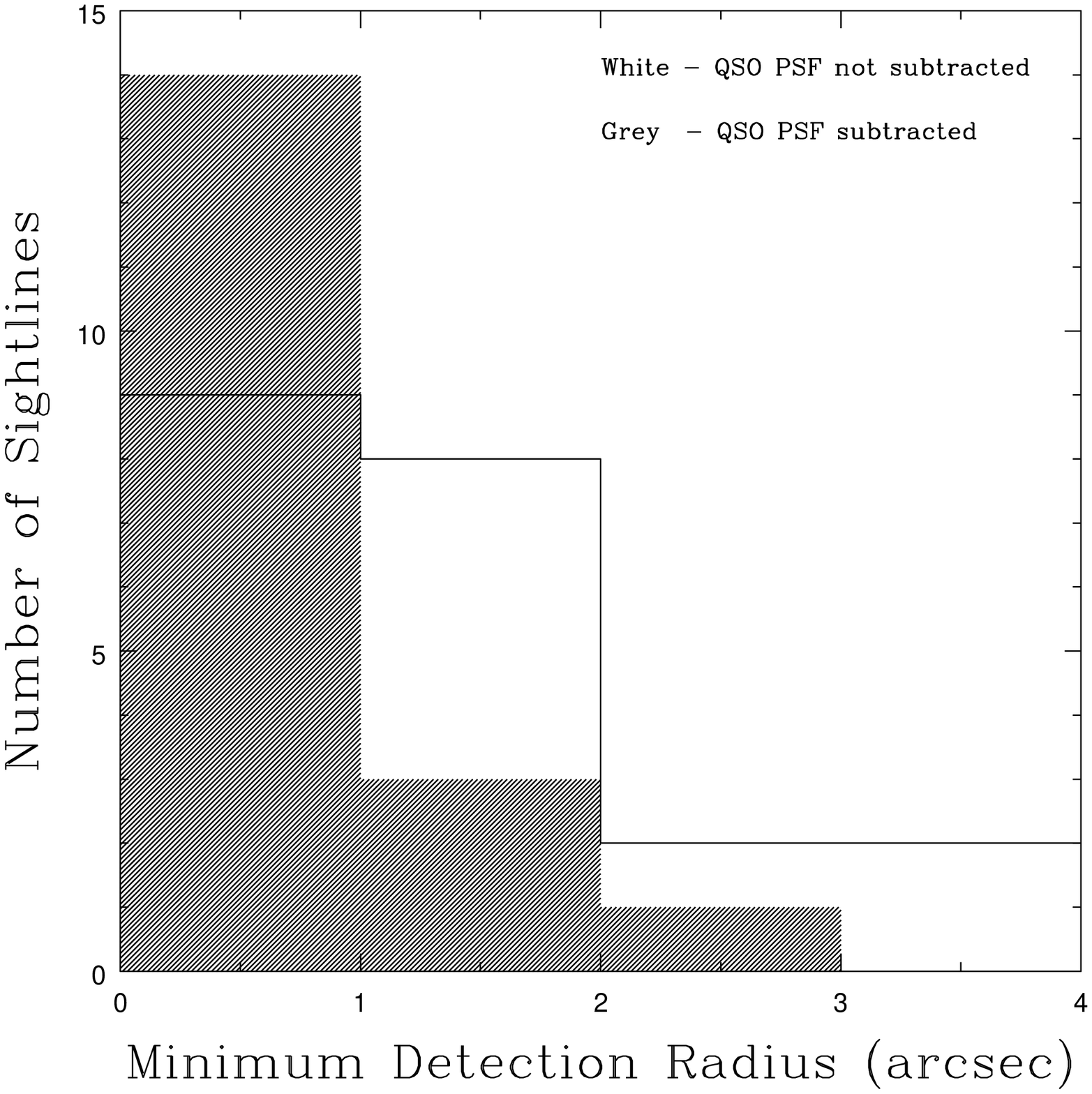} \hfil
\includegraphics[angle=0,width={0.45\textwidth}]{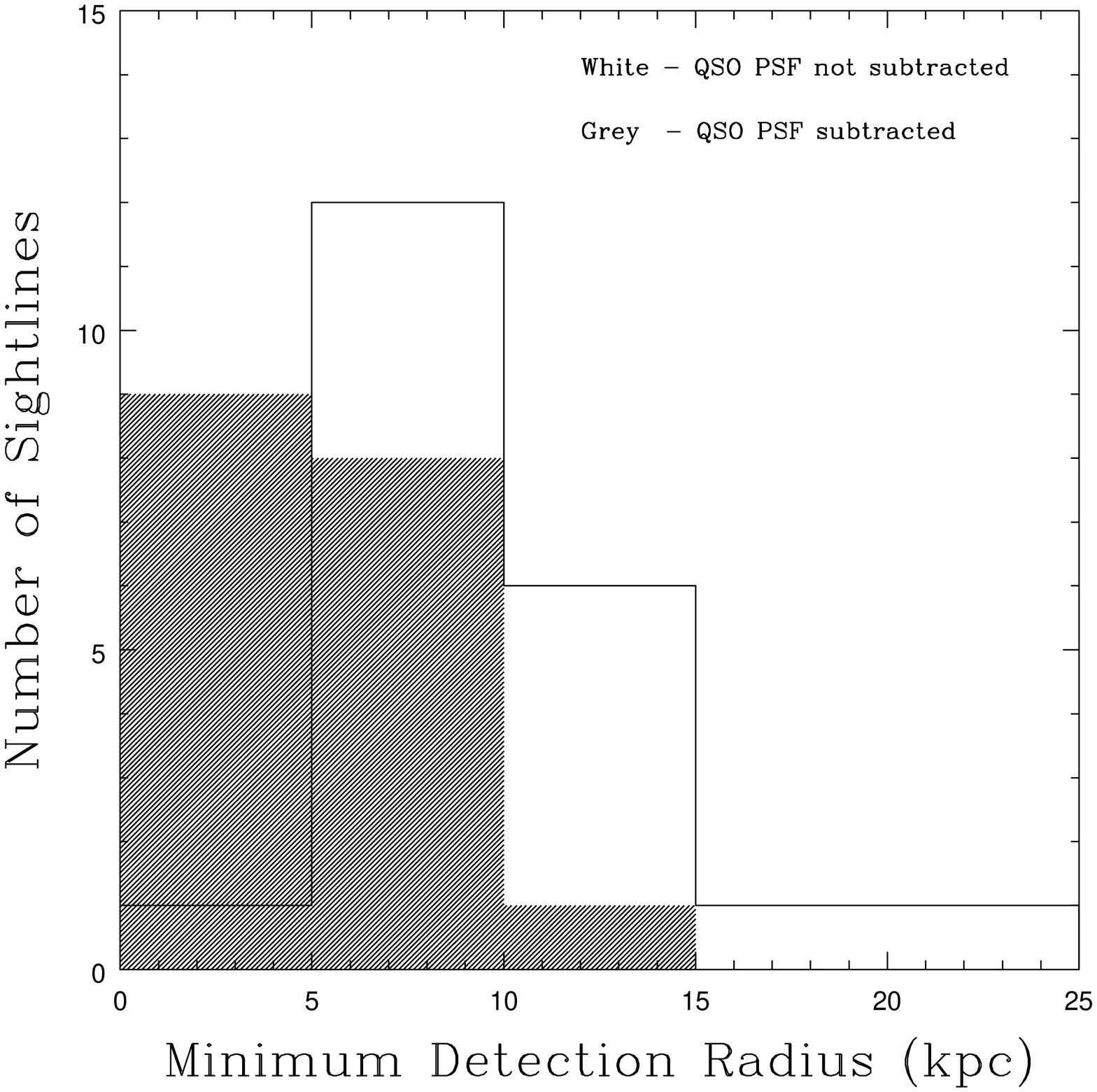}
\caption{Distribution of minimum impact parameters from the quasar
sightline  for detection of galaxies in arcsec (left) and in kpc
(right) at the redshift of the absorber.  Fields for which the quasar
PSF was subtracted are represented by the grey  histogram. Not all
fields had point sources (other than the quasar)  that could be used
to model the PSF, therefore, the quasar PSF was not subtracted in
these fields. The minimum impact parameter for galaxy detection is,
therefore, generally larger. This sample is represented by the
unshaded histogram.}
\label{impar}
\end{figure*}

The distribution of seeing values for our data, expressed in terms of
the FWHM of a point source, is shown in Figure \ref{allseehist}.  The
seeing is a particularly important parameter for this study because
the quasar point spread function (PSF) limits our ability to study the
smallest impact parameters. For ground-based  imaging, techniques such
as adaptive optics (AO) achieve seeing values down to a few tenths of
an arcsecond. However, the presence of a bright (12$^{\rm th}$  to
15$^{\rm th}$ magnitude), nearby (within 30\arcsec), point source is
generally required for implementing AO techniques. Since the (faint)
quasar is the brightest object  in the majority of our fields, we
could not take advantage of AO. From Figure \ref{allseehist} it  can
be seen that the optical dataset has seeing values generally $\ga$
1\arcsec, while for the infrared data, seeing values $\la$ 1\arcsec\
were often achieved.  We were able to probe smaller impact parameters
than the seeing radius for images where the quasar PSF could be
subtracted. However, this was not possible for all fields, because
suitable PSF stars were not always available within the image, and we
did not observe PSF stars separately.  In Figure \ref{impar} we plot
minimum detection impact parameter histograms for the {\it K}-band
images in arcsec as well as in kpc at the redshift of the
absorbers. The minimum detection impact parameter for fields where the
quasar PSF could not be subtracted is conservatively taken to be the
radius at which the PSF blended with the background. This  estimate
depends on the brightness of the quasar as well as on the seeing.  For
PSF-subtracted fields, the minimum detection impact parameter is
measured as the  radius of the mask that was applied to the
subtraction residuals.  The average minimum impact parameters for all
fields is $7.3\pm2.7$ kpc. The average minimum impact parameter for
PSF-subtracted fields is $5.2\pm2.3$ kpc, and for non-PSF subtracted
fields, it is $9.1\pm3.0$ kpc.  We note here that whether the quasar
PSF was, or was not,  subtracted in our ground-based data has not
influenced the identification of  absorber galaxies (\S 3 and \S
4). For our $K$-band sample, we find that the  distribution of impact
parameters of absorber galaxies identified in  our PSF-subtracted
fields and non-PSF-subtracted fields are similar, with
Kolmogorov-Smirnov (KS) test probability $P_{KS}=0.93$.

The identification of galaxies causing absorption in quasar spectra
is, undoubtedly, best achieved from space. Not being able to probe to
within 5 or 10 kpc of the quasar sightline is the most severe
limitation  of a groundbased imaging programme. Low luminosity dwarf
galaxies directly along the quasar sightline will most likely be
missed, resulting in an incorrect identification of the absorbing
galaxy. Below, we attempt to quantify the possible number of missed
galaxies in our sample due to this  bias.

\section{Identification of Absorbing Galaxy Candidates}

The detection and photometry of sources were carried out using the
automated software SExtractor (Bertin \& Arnouts 1996).   The
SExtractor input parameter that defines the detection threshold for
source identification was set to 1$\sigma$ above the sky background,
and the  minimum detection area  was set to 5 adjoining pixels.
``Adjoining'' as implemented in  SExtractor refers to any pixels
touching at corners or sides. A source is considered to be a confident
detection if it was detected at the $2\sigma$ or higher level through
more than one filter. Its position was determined using the  image
with the best seeing.

\begin{figure*}
\includegraphics[angle=0,width={0.9\columnwidth}]{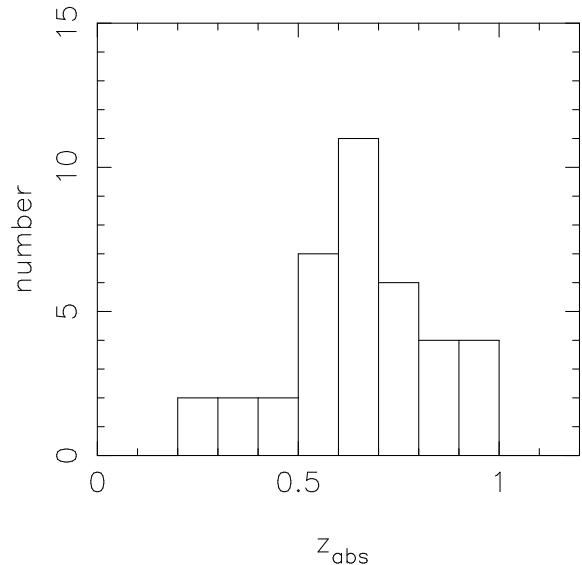}
\caption{Redshift distribution of absorbers in quasar fields with
$K$-band  surface brightness limits between 21.5 and 22.5 magnitudes
per square arcsec.}
\label{zabsK}
\end{figure*}

\begin{figure*}
\includegraphics[angle=0,width={0.9\columnwidth}]{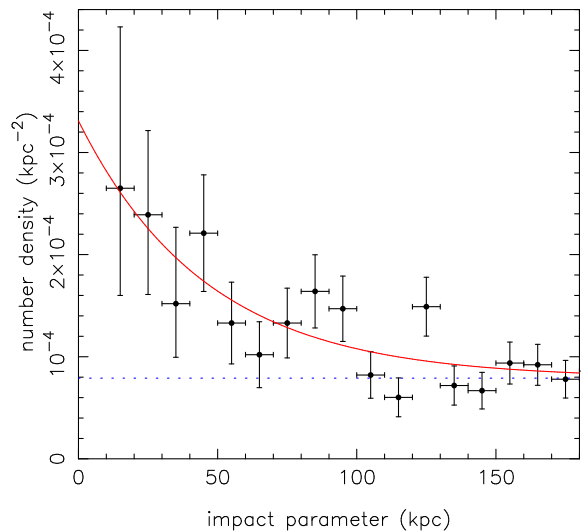}
\caption{The number density of galaxies as a function of impact
parameter from the quasar, calculated in annuli of width 10 kpc. Only
$K$-band images with surface brightness limits between $21.5$ and
$22.5$ magnitudes per square arcsec for absorbers between redshifts 0.5
and 0.8, where most of the  absorption systems lie, were used.  The
red line is an exponential fit to the data points with an e-folding
length of 46.1 kpc. The blue dotted line shows the background plus 
foreground galaxy  number
density of $7.9 \times 10^{-5}$ kpc$^{-2}$. The discrepant point at
an impact parameter of $\approx$ 125 kpc notwithstanding, galaxies beyond 
$\approx$ 100 kpc can, statistically, be considered background or foreground
galaxies that are not associated with the absorption system.}
\label{Knumden}
\end{figure*}

Figure \ref{allSBhist} shows that most of the $K$-band data (38
fields) reach surface brightnesses between 21.5 and 22.5 $K$
magnitudes per square  arcsec at the 3$\sigma$ level. Figure
\ref{zabsK} gives the redshift distribution of the 38 absorbers in
these fields, 24 of which have redshifts $0.5<z<0.8$. This is  a small
enough redshift interval that we use a single value for the angular
diameter distance to estimate the surface density of galaxies.  We
then use this sample to estimate: (1) the background (and foreground)
number density of galaxies, (2) the  excess around the quasar line of
sight that can be attributed to the presence of an absorbing galaxy or
galaxies associated with it, and (3) the number of absorbing galaxies
that might have been missed due to the glare of the quasar PSF. Figure
\ref{Knumden} shows the number  of galaxies per square kpc as a
function of impact parameter from the  quasar calculated in annuli of
width 10 kpc. The red line is the best-fit exponential profile to the
data with an {\it e}-folding length of 46.1 kpc.  The horizontal asymptote,
which occurs at $7.9\times 10^{-5}$ kpc$^{-2}$,  is shown by the blue
dotted line. It represents the background plus foreground  galaxy
number density. Galaxies beyond $b \approx 100$ kpc can  be considered
background or foreground galaxies that are not associated with the
absorption systems. Moreover, assuming that the  distribution can be
extrapolated to impact parameter $b=0$ gives an  estimate of the
number of galaxies unaccounted for  due to the presence of the  quasar
PSF and the inability to subtract it perfectly. This suggests that the
expected number density at $b<10$ kpc is $\approx 3.1(10^{-4})$
galaxies per square kpc per quasar field, and that  $\approx 0.07$
galaxy candidates per field might have been missed. Or, on average,
one in every 14 fields may have a galaxy at $b<10$ kpc that is not
identify in our groundbased imaging survey. This amounts to
approximately four among the 55 identified candidate  galaxies in our
survey (\S 4).

Based on the above analysis, only objects within an impact parameter
$b=100$ kpc from the quasar at the absorber redshift (or lowest
absorber redshift in the case  of multiple absorbers per quasar sightline) 
are catalogued for each field,  since galaxies farther away
can statistically be considered background or foreground galaxies.

The absorbing galaxy has not been confirmed spectroscopically for any
of the fields presented here. A spectroscopic redshift that matches
the absorption redshift would, of course, lead to a more confident
identification of the absorbing galaxy (or a parcel of gas associated
with it).  In the absence of spectroscopic data, we assign a galaxy as
a ``candidate absorber'' with varying levels of confidence based on
several criteria.  The highest level of confidence is achieved when a
galaxy's photometric  redshift matches that of the \mgii\
absorption-line system within the uncertainties. Photometric redshifts
were determined for galaxies that were detected in four or more
filters.\footnote{We sometimes used SDSS  photometry to supplement our
IR measurements. Details of our {\it photo-z} technique and  the
galaxy templates used are described in the Appendix.} If more than one
galaxy in the field was determined to have a photometric redshift that
matched the absorption redshift, then the one closest to the quasar
sightline was selected as the candidate absorber. Next, if photometric
redshifts could not be determined (e.g., if a galaxy is detected in
fewer than  four filters), then we judged whether or not  the galaxy's
colours were consistent with it being at the absorption redshift. This
was done by comparing our measured galaxy  colours with the colours
derived from the redshifted 'hyperz'  galaxy templates of Hewett et
al. (2006), after converting  our AB magnitudes to Vega
magnitudes. Lastly, if no colour information was available, or if a
galaxy's colours were inconclusive, then the ``proximity criterion''
was used, whereby  the galaxy closest to the quasar sightline was
selected as the candidate absorber.  For sightlines with two
absorbers, assignment of the absorbing galaxies was often
ambiguous. Depending on the specifics of the field, we were sometimes
unable to assign a galaxy to the absorber. In addition, some fields
were observed under non-photometric conditions, while for others no
calibration information was available.  Although photometry could not
be carried out for the objects in these fields,  impact parameter
information could nevertheless be extracted. The proximity criterion
was employed in these cases as well. These galaxies are  not part of
the statistical sample analyzed here since no luminosity information
exists for them.

We assign a ``CL'' value, or confidence level, for each galaxy
identification. Galaxies which have been confidently identified
through  photometric redshifts that match the absorption redshift  are
labeled as having CL = 1. Identifications which were made based on
colours that were consistent with a galaxy being at the absorption
redshift, the proximity criterion, or photometric-redshift matches
that were only marginally consistent with the absorption redshift are
assigned confidence level CL =  2 or 3, with 2 being the more
confident identification.  No galaxy identification was possible for a
few fields.  For example, this may happen if an absorber redshift does
not match the photometric redshift of any of the galaxies in a given
field, or when galaxy colours are ambiguous or are consistent with a
large redshift range. These fields are not assigned a CL value.

\subsection{Examples}

We now illustrate our process of absorbing galaxy identification with
a few  representative examples that include most of the issues we
faced while assigning galaxies to absorption systems. The images and
photometry  for all objects in our sample are available on line at
http://enki.phyast.pitt.edu/Imaging.php.  We also provide results from
photometric redshift and stellar population synthesis  template fits,
details of which are explained in the Appendix.  Readers who are not
interested in the details of galaxy selection can  skip to \S4.

Here we provide our reduced images, photometry tables, and photometric
redshift fits and derived stellar population synthesis parameters for
four fields. Sources detected within 100 kpc  of the quasar sightline
at the absorption redshift (or smallest absorption redshift for
multiple absorbers along the same sightline) are numbered in order of
increasing impact parameter from the quasar, and ellipses are drawn
around sources in each image only to guide the eye. Photometry tables
give positions relative to the quasar, AB magnitudes, and the
detection significance, ``DS'', which is defined as the number of
standard deviations the source is detected above the background.  $DS
= S/(B\times N_{pix})$,  where $S$ is the net source counts, $B$ is
the counts per pixel that correspond to a source detected at 1$\sigma$
above the background, and $N_{pix}$ is the number of pixels within the
detection isophote. A source is considered to be a detection if $DS
\geq 2$ and $N_{pix} \geq 5$. Tables describing  photometric redshift
fits give details of the stellar population templates that best fit
the photometry.  The information provided includes object number as
marked on the images  and its projected distance from the quasar in
arcsec and kpc assuming that the galaxy is at the absorption redshift,
age of the stellar population, star formation rate {\it e}-folding
time, $\tau$, extinction, $E(B-V)$, metal mass fraction, $Z$ ($Z_\odot
= 0.2$),  and the photometric redshift and error.

\subsubsection{Example 1: the 0153+0009 field}

This is an example of our highest level of confidence for absorbing
galaxy  identification, where the photometric  redshift of the galaxy
with the smallest impact parameter to the quasar matches the redshift
of the absorption-line system.

\begin{figure*}
\includegraphics[height=2.95in,height=2.95in]{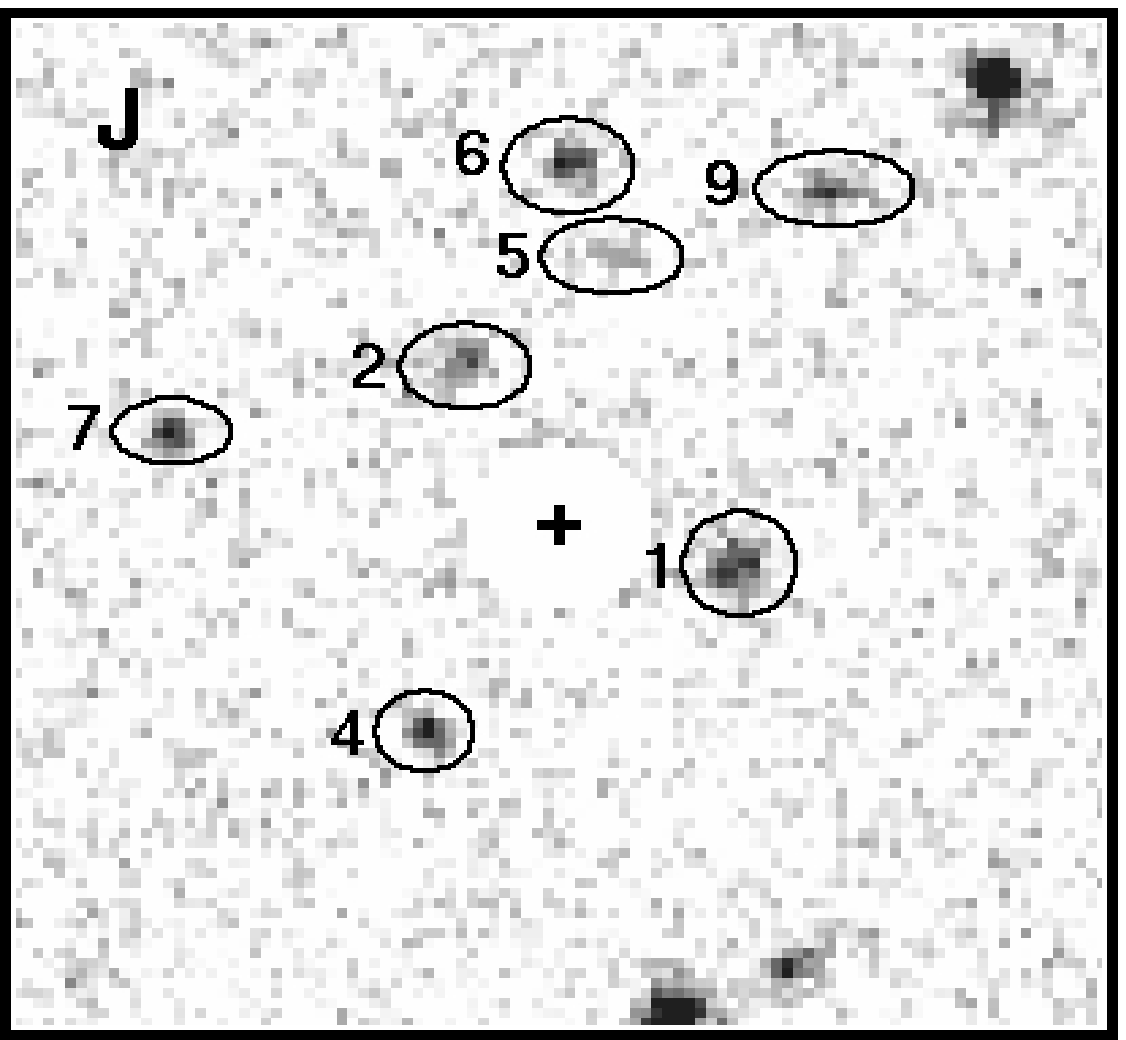}
\includegraphics[height=2.95in,height=2.95in]{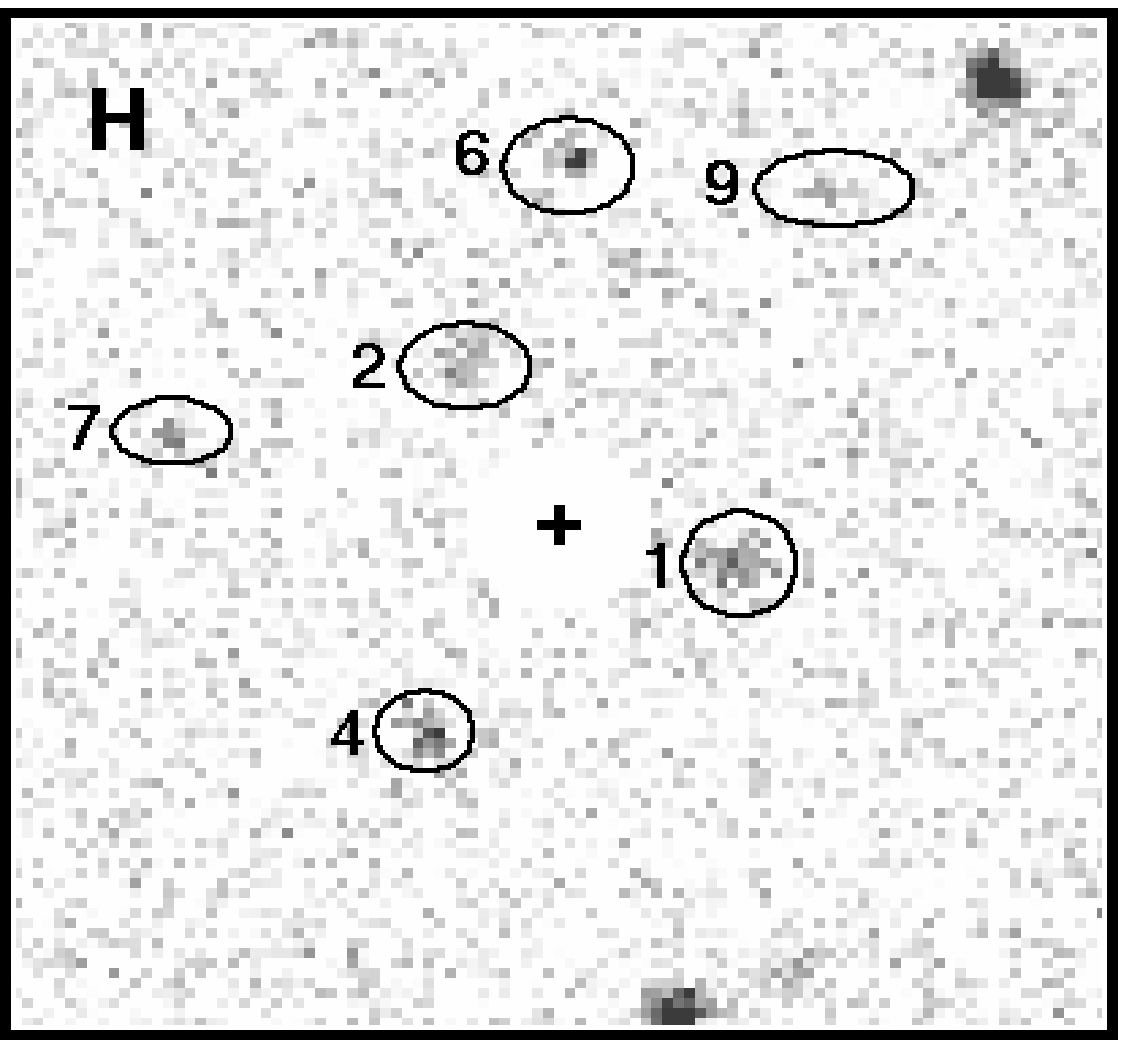}
\includegraphics[height=2.95in,height=2.95in]{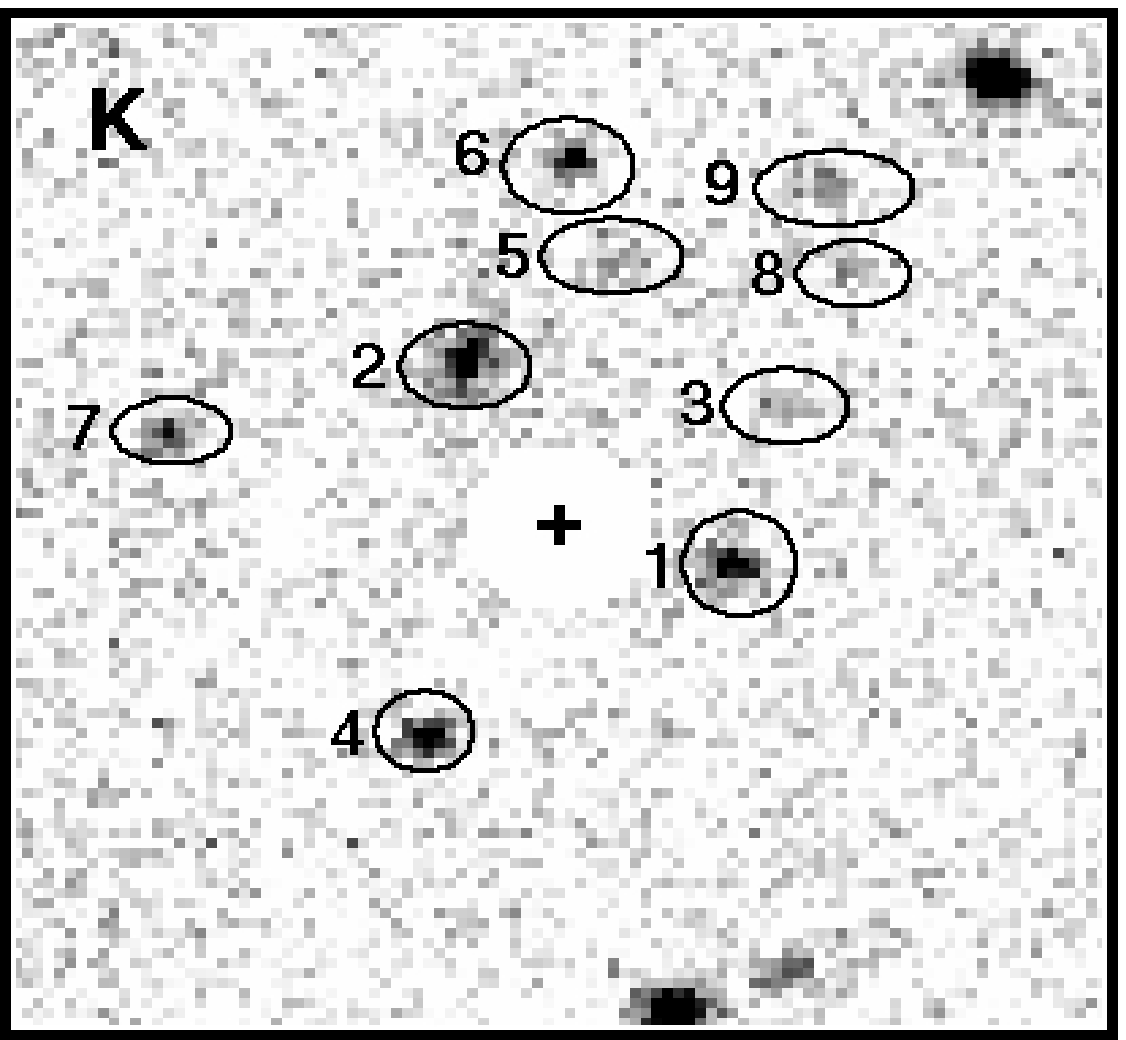}
\caption{30$\arcsec \times$ 30$\arcsec$ $J, H, K$ images of the
0153+0009 field.  This field has a subDLA system at $z_{abs}$ =
0.7714. As discussed in the text (\S3.1.1), we identify Object 1 as the absorbing 
galaxy. The images shown above correspond to $\approx$ 222 $\times$
222 kpc$^{2}$ at  the absorber redshift. The quasar has been masked in
all the frames, and its position is marked by a ``+''.  The quasar PSF
could not be subtracted, as there were no suitable PSF stars in the
field. North is up and east is to the left. Photometry for all labeled
objects is given in Table 2. Ellipses are drawn only to guide the
eye. Objects that are unmarked have impact parameters greater than 100
kpc, and are not considered to be candidate absorbers.}
\label{0153Images}
\end{figure*}

The sightline towards the quasar 0153+0009 (SDSS J015318.19+000911.3)
contains a subDLA system at $z_{abs} = 0.7714$ with a column density
of $\log {N_{HI}} = 19.70^{+0.08}_{-0.10}$ $\rm cm^{-2}$ (RTN06).  We
obtained  $J, H,$ and $K$ images of this field (see Figure
\ref{0153Images}). We have used SDSS optical photometry for this field
to supplement our infrared data; together the data were used to
determine its photometric redshift.  Our measured photometry is given
in Table \ref{0153mags}.  The quasar PSF could not be subtracted as
there were no suitable PSFs stars in the field, and so the quasar has
been masked out. We detect nine objects within 100 kpc of the quasar.

\begin{figure*}
\includegraphics[width=3.0in]{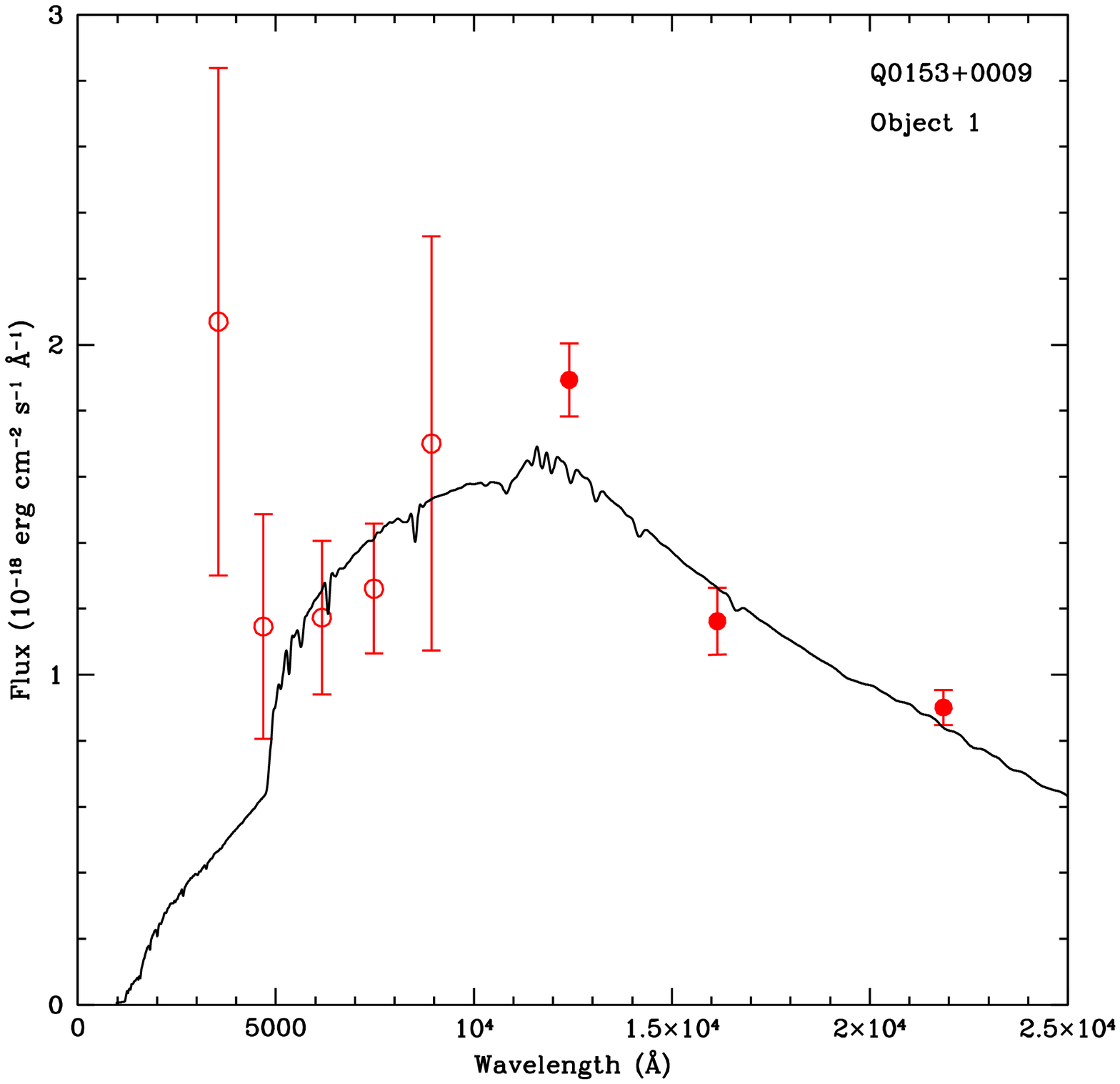}
\includegraphics[width=3.0in]{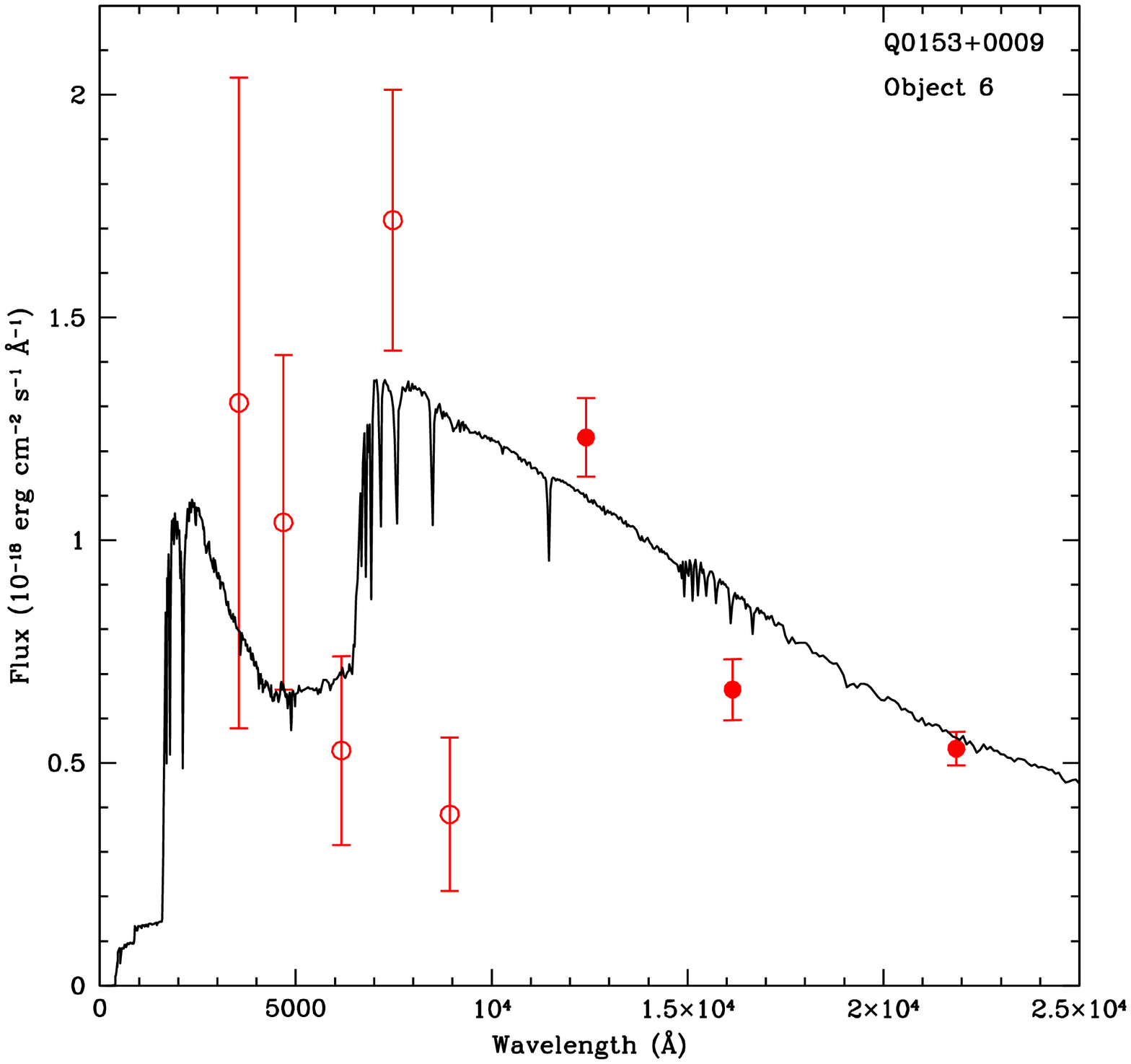}
\includegraphics[width=3.0in]{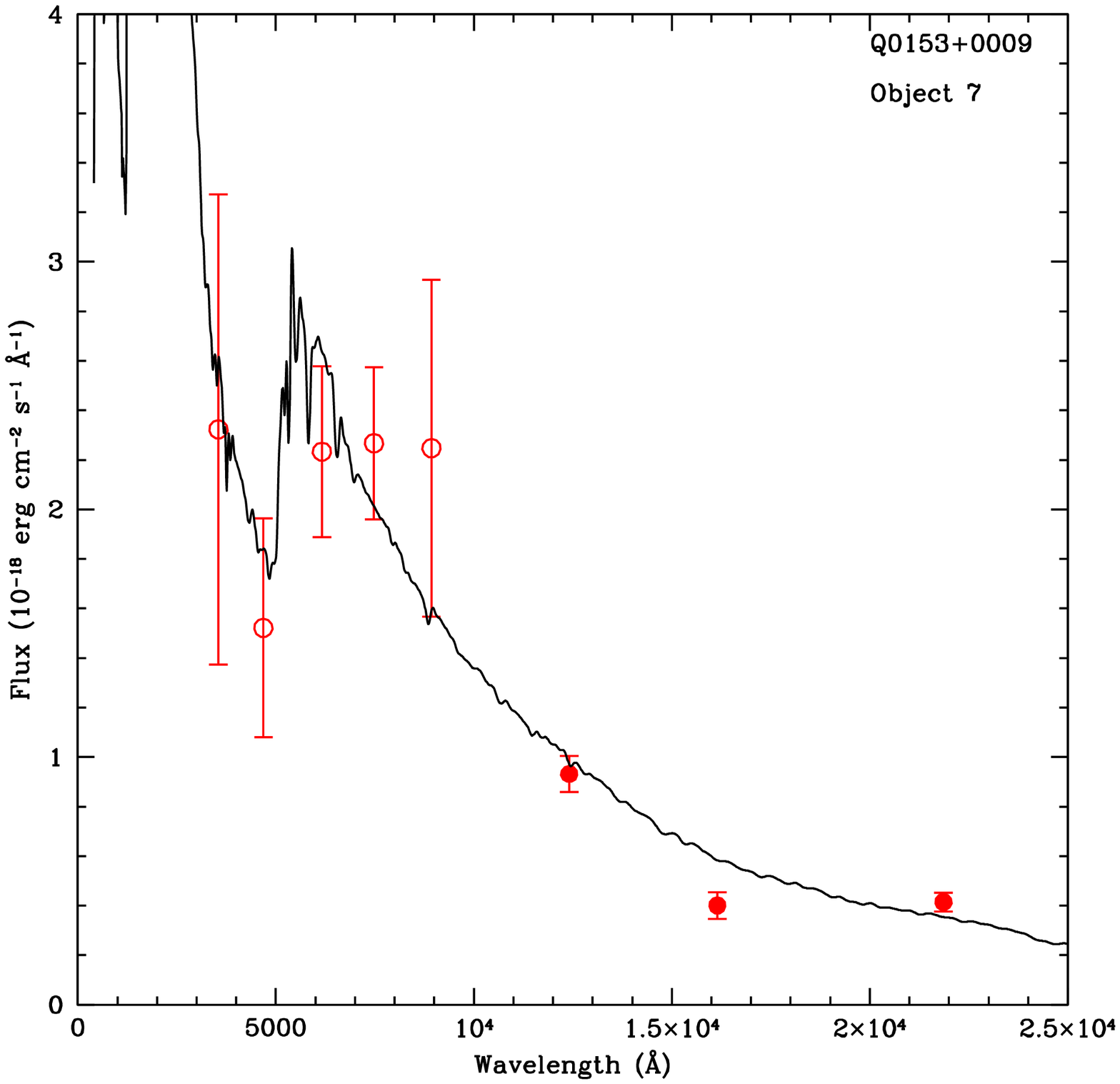}
\includegraphics[width=3.0in]{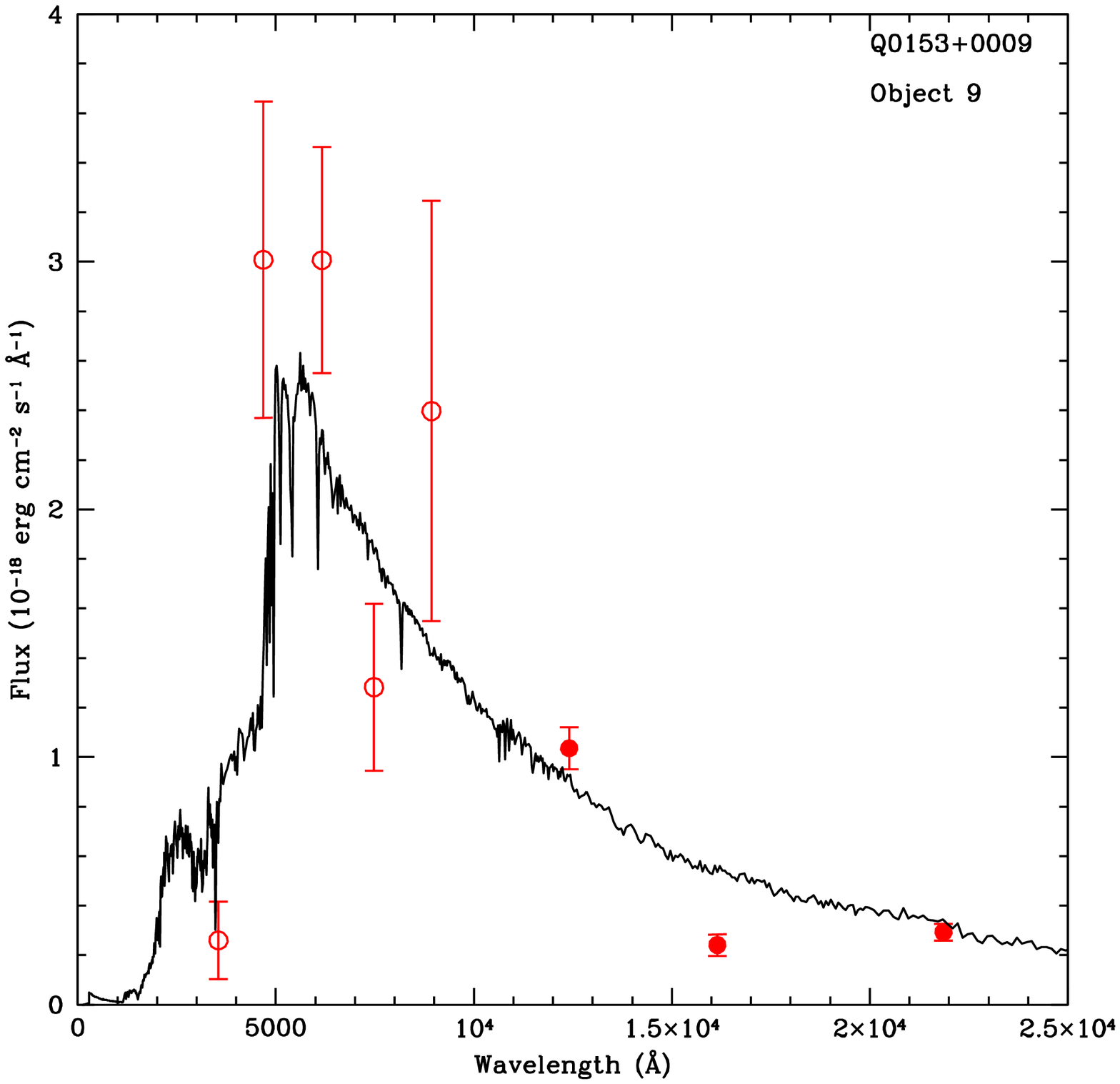}
\caption{The curves are the best-fit stellar population synthesis
models to the  photometry for Objects 1, 6, 7, and 9 in  the 0153+0009
field. The  SDSS data are shown as open circles and our infrared
photometric data, from Table \ref{0153mags}, are shown as solid
circles. See Table \ref{0153photozdata} for model details. As discussed 
in the text (\S3.1.1), we identify Object 1 as the absorbing 
galaxy. }
\label{0153photoz}
\end{figure*}

Object 1 is at $\theta$ = 4.9$\arcsec$, which is equivalent to  
36.6 kpc at the absorber redshift.
SDSS photometry for Object 1 was obtained from S. Zibetti (private
communication) since it is not in the ``{\tt photoObj}'' catalogue made
available in the SDSS database.  S. Zibetti ran his PSF subtraction
software on the SDSS image (Zibetti et al. 2007), and provided us with
the photometry of Object 1 in all five SDSS bands (Table
\ref{0153zib}). A photometric redshift of $z_{phot}$ = 0.745$\pm$0.040
is derived for Object 1 by supplementing these magnitudes with our
infrared data (see Table \ref{0153photozdata}). The stellar population
template fit is shown in Figure
\ref{0153photoz}.  This is consistent with the absorption redshift to
within the errors, and therefore, Object 1 is considered to be the
absorbing galaxy.

The $J - K$ = 1.38 colour of Object 2 is not consistent
with it being at the absorption redshift. Object 3 is
included in the photometry table because it looks real by eye.
However, based on its detection significance, {\it DS}, we do not
consider it to be a confident detection.  No
redshift information could be extracted from the IR data on Objects 4,
5, and 8; they are not detected in the SDSS images. 

 Object 6 is identified as a star in the SDSS database,
however, it is extended in our images. Objects 7 and 9 have  $z_{phot}
= 0.385 \pm 0.157$, and $0.073 \pm 0.046$ respectively, according to
the SDSS database\footnote{When available, the SDSS photometric
redshift we report is {\tt photozcc2}
(Oyaizu et al. 2008), otherwise, the SDSS photometric redshift 
labeled ``PhotoZ'' is reported.}.   The best-fit
stellar population synthesis model to our IR photometry and SDSS
optical photometry for Objects 6, 7, and 9 are shown  in Figure
\ref{0153photoz}, and the stellar population fit parameters are given in 
Table \ref{0153photozdata}. 

In summary, due to its matching photometric redshift as well as proximity, 
Object 1 is selected as the absorbing galaxy with CL = 1.
The photometric redshift derived for Object
6 is consistent with the absorption redshift, making it likely that
Objects 1 and 6 are members of the same galaxy cluster or group.

\begin{landscape}
\begin{table}
  \begin{minipage}[]{0.95\textwidth}
  \caption{0153+0009: Infrared Photometry}
  \begin{tabular}{crrrcccccc}
\hline
Object & $\Delta\alpha$\tablenotemark{a} & $\Delta\delta$\tablenotemark{a} & $\theta$\tablenotemark{a} &
$J\pm\sigma_{J}$ & DS $(N_{pix})$\tablenotemark{b} & 
$H\pm\sigma_{H}$ & DS $(N_{pix})$\tablenotemark{b} & 
$K\pm\sigma_{K}$ & DS $(N_{pix})$\tablenotemark{b}\\
 & $\arcsec$  & $\arcsec$  & $\arcsec$ & & & & & & \\
\hline
QSO& 0.0 & 0.0 & 0.0 & 17.18 $\pm$ 0.003 & 31.9 (188) & 17.58 $\pm$ 0.01 & 16.8 (120) & 17.45 $\pm$ 0.004 & 20.5 (138) \\  
1 & $-$4.7 & $-$1.4 & 4.9 & 21.43 $\pm$ 0.07 & 2.9 (41) & 21.39 $\pm$ 0.10 & 2.7 (22) & 21.01 $\pm$ 0.07 & 3.0 (35) \\ 
2 & +2.5 & +4.7 & 5.3 & 21.88 $\pm$ 0.09 & 2.3 (34) & 21.82 $\pm$ 0.13 & 1.9 (21) & 20.50 $\pm$ 0.05 & 2.7 (63) \\ 
3 & $-$6.2 & +3.4 & 7.1 & \nodata & \nodata & \nodata & \nodata & 23.22 $\pm$ 0.22 & 1.7 (8) \\
4 & +3.6 & $-$6.5 & 7.4 & 22.10 $\pm$ 0.08 & 2.9 (22) & 21.55 $\pm$ 0.11 & 2.6 (20) & 21.19 $\pm$ 0.07 & 3.5 (26) \\ 
5 & $-$1.7 & +7.7 & 7.9 & 23.09 $\pm$ 0.16 & 2.0 (13) & \nodata & \nodata & 22.95 $\pm$ 0.21 & 2.0 (9) \\ 
6 & $-$0.4 & +10.8 & 10.8 & 21.90 $\pm$ 0.08 & 3.0 (26) & 22.00 $\pm$ 0.12 & 2.9 (12) & 21.58 $\pm$ 0.08 & 3.3 (19) \\ 
7 & +10.8 & +2.6 & 11.1 & 22.20 $\pm$ 0.09 & 2.9 (20) & 22.54 $\pm$ 0.16 & 2.6 (8) & 21.85 $\pm$ 0.10 & 2.6 (19) \\ 
8 & $-$8.1 & +7.6 & 11.1 & \nodata & \nodata & \nodata & \nodata & 23.12 $\pm$ 0.20 & 2.2 (7) \\
9 & $-$7.3 & +10.0 & 12.4 & 22.09 $\pm$ 0.09 & 2.6 (25) & 23.10 $\pm$ 0.21 & 2.1 (6) & 22.23 $\pm$ 0.13 & 2.2 (16) \\
\hline
\vspace{-0.7cm}
\tablenotetext{a}{Relative to the quasar.}  
\tablenotetext{b}{DS is the
``detection significance'', and is defined as the number of sigma
above the background that the source is detected. $DS = S/(B\times N_{pix})$, 
where $S$ is the net source counts, $B$ is the counts per pixel of a source
that could be detected at 1$\sigma$ above the background, and $N_{pix}$ is the
number of pixels within the detection isophote. A source is considered
to be a detection if DS $\geq 2$ and $N_{pix} \geq 5$.}
\label{0153mags}
\end{tabular}
\end{minipage}
\end{table}

\begin{table}
  \begin{minipage}[]{0.95\columnwidth}
  \caption{0153+0009: Supplemental Photometry\tablenotemark{a} }
  \begin{tabular}{cc}
\hline
\multicolumn{2}{c}{Object 1} \\
\hline
$u'\pm\sigma_{u'}$ & 22.20$\pm$0.50 \\ 
$g'\pm\sigma_{g'}$ & 24.30$\pm$0.38 \\ 
$r'\pm\sigma_{r'}$ & 23.18$\pm$0.24 \\
$i'\pm\sigma_{i'}$ & 22.24$\pm$0.18 \\ 
$z'\pm\sigma_{z'}$ & 21.18$\pm$0.50 \\ 
\hline
\vspace{-0.7cm}
\tablenotetext{a}{SDSS photometry provided by S. Zibetti, private
communication.}
\label{0153zib}
\end{tabular}
\end{minipage}
\end{table}

\begin{table}
  \begin{minipage}[]{0.95\textwidth}
  \caption{0153+0009: Photometric Redshift Fits\tablenotemark{a}}
  \begin{tabular}{crrccccc}
\hline
\multicolumn{3}{c}{Galaxy} & \multicolumn{5}{|c}{Stellar Population Synthesis Model Parameters}\\
\hline
\# & $\theta$\tablenotemark{b} & $b$ & Age & $\tau$ & $E(B -V)$ & $Z$ & $z_{phot}\pm\sigma_{z_{phot}}$ \\
 & $\arcsec$ & kpc & Gyr & Gyr & & \\
\hline
1 & 4.9 & 36.6 & 12.0 & 5.00 & 1.00 & 0.0040 & 0.745$\pm$0.040 \\ 
6 & 10.8 & 80.1 & 5.00 & 12.0 & 0.30 & 0.0004 & 0.745$\pm$0.113 \\ 
7 & 11.1 & 82.1 & 3.00 & 3.00 & 0.00 & 0.0040 & 0.344$\pm$0.294 \\  
9 & 12.4 & 92.2 & 1.00 & 0.10 & 0.00 & 0.0080 & 0.244$\pm$0.157 \\
\hline
\label{0153photozdata}
\vspace{-0.7cm}
\tablenotetext{a}{$z_{abs} = 0.7714$}
\tablenotetext{b}{Relative to the quasar}
\end{tabular}
\end{minipage}
\end{table}
\end{landscape}

\subsubsection{Example 2: the 0735+178 field}

This is an example of a field where the three objects with the
smallest impact parameters are ruled out as absorbing galaxy
candidates. The fourth closest object is the best candidate for the
absorbing galaxy.

\begin{figure*} 
\begin{center}
\includegraphics[height=2.95in,height=2.95in]{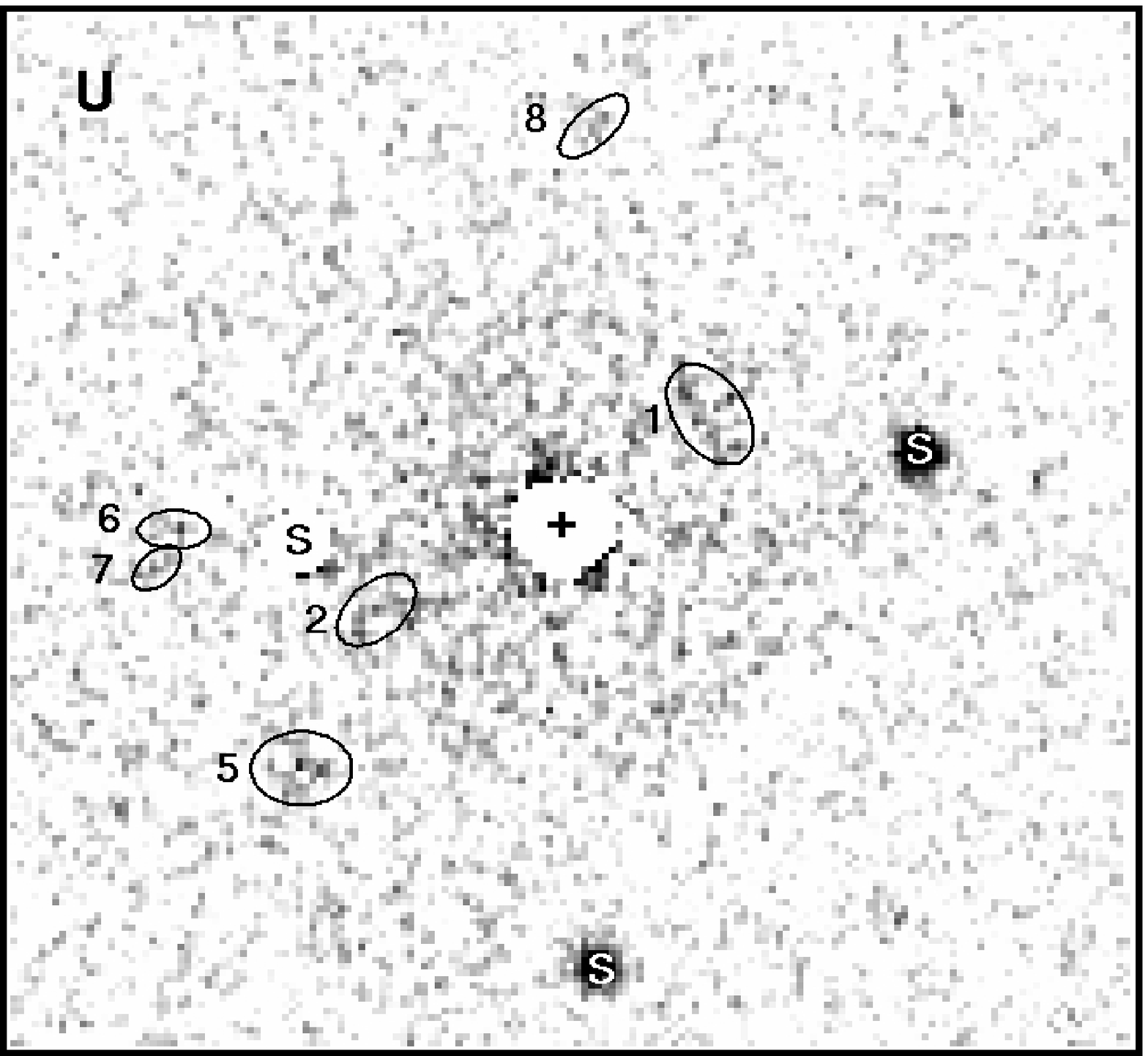}
\includegraphics[height=2.95in,height=2.95in]{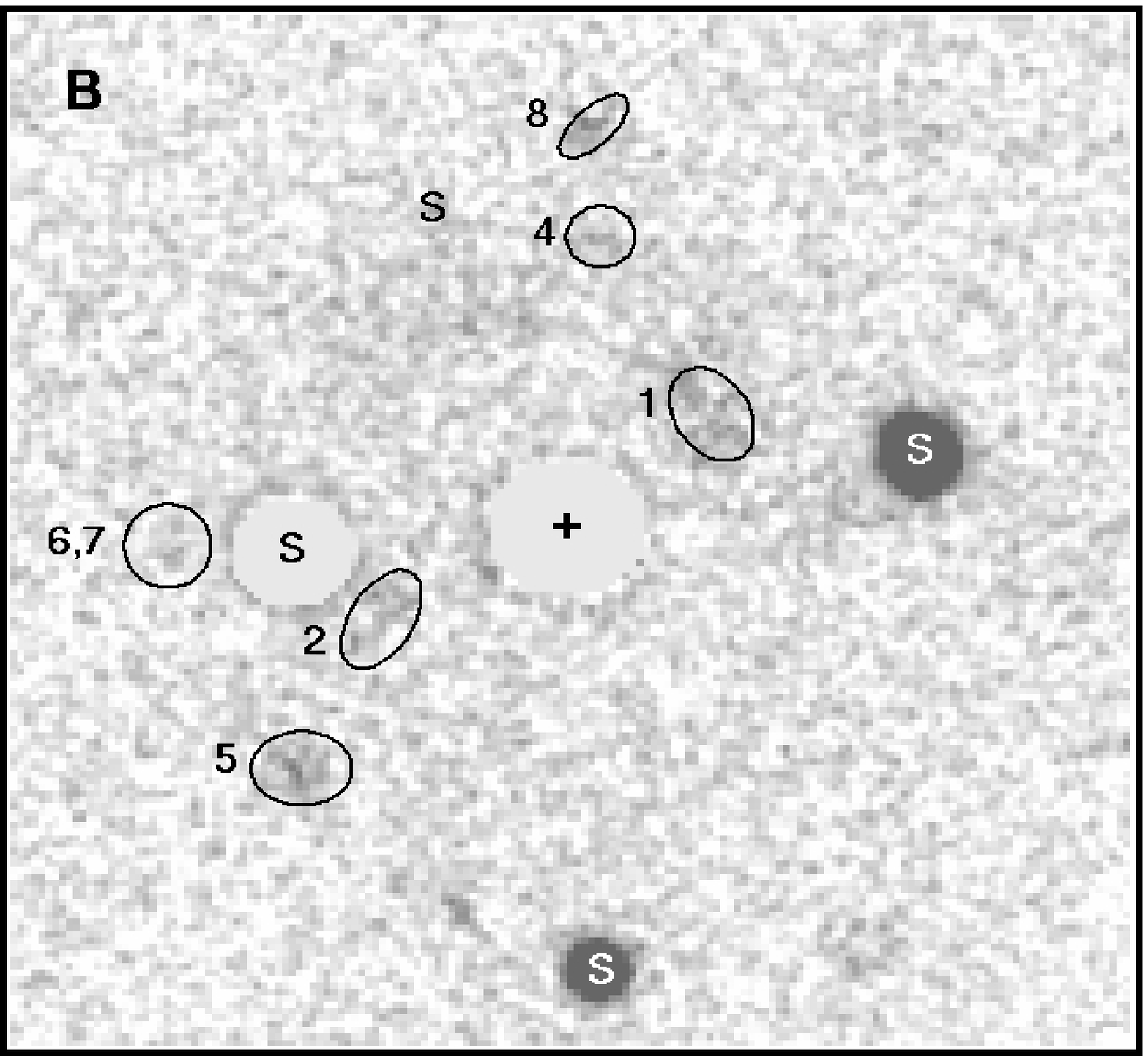}
\includegraphics[height=2.95in,height=2.95in]{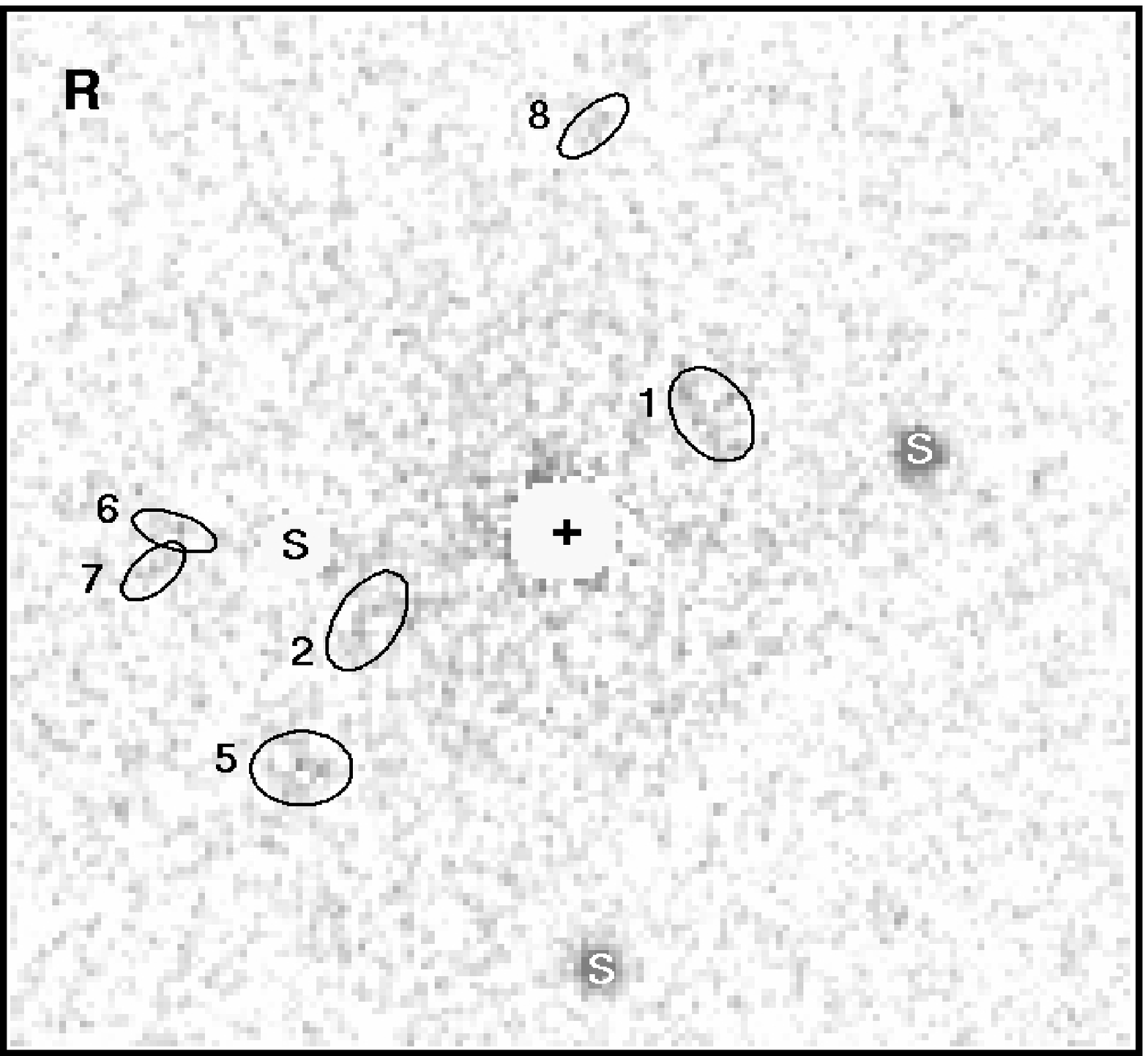}
\includegraphics[height=2.95in,height=2.95in]{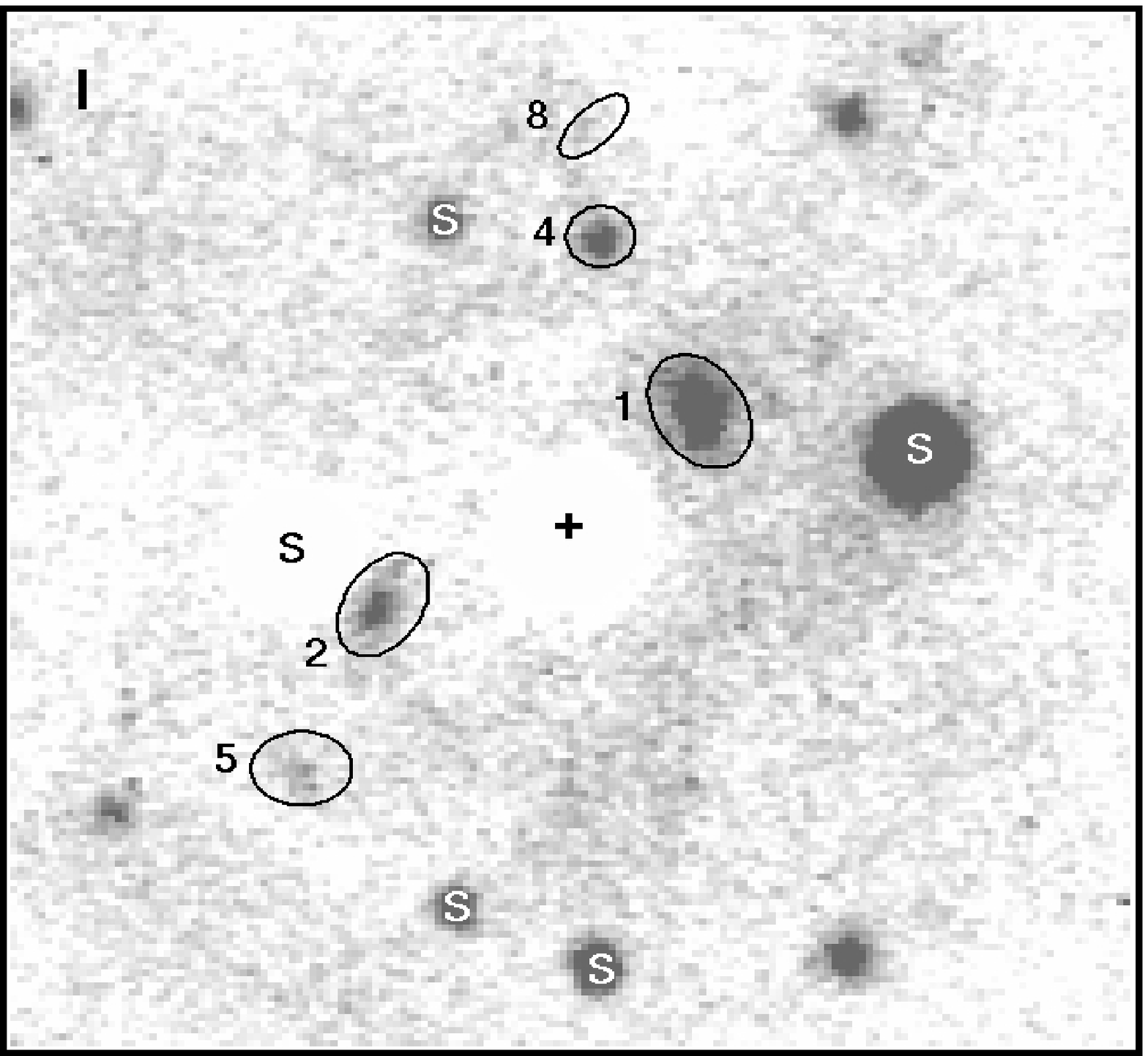}
\caption{44$\arcsec \times$ 44$\arcsec$ PSF-subtracted $U, B, R, I$
images  of the 0735+178 field. This field has a LLS at
$z_{abs}$ = 0.4240. As discussed in the text (\S3.1.2), we 
identify Object 4 as the absorbing 
galaxy. The images shown above correspond to $\approx$
245 $\times$ 245 kpc$^{2}$  at the absorber redshift.  The central
pixels of the quasar PSF subtraction  residuals have been masked, and
the position of its center is marked by a ``+''.  A nearby star,
10.9$\arcsec$ east of the quasar, was also subtracted. All stars in
the field are indicated by an ``S''.  North is up and east is to the
left.}
\label{0735optImages}
\end{center}
\end{figure*}

\begin{figure*}
\includegraphics[height=2.95in,height=2.95in]{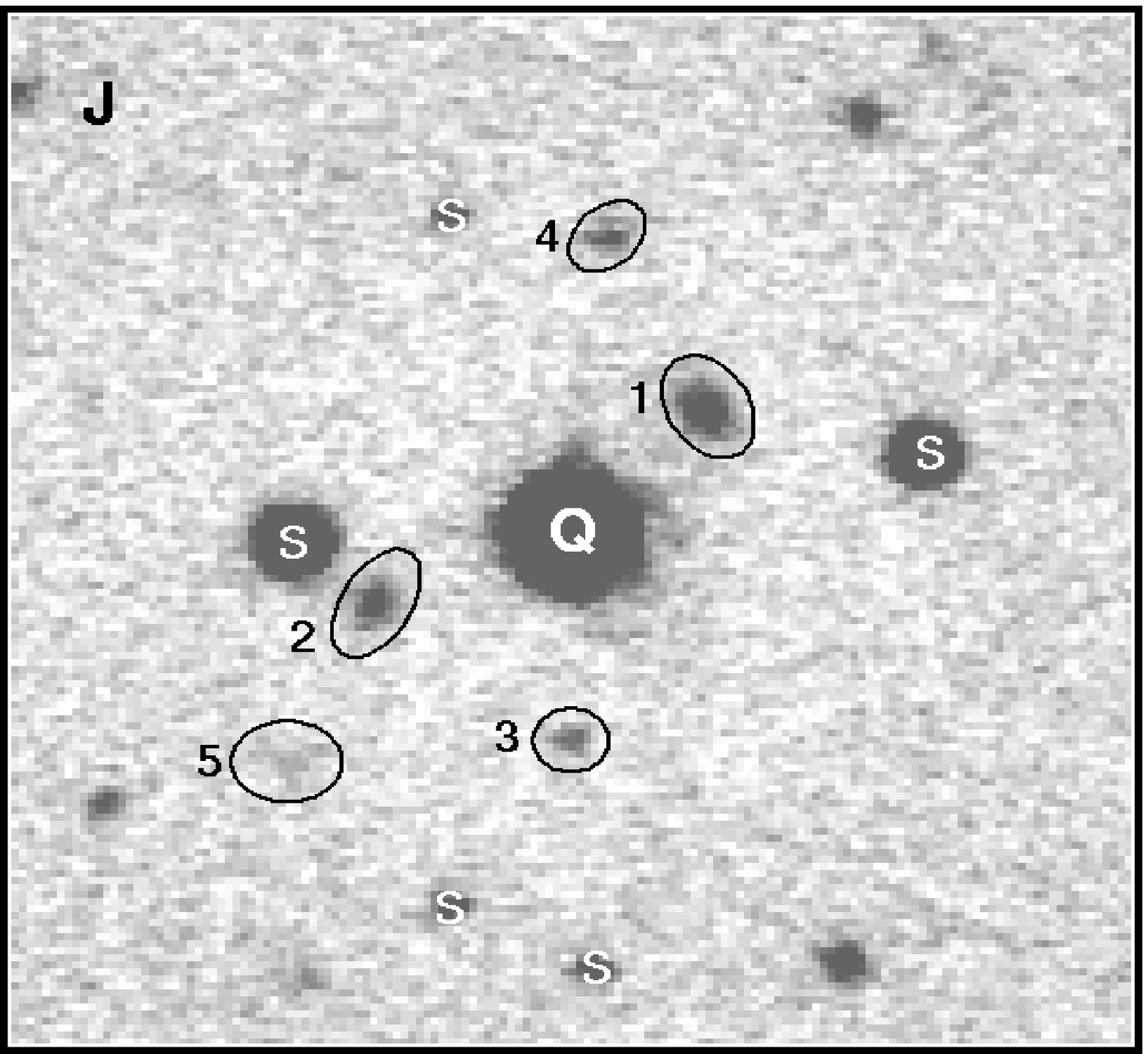}
\includegraphics[height=2.95in,height=2.95in]{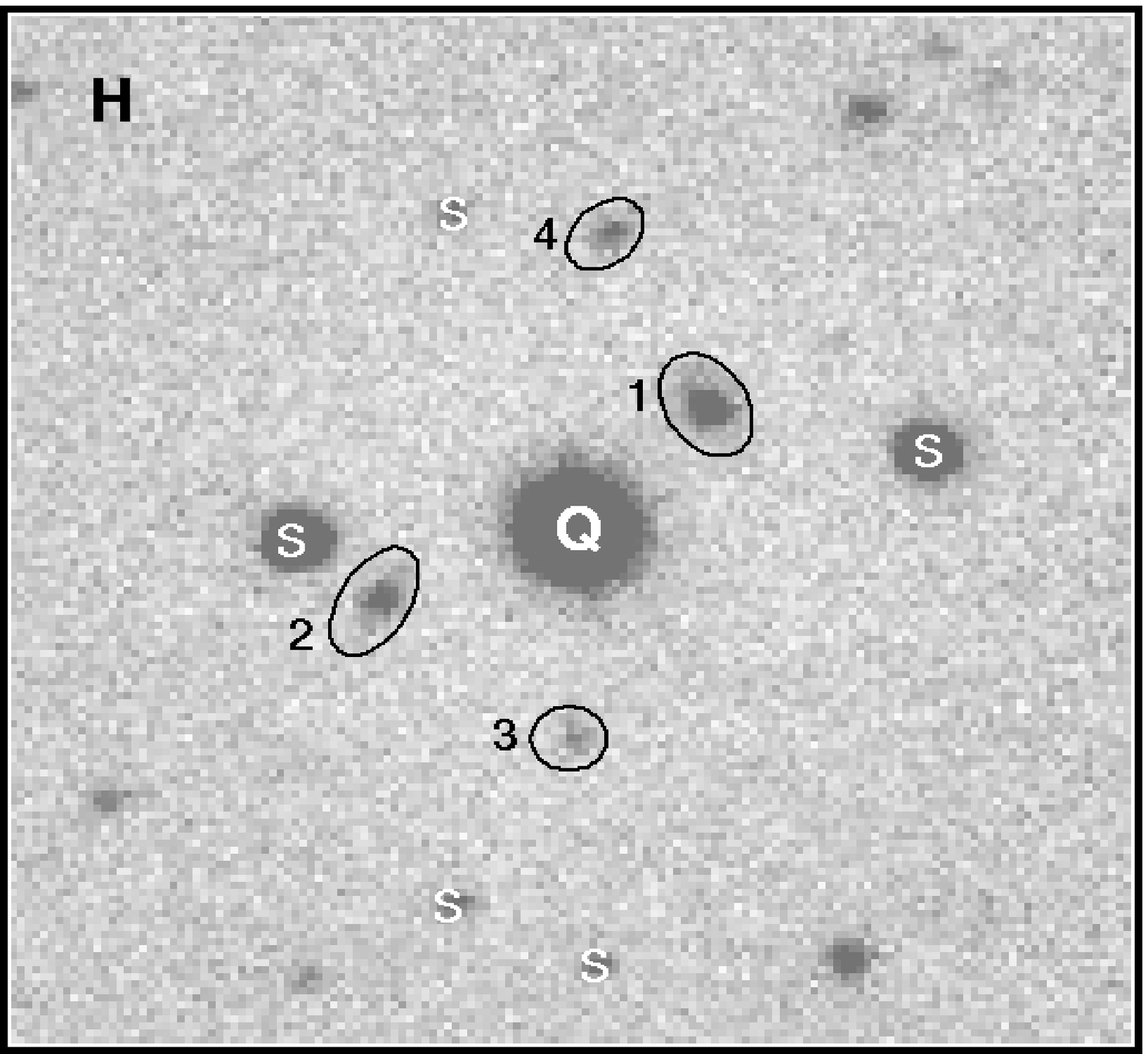}
\includegraphics[height=2.95in,height=2.95in]{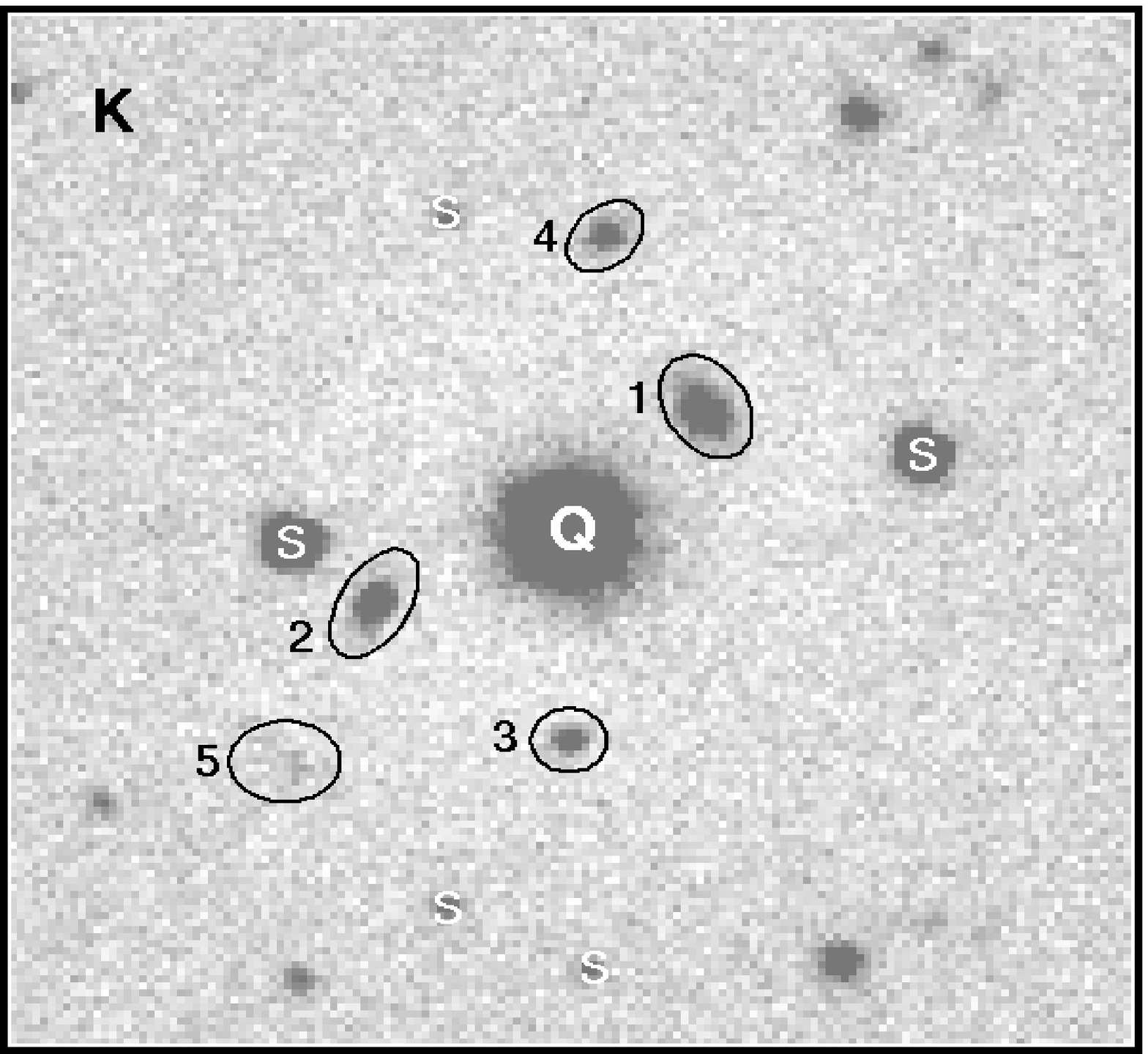}
\caption{Same as Figure \ref{0735optImages}, but for $J, H,$ and $K$.
 The quasar is marked
 by the letter ``Q''.  The quasar PSF could not be subtracted as there were no
 suitable PSF stars in the field. As discussed in the text (\S3.1.2), we 
identify Object 4 as the absorbing galaxy.}
\label{0735IRImages}
\end{figure*}

The sightline towards the quasar 0735+178 contains a LLS at
$z_{abs}$ = 0.4240 with a column density $\log N_{HI} < 19$ $\rm
cm^{-2}$ (RTN06). This is an interesting system because it has
relatively strong Fe II $\lambda 2600$ and Mg I $\lambda
2852$ absorption (see RTN06).  These systems generally tend to have higher \hi
column densities ($\log N_{HI} > 19$, RTN06), and therefore the
identification of the galaxy causing this unusual absorption-line
pattern might be illuminating.  A complete optical and infrared set of
images  is available for this field. The images are shown in Figures 
\ref{0735optImages} and \ref{0735IRImages}.   PSF subtractions were
carried out on the optical data and no objects were detected within
the subtracted region. The quasar PSF could not be subtracted on the
infrared images as there were no suitable PSF stars in the
field. Photometric measurements for the eight objects detected in this
field  are given in Tables \ref{0735optmags} and \ref{0735IRmags}.

\begin{figure*}
\includegraphics[width=3.0in]{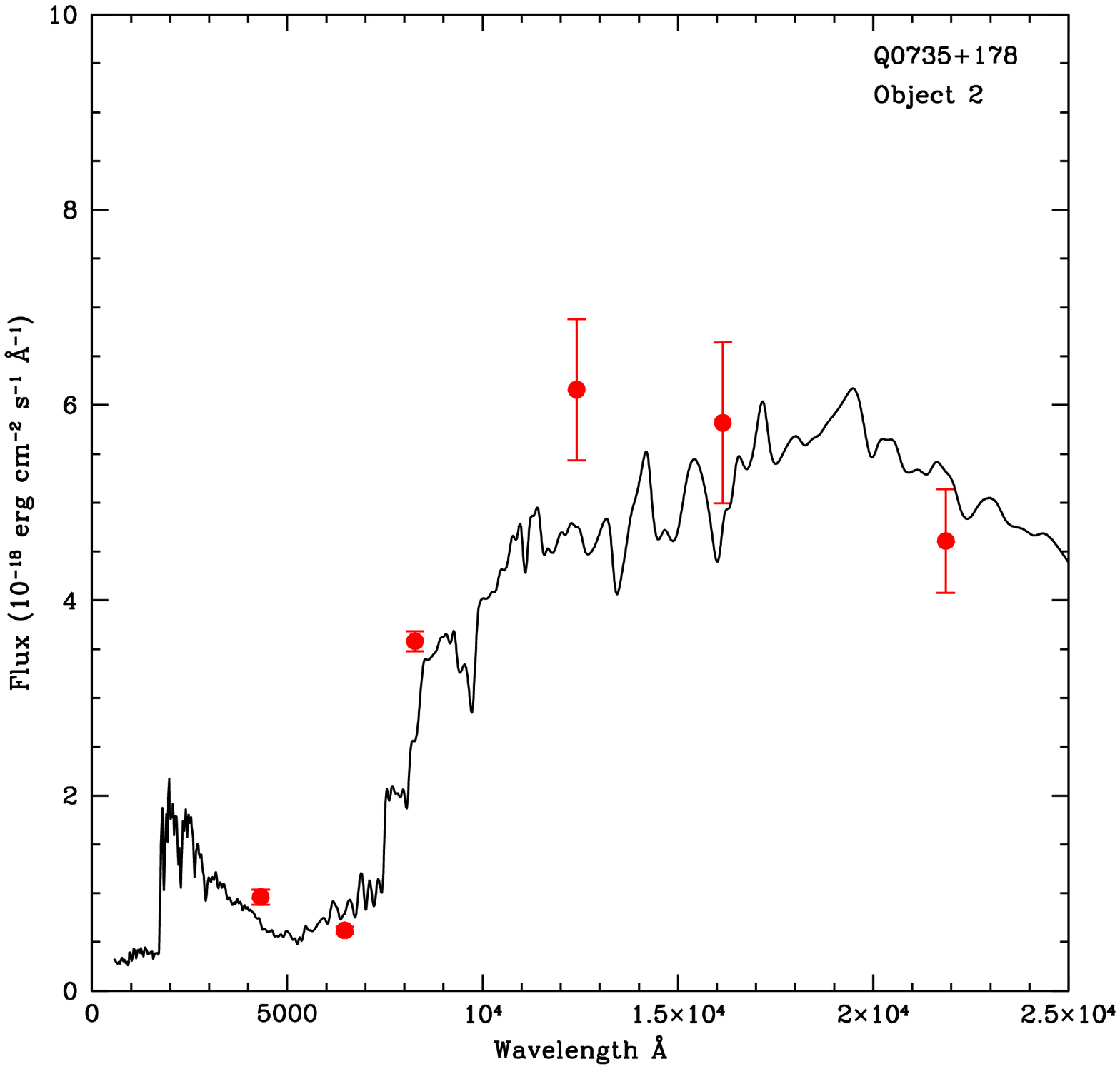}
\includegraphics[width=3.0in]{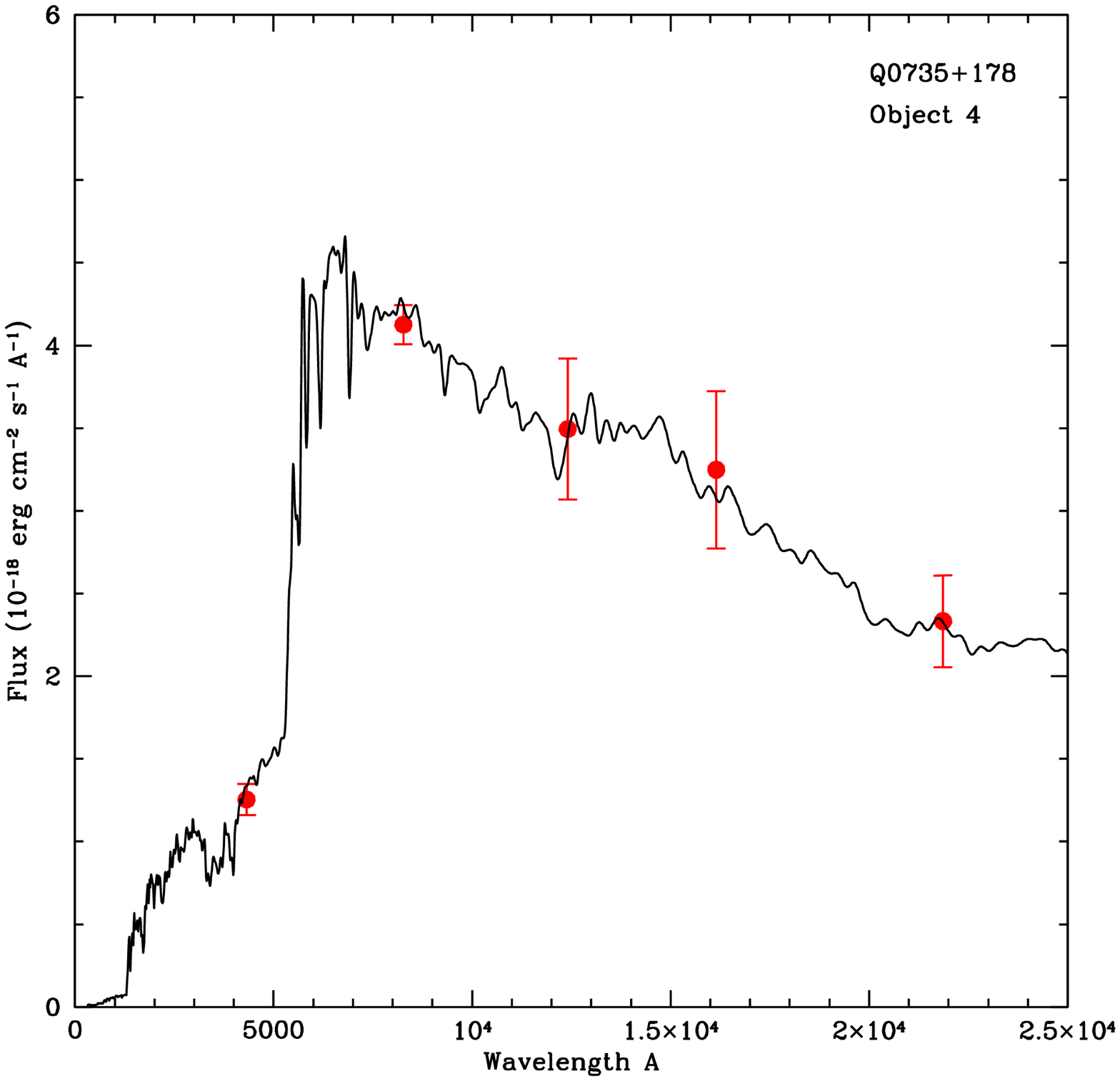}
\includegraphics[width=3.0in]{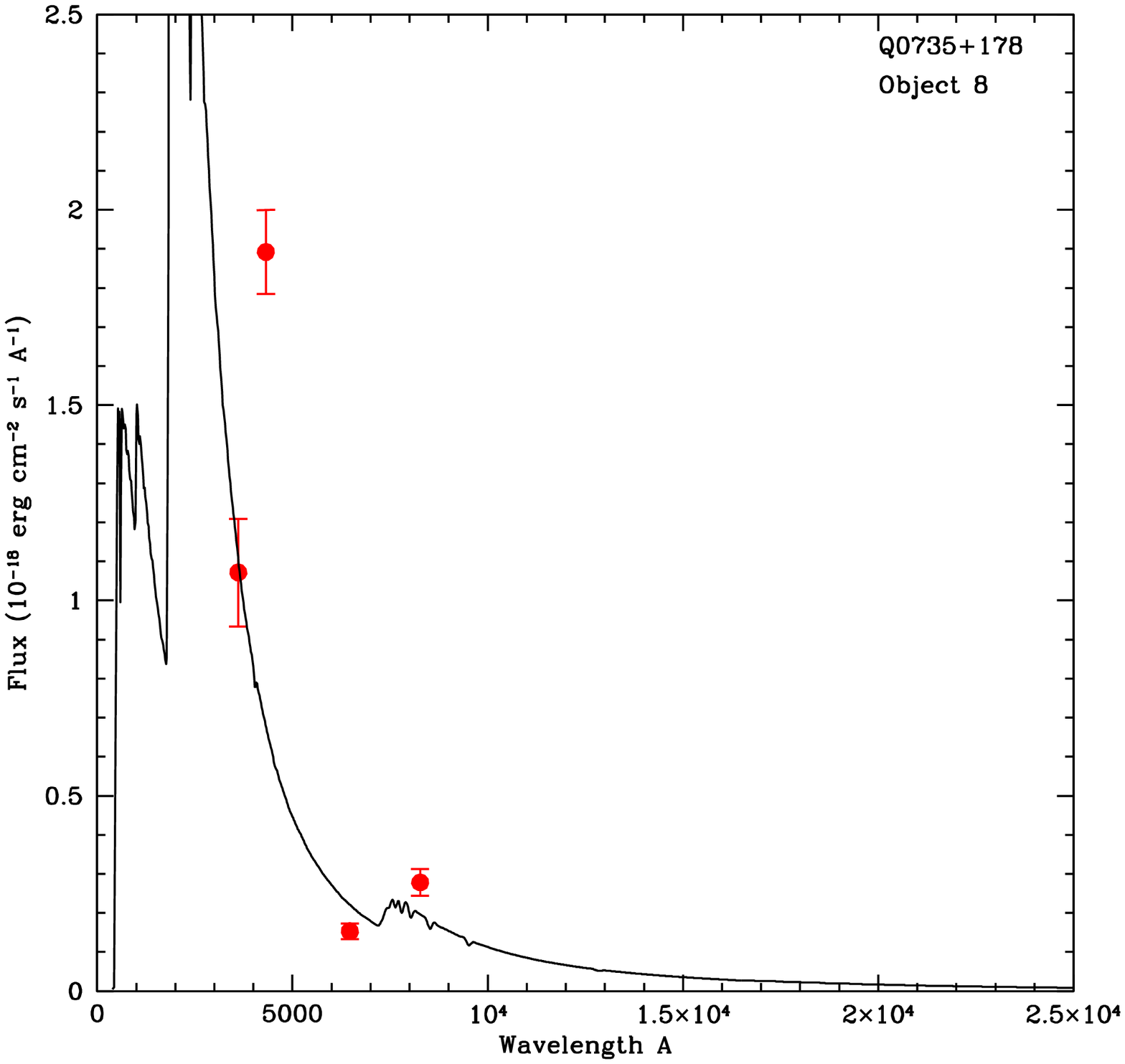}
\caption{The curves are the best-fit stellar population synthesis
models to our photometry  (solid circles) for the objects in the
0735+178 field.  See Table \ref{0735photozdata} for the model
parameters. As discussed in the text (\S3.1.2), we  identify Object 4
as the absorbing galaxy.}
\label{0735photoz}
\end{figure*}

Objects 1, 2, and 5 are in the SDSS database as having $z_{phot} =
0.662\pm 0.057$, $0.451\pm 0.166$, and  $0.238\pm 0.144$
respectively. Stickel et al. (1993) obtained a spectrum of Object 1,
and determined it to be at redshift $z$ = 0.645.  It is therefore
ruled out as the absorber candidate.  The best-fit stellar population
synthesis models to our photometry for Objects 2, 4, and 8 are shown
in Figure \ref{0735photoz}, and the best-fit template parameters are
listed in Table \ref{0735photozdata}. We derive a photometric redshift
$z_{phot} = 0.878\pm 0.038$ for Object 2, which is inconsistent with
that listed in the SDSS database. The addition of infrared photometric
measurements provides stronger constraints on the fit, making our
redshift determination more reliable than that reported in the SDSS
database.  Object 3 is only detected in the infrared, and its $J - H$
= 0.29 and $H - K$ = 0.81 colours are not consistent
with it being at the absorption redshift. We note that Object 3 has $R
- K >$ 7.6, which makes it an ``extremely red object'' (ERO).  Object
4 is classified as a star in the  SDSS database, however, it is
extended on our images. The photometric redshift that we derive for
Object 4 is consistent with the absorption redshift, and it is
identified as the candidate absorber in this field.  A stellar
population synthesis model could not be fit to the photometry of
Object 5 probably because of its low surface brightness and low
detection significance in all our images.  Its photometry is therefore
highly uncertain.

In summary, Object 4 is selected as the absorbing galaxy in this field
since its photometric redshift, $0.423 \pm 0.179$, matches the
absorption redshift, and the galaxies with smaller impact parameters
are ruled out as candidate absorbers. This identification is assigned
CL = 1.

\begin{landscape}
\begin{table}
\scriptsize
  \caption{0735+178: Optical Photometry}
  \begin{tabular}{c r r r c c c c c c c c}
\hline
Object & $\Delta\alpha$\tablenotemark{a} & $\Delta\delta$\tablenotemark{a} & $\theta$\tablenotemark{a} &
$U\pm\sigma_{U}$ & DS $(N_{pix})$\tablenotemark{b} & 
$B\pm\sigma_{B}$ &  DS $(N_{pix})$\tablenotemark{b} & 
$R\pm\sigma_{R}$ & DS $(N_{pix})$\tablenotemark{b} &
$I\pm\sigma_{I}$ & DS $(N_{pix})$\tablenotemark{b} \\
 & $\arcsec$  & $\arcsec$  & $\arcsec$ & \multicolumn{8}{c}{} \\
\hline
QSO & +0.0 & 0.0 & 0.0 & 15.71 $\pm$ 0.001 & 74.8 (2007) & 15.50 $\pm$ 0.001 & 138.8 (2187) & 15.90 $\pm$
0.001 & 225.1 (1231) & 15.20 $\pm$ 0.001 & 37.7 (4057) \\ 
1 & $-$5.2 & +5.1 & 7.3 & 23.04 $\pm$ 0.06 & 2.6 (67) & 23.80 $\pm$ 0.07 & 2.8 (51)
& 23.89 $\pm$ 0.06 & 2.6 (67) & 19.81 $\pm$ 0.01 & 3.5 (633) \\ 
2 & +7.7 & $-$3.2 & 8.4 & 23.21 $\pm$ 0.07 & 2.3 (66) & 24.46 $\pm$ 0.09 &
2.8 (28) & 24.06 $\pm$ 0.07 & 2.3 (66) & 21.62 $\pm$ 0.03 & 3.3 (125) \\
3 & +0.1 & $-$9.0 & 9.0 & \nodata & \nodata & \nodata & \nodata &
\nodata & \nodata & \nodata & \nodata \\ 4 & $-$1.4 & +12.6 & 12.7 &
\nodata & \nodata & 24.17 $\pm$ 0.08 & 2.9 (36) & \nodata & \nodata &
21.47 $\pm$ 0.03 & 3.4 (140) \\ 
5 & +10.8 & $-$10.1 & 14.8 & 23.57 $\pm$ 0.08 & 2.5 (43) & 23.16 $\pm$ 0.05 & 3.0 (86) & 24.41 $\pm$ 0.08
& 2.5 (43) & 22.93 $\pm$ 0.07 & 2.4 (52) \\ 
6 & +15.6 & +0.2 & 15.6 & 25.33 $\pm$ 0.18 & 2.1 (10) & 24.88 $\pm$ 0.11 & 4.5 (12) & 26.17
$\pm$ 0.18 & 2.1 (10) & \nodata & \nodata \\ 
7 & +16.4 & $-$1.5 & 16.4 & 25.14 $\pm$ 0.17 & 2.1 (12) & \nodata & \nodata & 25.99 $\pm$ 0.17 &
2.1 (12) & \nodata & \nodata \\ 
8 & $-$1.0 & +17.3 & 17.3 & 24.73 $\pm$ 0.15 & 3.4 (11) & 23.72 $\pm$ 0.06 & 2.4 (66) & 25.58 $\pm$ 0.15
& 3.4 (11) & 24.39 $\pm$ 0.14 & 2.3 (14) \\ 
\hline
\label{0735optmags}
\vspace{-0.5cm}
\tablenotetext{a}{Same as for Table \ref{0153mags}.}
\tablenotetext{b}{Same as for Table \ref{0153mags}.}
\end{tabular}
\end{table}

\begin{table}
\scriptsize
  \caption{0735+178: Infrared Photometry}
  \begin{tabular}{c r r r c c c c c c }
\hline
Object & $\Delta\alpha$\tablenotemark{a} & $\Delta\delta$\tablenotemark{a} & $\theta$\tablenotemark{a} &
$J\pm\sigma_{J}$ & DS $(N_{pix})$\tablenotemark{b} & 
$H\pm\sigma_{H}$ & DS $(N_{pix})$\tablenotemark{b} & 
$K\pm\sigma_{K}$ & DS $(N_{pix})$\tablenotemark{b}\\
 & $\arcsec$  & $\arcsec$  & $\arcsec$ & \multicolumn{6}{c}{} \\
\hline
QSO & +0.0 & 0.0 & 0.0 & 14.04 $\pm$ 0.13 & 116.3 (578) & 13.58 $\pm$ 0.16 & 116.3 (433) & 13.35 $\pm$ 0.13 & 142.8 (535) \\ 
1 & $-$5.2 & +5.1 & 7.3 & 19.40 $\pm$ 0.13 & 4.3 (114) & 18.99 $\pm$ 0.16 & 4.1 (83) & 18.66 $\pm$ 0.13 & 5.6 (102) \\ 
2 & +7.7 & $-$3.2 & 8.4 & 20.15 $\pm$ 0.14 & 3.6 (67) & 19.64 $\pm$ 0.17 & 3.7 (51) & 19.24 $\pm$ 0.13 & 6.4 (53) \\ 
3 & +0.1 & $-$9.0 & 9.0 & 21.31 $\pm$ 0.15 & 3.3 (25) & 21.02 $\pm$ 0.19 & 2.6 (20) & 20.21 $\pm$ 0.14 & 4.9 (28) \\ 
4 & $-$1.4 & +12.6 & 12.7 & 20.77 $\pm$ 0.14 & 3.4 (40) & 20.27 $\pm$ 0.17 & 3.4 (31) & 19.98 $\pm$ 0.14 & 4.3 (40)\\
5 & +10.8 & $-$10.1 & 14.8 & 21.88 $\pm$ 0.18 & 2.1 (24) & \nodata & \nodata & 22.17 $\pm$ 0.21 & 2.5 (9) \\ 
6 & +15.6 & +0.2 & 15.6 & \nodata & \nodata & \nodata & \nodata & \nodata & \nodata \\ 
7 & +16.4 & $-$1.5 & 16.4 & \nodata & \nodata & \nodata & \nodata & \nodata & \nodata \\ 
8 & $-$1.0 & +17.3 & 17.3 & \nodata & \nodata & \nodata & \nodata & \nodata & \nodata \\ 
\hline
\label{0735IRmags}
\vspace{-0.5cm}
\tablenotetext{a}{Same as for Table \ref{0153mags}.}
\tablenotetext{b}{Same as for Table \ref{0153mags}.}
\end{tabular}
\end{table}

\begin{table}
  \caption{0735+178: Photometric Redshift Fits\tablenotemark{a}}
  \begin{tabular}{crrccccc}
\hline
\multicolumn{3}{c}{Galaxy} & \multicolumn{5}{|c}{Stellar Population Synthesis Model Parameters}\\
\hline
\# & $\theta$\tablenotemark{b} & $b$ & Age & $\tau$ & $E(B -V)$ & $Z$ & $z_{phot}\pm\sigma_{z_{phot}}$ \\
 & $\arcsec$ & kpc & Gyr & Gyr & & \\
\hline
2 &  8.4 & 46.5 & 15.0 & 3.00 & 0.10 & 0.0500 & 0.878$\pm$0.038 \\ 
4 & 12.7 & 70.5 &  0.50 & 0.10 & 0.30 & 0.0500 & 0.423$\pm$0.179 \\ 
8 & 17.3 & 96.5 &  0.10 & 0.10 & 0.00 & 0.0001 & 0.959$\pm$0.644 \\ 
\hline
\label{0735photozdata}
\vspace{-0.5cm}
\tablenotetext{a}{$z_{abs}=0.4240$}
\tablenotetext{b}{Relative to the quasar}
\end{tabular}
\end{table}
\end{landscape}

\subsubsection{Example 3: the 1109+0051 field}

\begin{figure*}
\includegraphics[height=2.95in,height=2.95in]{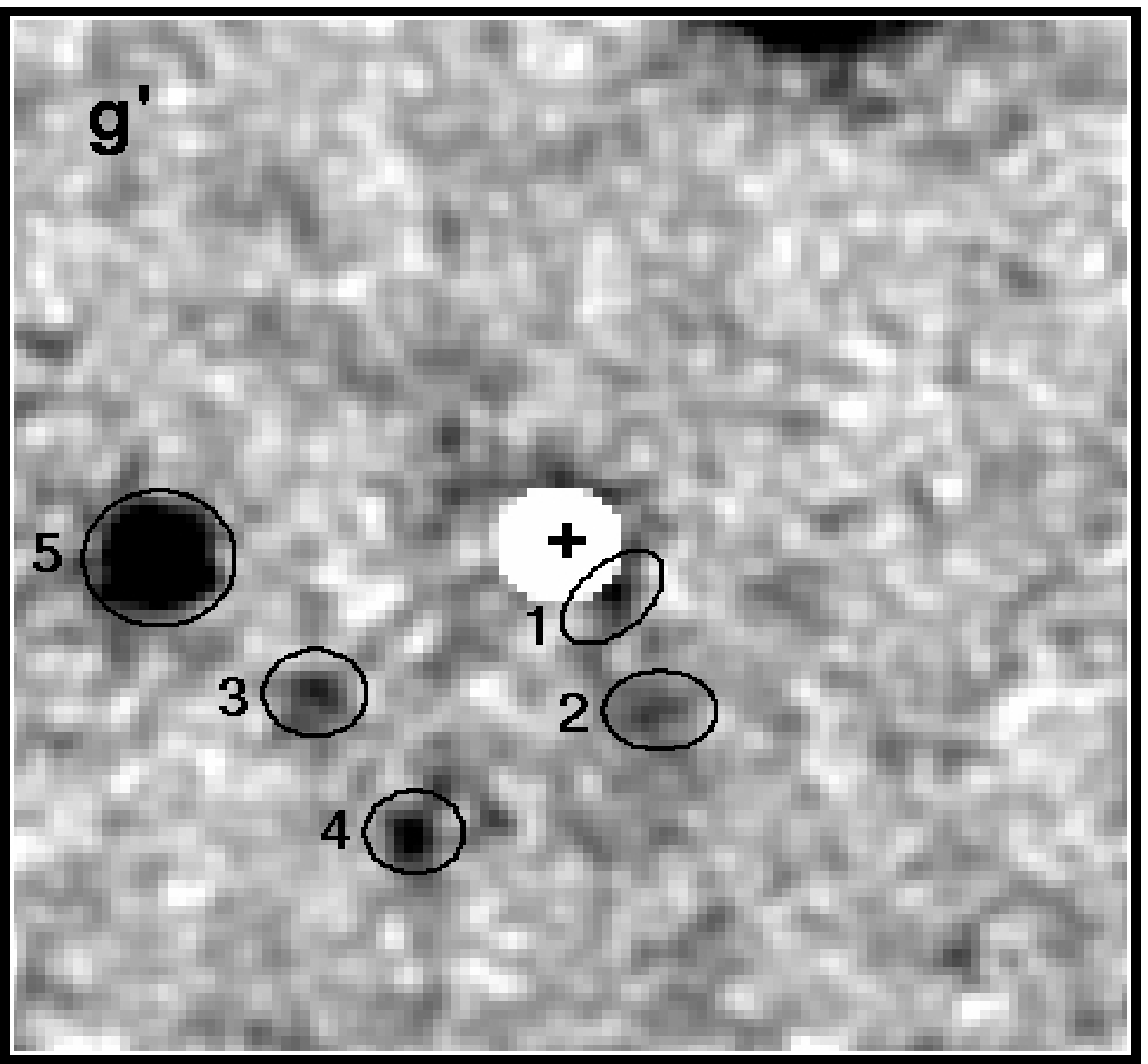}
\includegraphics[height=2.95in,height=2.95in]{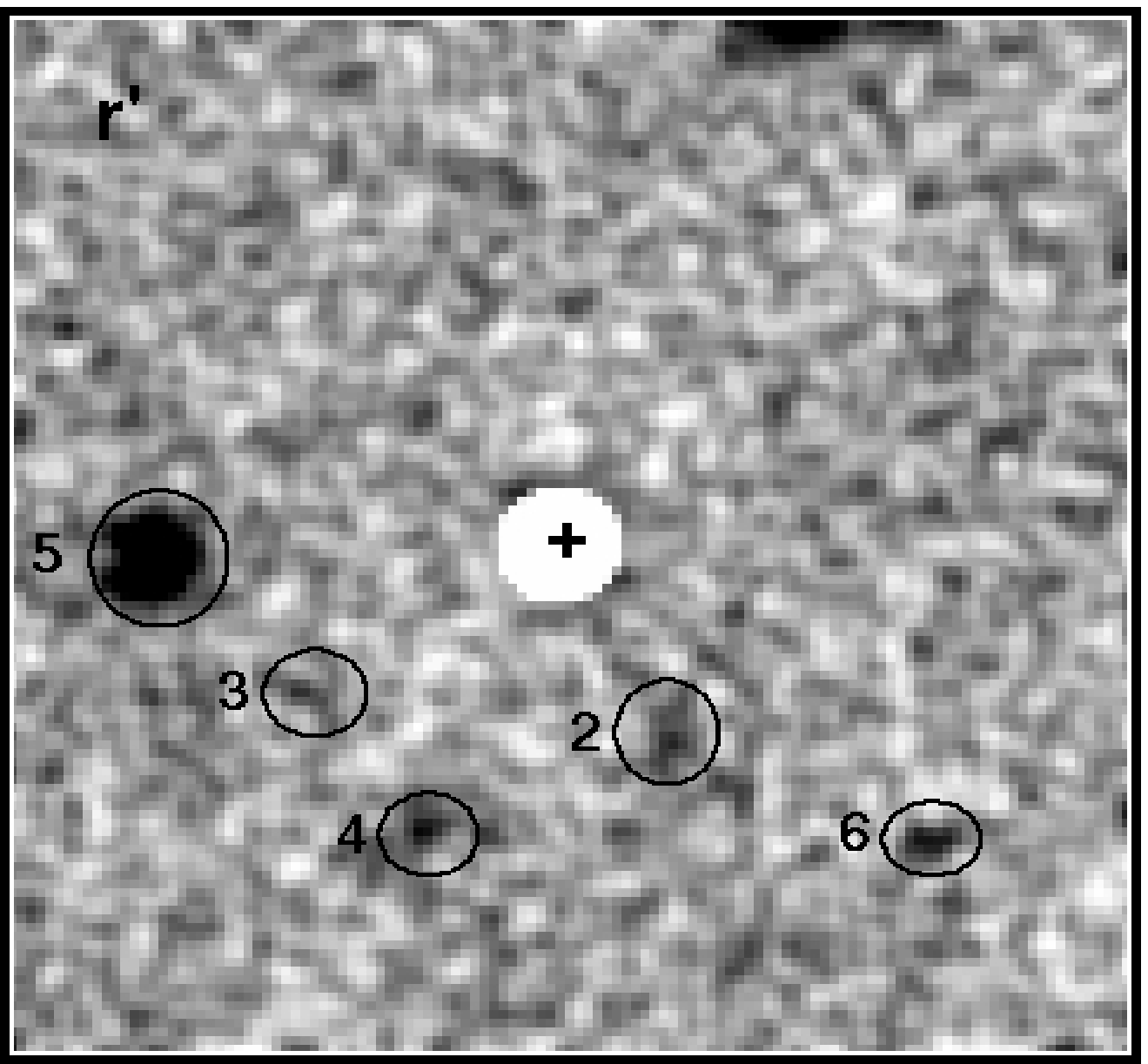}
\caption{36$\arcsec \times$ 36$\arcsec$ PSF-subtracted $g'$ and $r'$
images of the 1109+0051 field.  This field has two subDLA systems,
one at $z_{abs}$ = 0.4180 and the other at $z_{abs}$ = 0.5520. As 
discussed in the text (\S3.1.3), we 
identify Object 1 as the absorbing galaxy at $z_{abs}=0.4180$ and Object
3 as the absorbing galaxy at $z_{abs}$ = 0.5520. The
images shown above correspond to  $\approx$ 199 $\times$ 199 kpc$^{2}$
and $\approx$ 231 $\times$ 231 kpc$^{2}$ at the two redshifts,
respectively. The frames  are smoothed to bring out LSB features. The
PSF residuals have been masked.  The quasar position is marked by a
''+''.  North is up and  east is to the left.}
\label{1109optImages}
\end{figure*}

\begin{figure*}
\includegraphics[height=2.95in,height=2.95in]{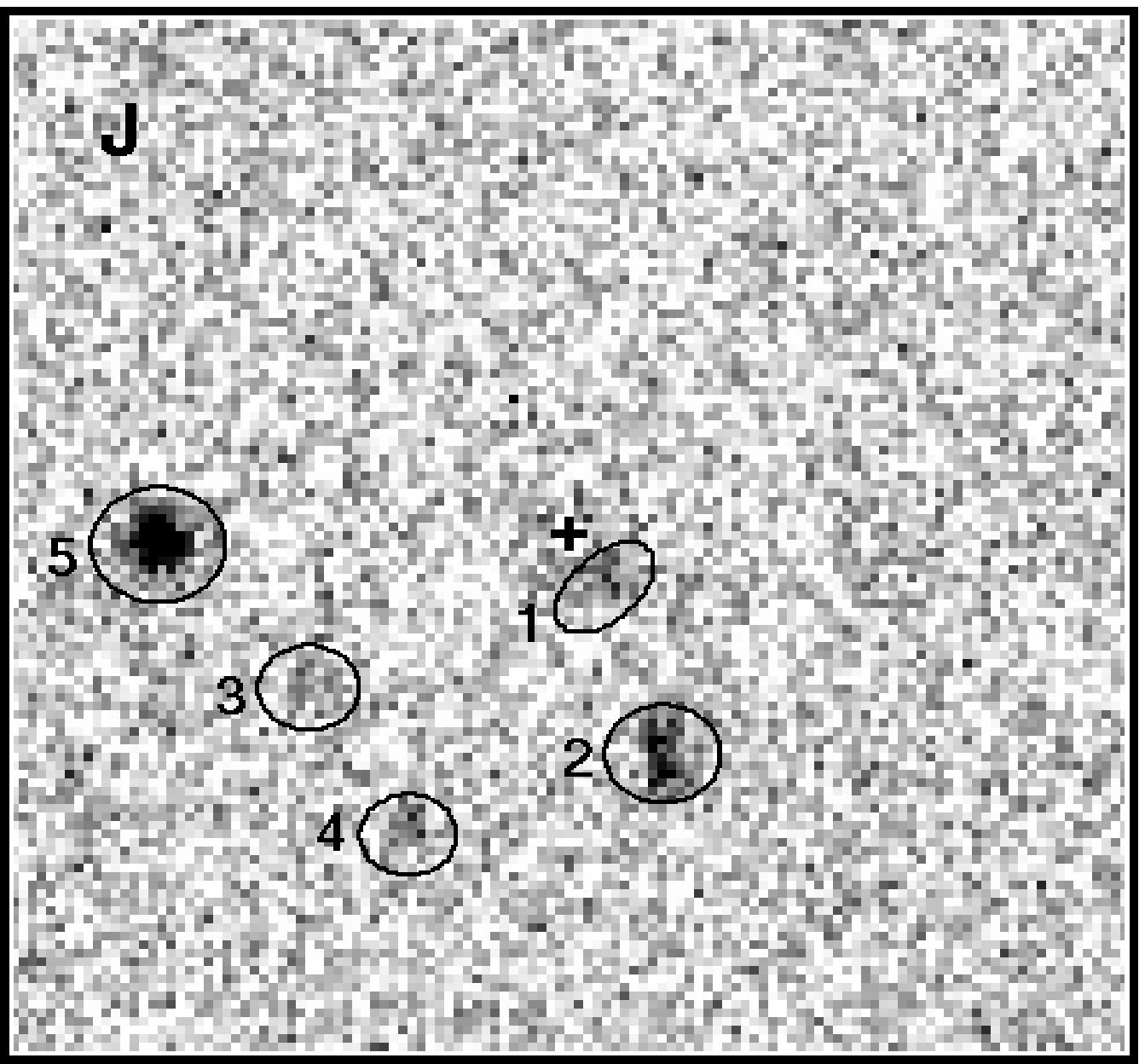}
\includegraphics[height=2.95in,height=2.95in]{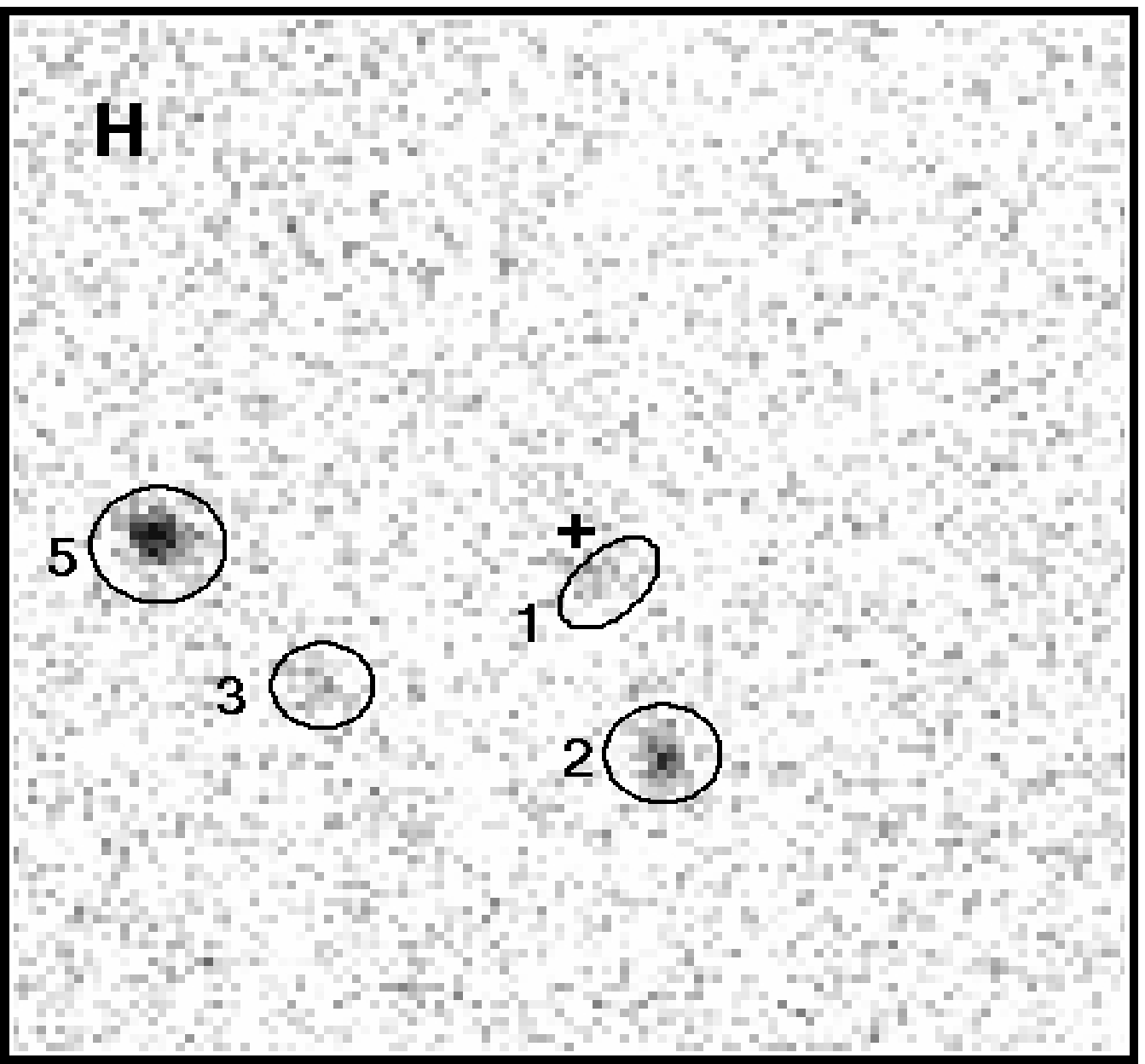}
\includegraphics[height=2.95in,height=2.95in]{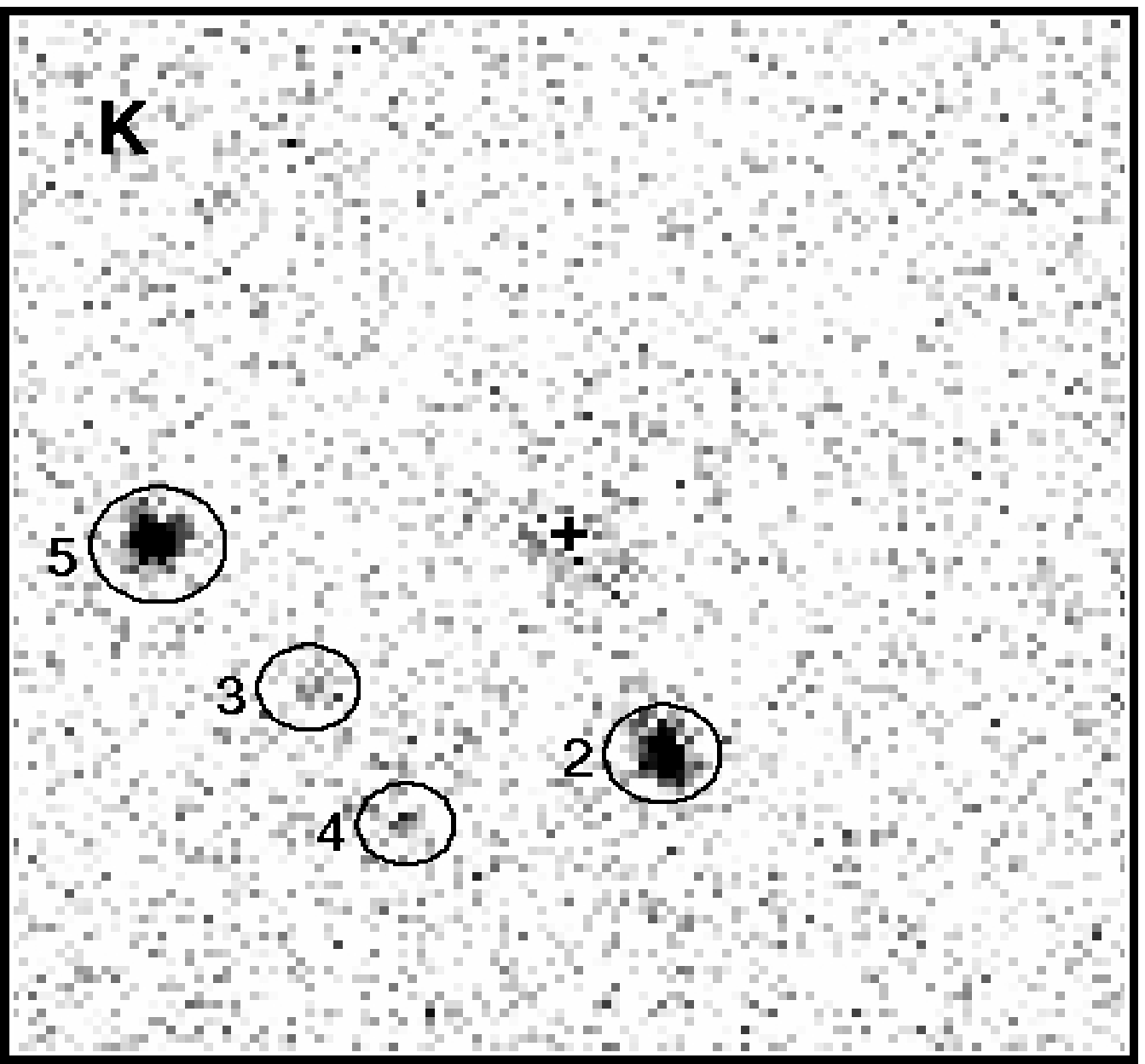}
\caption{Same as Figure \ref{1109optImages}, but for {\it J, H,} and
{\it K}.  There is evidence for Object 1 in the $K$ frame, however, it
does not meet the ``5 contiguous pixels above 1$\sigma$'' detection
criterion. As discussed in the text (\S3.1.3), we 
identify Object 1 as the absorbing galaxy at $z_{abs}=0.4180$ and Object
3 as the absorbing galaxy at $z_{abs}$ = 0.5520.}
\label{1109IRImages}
\end{figure*}

This field is not as straightforward as the previous ones.  The
sightline towards the quasar 1109+0051 (SDSS J110936.35+005111.3)
contains two subDLA systems, one  at $z_{abs}$ = 0.4181 with a column
density of $\log N_{HI} = 19.08^{+0.22}_{-0.38}$ $\rm cm^{-2}$ and the
other at  $z_{abs}$ = 0.5520 with a column density of $\log N_{HI} =
19.60^{+0.10}_{-0.12}$ $\rm cm^{-2}$ (RTN06).   Images of this field
were obtained in {\it g', r', J, H}, and {\it K}, from which six
objects are detected  (Figures \ref{1109optImages} and
\ref{1109IRImages}).  The quasar PSF subtraction revealed no object
within the subtracted region. The optical and infrared photometry are
given in Tables \ref{1109optmags} and \ref{1109IRmags}, respectively.

\begin{figure*}
\begin{center}
\includegraphics[width=3.0in]{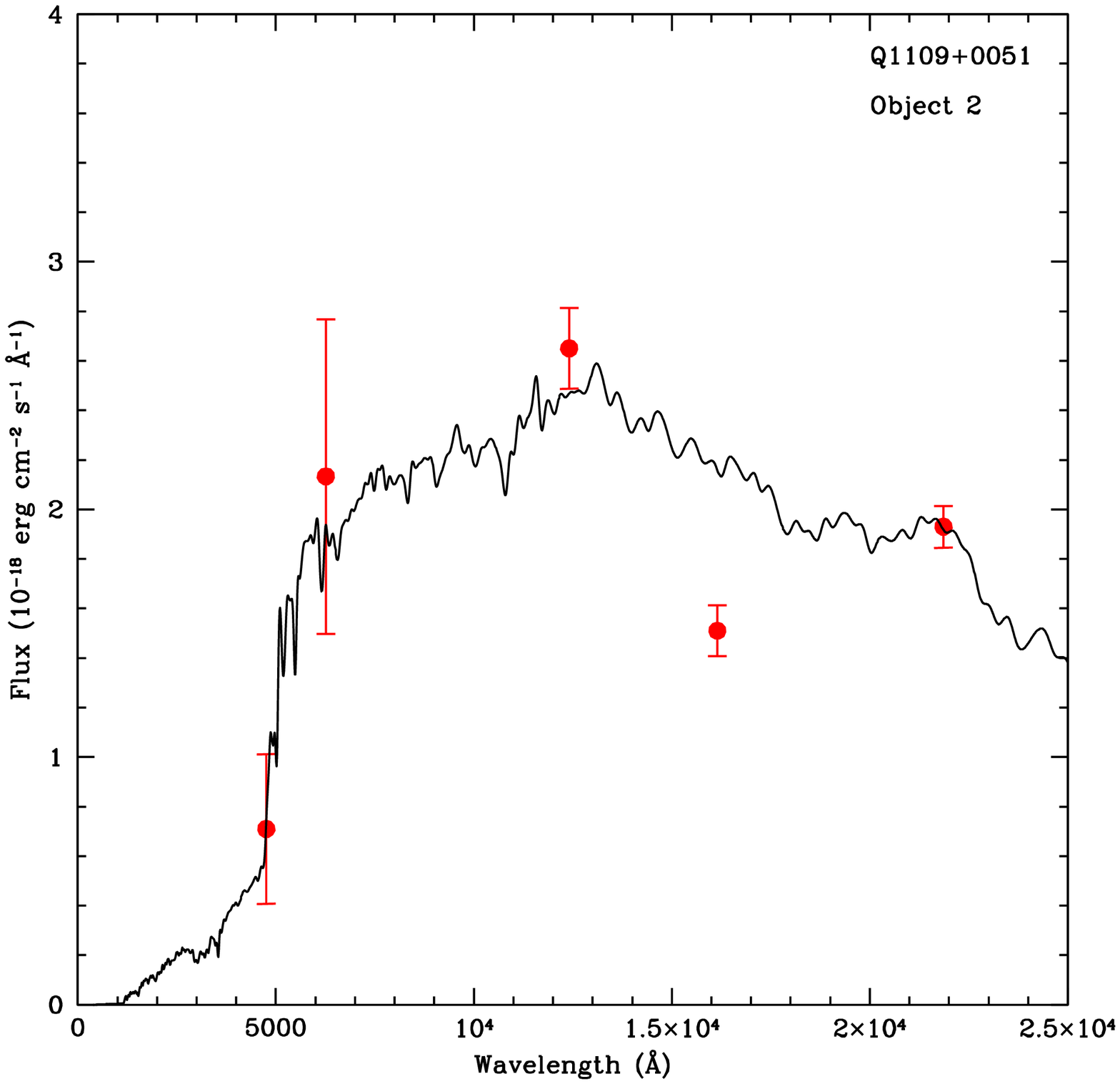}
\includegraphics[width=3.0in]{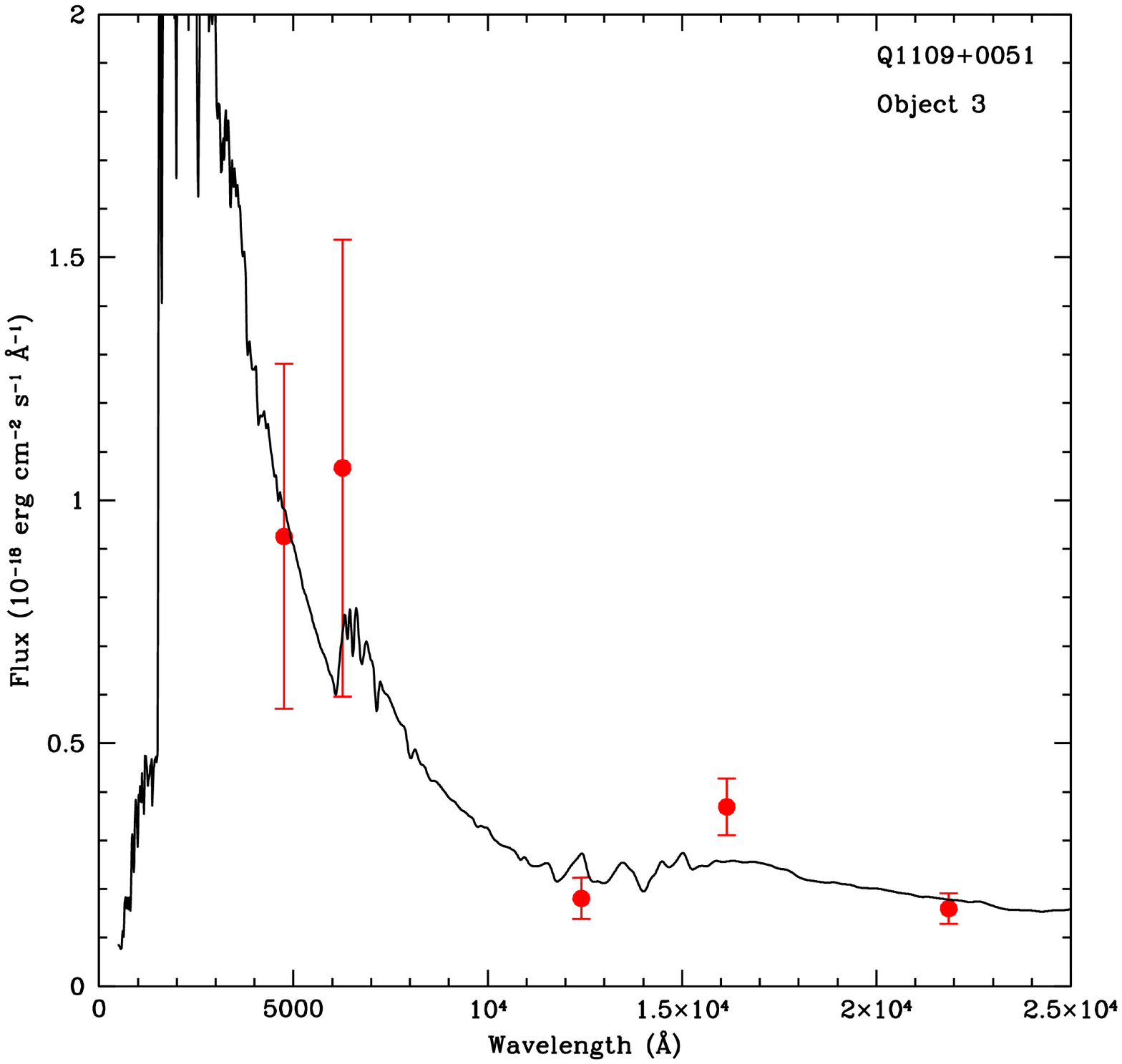}
\includegraphics[width=3.0in]{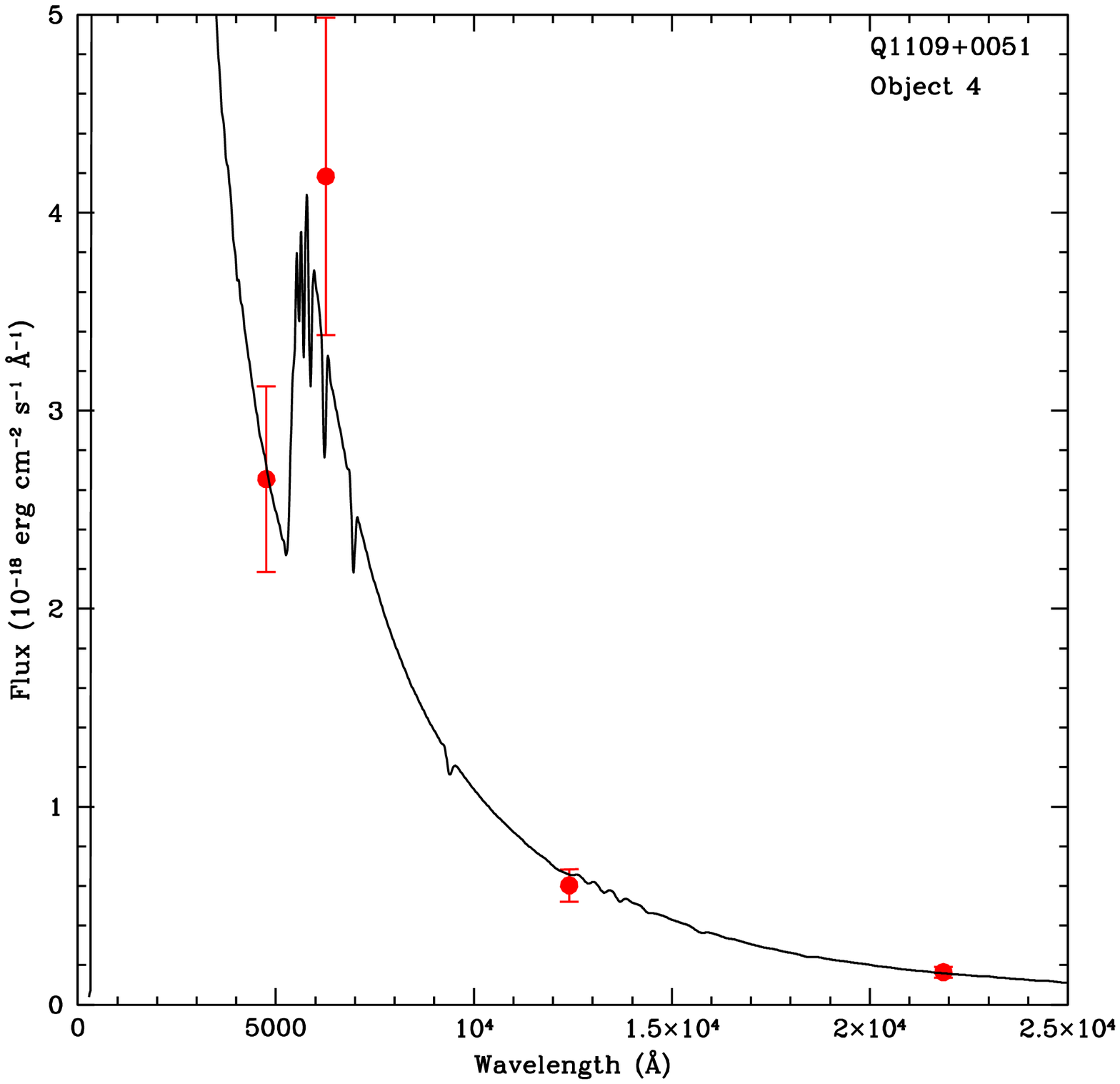}
\includegraphics[width=3.0in]{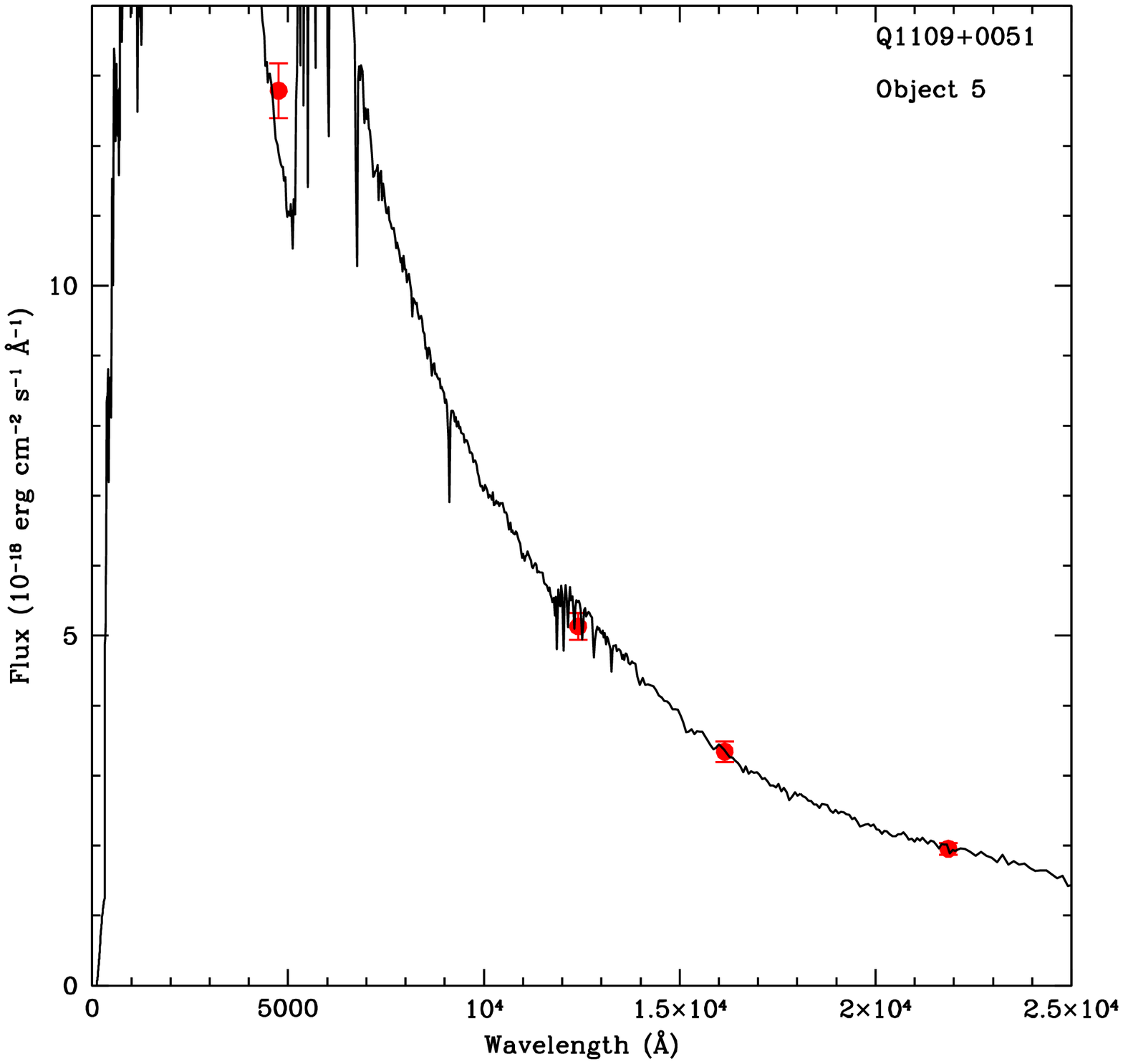}
\caption[Q1109+0051 stellar population synthesis fits]{The curves are
the best-fit stellar population synthesis models to our  photometry
(solid circles) for objects in the 1109+0051 field. The best-fit
parameters are listed in Table \ref{1109photozdata}. As discussed 
in the text (\S3.1.3), we identify Object 3 as the absorbing galaxy at 
$z_{abs}$ = 0.5520.}
\label{1109photoz}
\end{center}
\end{figure*}

Object 1 has an impact parameter $\theta = 1.3\arcsec$, which at the
two absorber redshifts corresponds to  7 kpc and 8 kpc,
respectively. It is detected in the $g', J$, and $H$-bands. Since it
overlaps with the quasar PSF, its photometry is uncertain. We consider
it to qualify as a candidate absorber due to its proximity to the
sightline.  The best-fit stellar population synthesis models to our
photometry for Objects 2, 3, 4, and 5 are shown in Figure
\ref{1109photoz}, and the model parameters are tabulated in Table
\ref{1109photozdata}.  The best-fit stellar population synthesis model
to the photometry of  Object 2 is only marginally consistent with (within
$2\sigma$ of) the
lower absorption redshift system,  $z$ = 0.4181. The photometry of
Object 3 results in a photometric redshift of $z = 0.645 \pm 0.157$,
which is  consistent with the absorption system at $z_{abs}$ =
0.5520. However, as can be seen from Table \ref{1109photozdata}, the
photometric redshift we derive for Object 4, while inconsistent with
the SDSS-derived photometric redshift of $0.138 \pm 0.044$, is
consistent with both absorption redshifts. In addition, the redshift
derived for Object 5 is consistent with the absorption system at
$z_{abs}$ = 0.4181 (but inconsistent with the SDSS redshift of $0.280
\pm 0.063$).  Therefore, Objects 1, 3, 4, and 5 are all  potential
absorber candidates for the absorption systems in this field, while
Object 2 is marginally consistent at the lower redshift.  Given this
ambiguity, we use the proximity criterion as the deciding factor, and
select Object 1 as the $z_{abs}$ = 0.4181 candidate and Object 3 as
the $z_{abs}$ = 0.5520 candidate, both with confidence level CL = 2.

\begin{landscape}
\begin{table}
\scriptsize
  \caption{1109+0051: Optical Photometry}
  \begin{tabular}{c r r r c c c c}
\hline
Object & $\Delta\alpha$\tablenotemark{a} & $\Delta\delta$\tablenotemark{a} & $\theta$\tablenotemark{a} &
$g'\pm\sigma_{g'}$ & DS $(N_{pix})$\tablenotemark{b} & 
$r'\pm\sigma_{r'}$ &  DS $(N_{pix})$\tablenotemark{b} \\
 & $\arcsec$  & $\arcsec$  & $\arcsec$ & \multicolumn{4}{c}{} \\
\hline
QSO& 0.0 & 0.0 & 0.0 &  18.45 $\pm$ 0.03 & 29.5 (422) & 18.45 $\pm$ 0.05 & 9.1 (263)  \\ 
1 & $-$0.6 & $-$1.1&1.3 &  23.79 $\pm$ 0.33 & 2.6 (26) & \nodata & \nodata  \\ 
2 & $-$3.4 & $-$7.4 &8.1&  24.58 $\pm$ 0.64 & 2.6 (12) & 22.79 $\pm$ 0.42 & 2.7 (14)  \\ 
3 & +8.4 & $-$5.4 & 10.0&  24.29 $\pm$ 0.56 & 2.6 (16) & 23.54 $\pm$ 0.67 & 2.3 (8)  \\ 
4 & +4.6 & $-$9.9 & 10.9&  23.15 $\pm$ 0.24 & 3.8 (34) & 22.05 $\pm$ 0.27 & 3.8 (20)  \\ 
5 & +13.4 & $-$0.7&13.4 &  21.44 $\pm$ 0.07 & 4.8 (142) & 20.52 $\pm$ 0.10 & 3.0 (111) \\ 
6 & $-$11.7&$-$10.6&15.8 &  \nodata & \nodata & 22.85 $\pm$ 0.42 & 2.2 (16)\\
\hline
\vspace{-0.5cm}
\tablenotetext{a}{Same as for Table \ref{0153mags}.}
\tablenotetext{b}{Same as for Table \ref{0153mags}.}
\label{1109optmags}
\end{tabular}
\end{table}

\begin{table}
\scriptsize
  \caption{1109+0051: Infrared Photometry}
  \begin{tabular}{c r r r c c c c c c}
\hline
Object & $\Delta\alpha$\tablenotemark{a} & $\Delta\delta$\tablenotemark{a} & $\theta$\tablenotemark{a} &
$J\pm\sigma_{J}$ & DS $(N_{pix})$\tablenotemark{b} & 
$H\pm\sigma_{H}$ & DS $(N_{pix})$\tablenotemark{b} & 
$K\pm\sigma_{K}$ & DS $(N_{pix})$\tablenotemark{b}\\
 & $\arcsec$  & $\arcsec$  & $\arcsec$ & \multicolumn{6}{c}{} \\
\hline
QSO& 0.0 & 0.0 & 0.0 & 18.06 $\pm$ 0.01 & 12.8 (151) & 18.41 $\pm$ 0.01 & 9.1 (113) & 18.03 $\pm$ 0.01 & 10.8 (119) \\ 
1 & $-$0.6 & $-$1.1 & 1.3 & 22.24 $\pm$ 0.13 & 2.6 (16) & 22.36 $\pm$ 0.17 & 2.1 (13) & \nodata & \nodata \\ 
2 & $-$3.4 & $-$7.4 0.& 8.1 & 21.07 $\pm$ 0.07 & 2.6 (47) & 21.10 $\pm$ 0.08 & 2.5 (34) & 20.18 $\pm$ 0.05 & 3.2 (55) \\ 
3 & +8.4 & $-$5.4 & 10.0 & 23.98 $\pm$ 0.29 & 1.7 (5) & 22.63 $\pm$ 0.19 & 2.1 (10) & 22.88 $\pm$ 0.23 & 2.4 (6) \\ 
4 & +4.6 & $-$9.9 & 10.9 & 22.67 $\pm$ 0.16 & 2.0 (14) & \nodata & \nodata & 22.86 $\pm$ 0.21 & 2.5 (6) \\ 
5 & +13.4 & $-$0.7 & 13.4 & 20.35 $\pm$ 0.04 & 3.5 (67) & 20.24 $\pm$ 0.05 & 3.2 (59) & 20.17 $\pm$ 0.05 & 3.3 (55) \\ 
6 & $-$11.7 & $-$10.6 & 15.8 & \nodata & \nodata & \nodata & \nodata & \nodata & \nodata \\ 
\hline
\label{1109IRmags}
\vspace{-0.5cm}
\tablenotetext{a}{Same as for Table \ref{0153mags}.}
\tablenotetext{b}{Same as for Table \ref{0153mags}.}
\end{tabular}
\end{table}

\begin{table}
\caption{1109+0051: Photometric Redshift Fits\tablenotemark{a}}
  \begin{tabular}{crrccccc}
\hline
\multicolumn{3}{c}{Galaxy} & \multicolumn{5}{|c}{Stellar Population Synthesis Model Parameters}\\
\hline
\# & $\theta$\tablenotemark{b} & $b$ & Age & $\tau$ & $E(B -V)$ & $Z$ & $z_{phot}\pm\sigma_{z_{phot}}$ \\
 & $\arcsec$ & kpc & Gyr & Gyr & & \\
\hline
 2 &  8.1 & 44.7 & 0.50 & 0.10 & 0.50 & 0.0500 & 0.266$\pm$0.109 \\ 
3 & 10.0 & 55.4 & 0.10 & 12.0 & 0.20 & 0.0500 & 0.645$\pm$0.157 \\ 
4 & 10.9 & 60.4 & 1.00 & 1.00 & 0.00 & 0.0001 & 0.433$\pm$0.222 \\ 
5 & 13.4 & 73.9 & 1.00 & 12.0 & 0.10 & 0.0080 & 0.388$\pm$0.050 \\ 
\hline
\label{1109photozdata}
\vspace{-0.5cm}
\tablenotetext{a}{$z_{abs}=0.4181$, 0.5520 }
\tablenotetext{b}{Relative to the quasar}
\end{tabular}
\end{table}
\end{landscape}

\subsubsection{Example 4: the 1715+5747 field}

\begin{figure*}
\includegraphics[height=2.95in,height=2.95in]{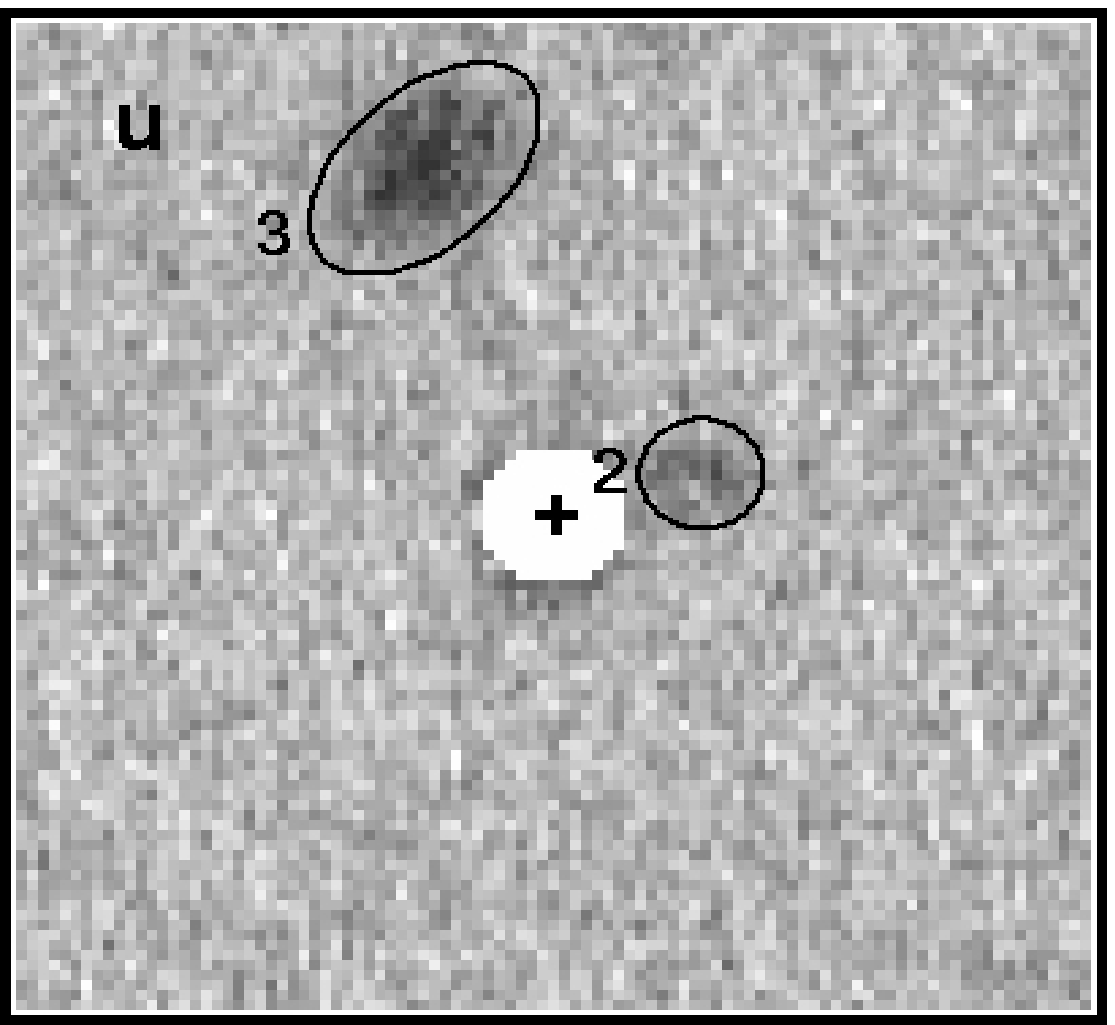}
\includegraphics[height=2.95in,height=2.95in]{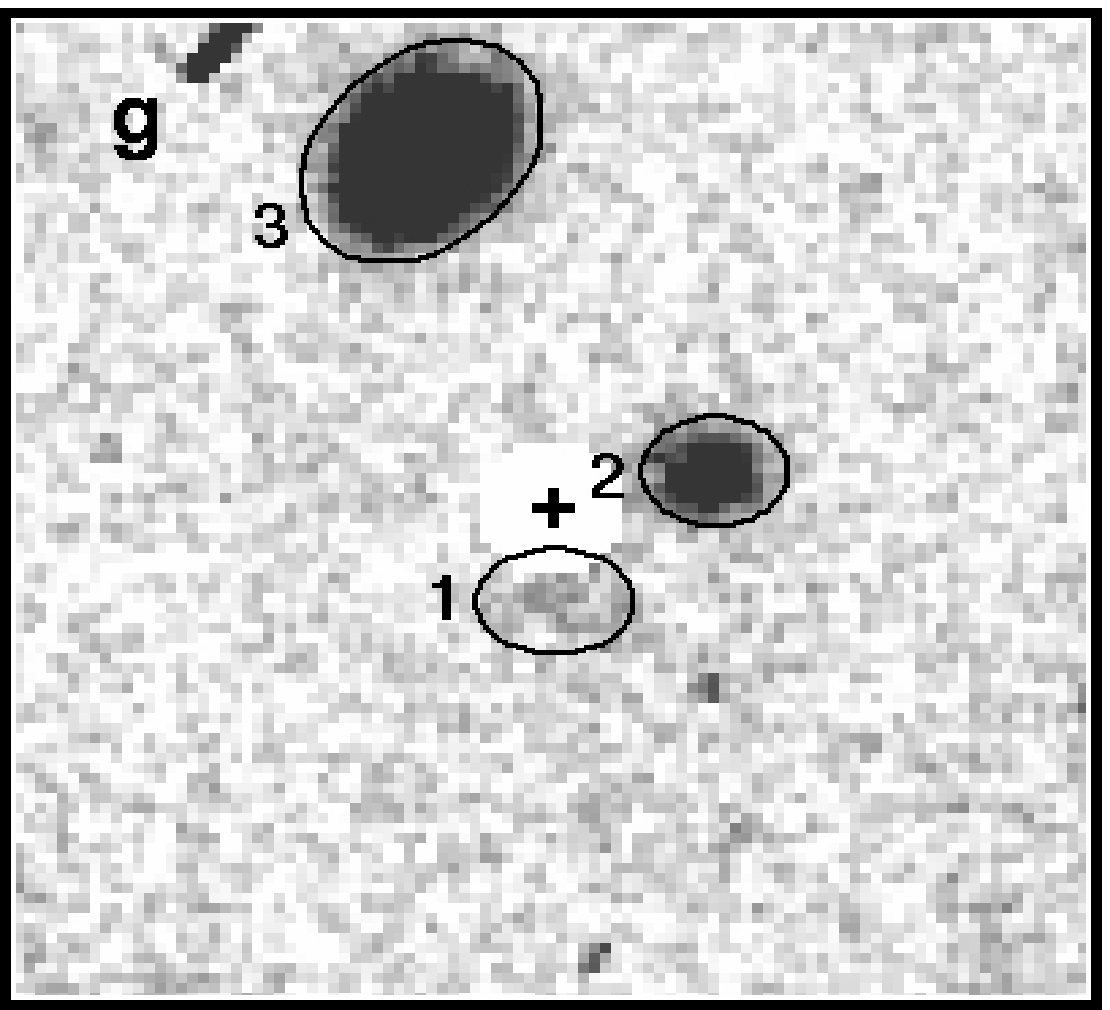}
\includegraphics[height=2.95in,height=2.95in]{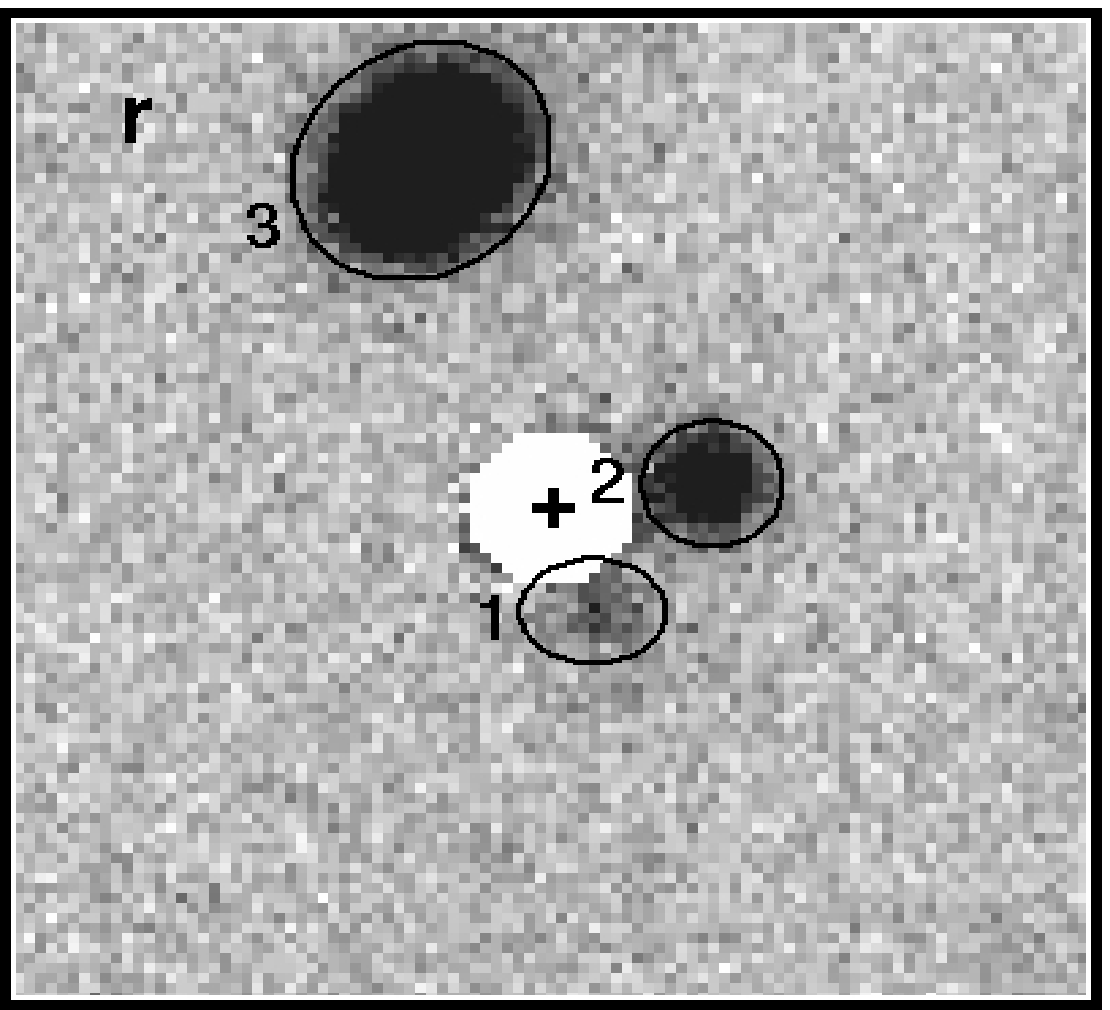}
\includegraphics[height=2.95in,height=2.95in]{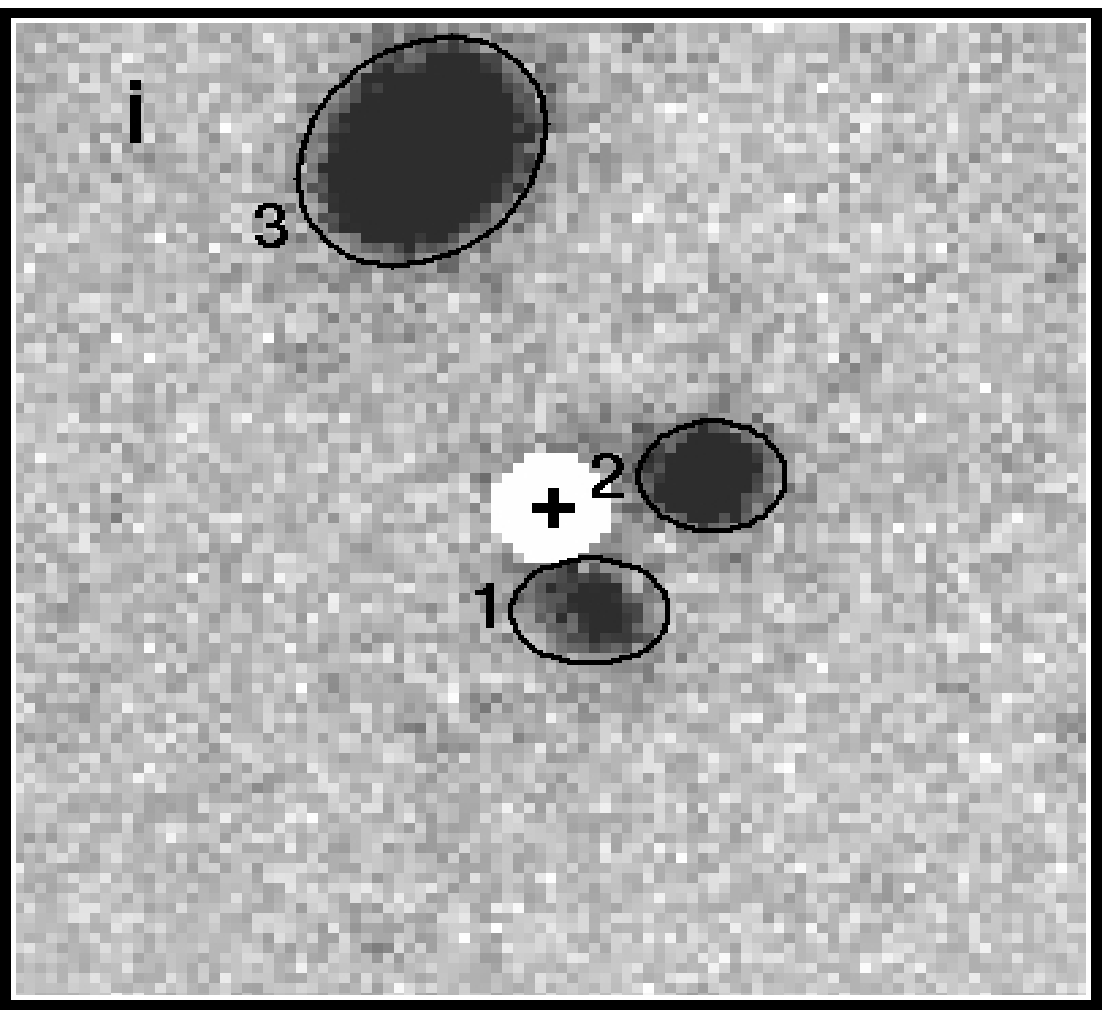}
\caption{30$\arcsec \times$ 30$\arcsec$ PSF subtracted $u', g', r',
i'$ images of the field  1715+5747. This field has a subDLA system at
$z_{abs}$ = 0.5579. None of the three objects in this field is a 
galaxy at the absorption redshift, and so the absorber galaxy in this
 field remains unidentified (\S3.1.4).
The image shown above corresponds to  $\approx$
220 $\times$ 220 kpc$^{2}$ at the absorber redshift.  The quasar PSF
subtraction residuals have been masked,  and the position of the
quasar is marked by a ``+''.  The track northeast of Object 3 is a
cosmic ray as are the two sources south and southwest of Object 1 in
the $g'$-band image.  North is up and east is to the left.}
\label{1715optImages}
\end{figure*}

\begin{figure*}
\includegraphics[height=2.95in,height=2.95in]{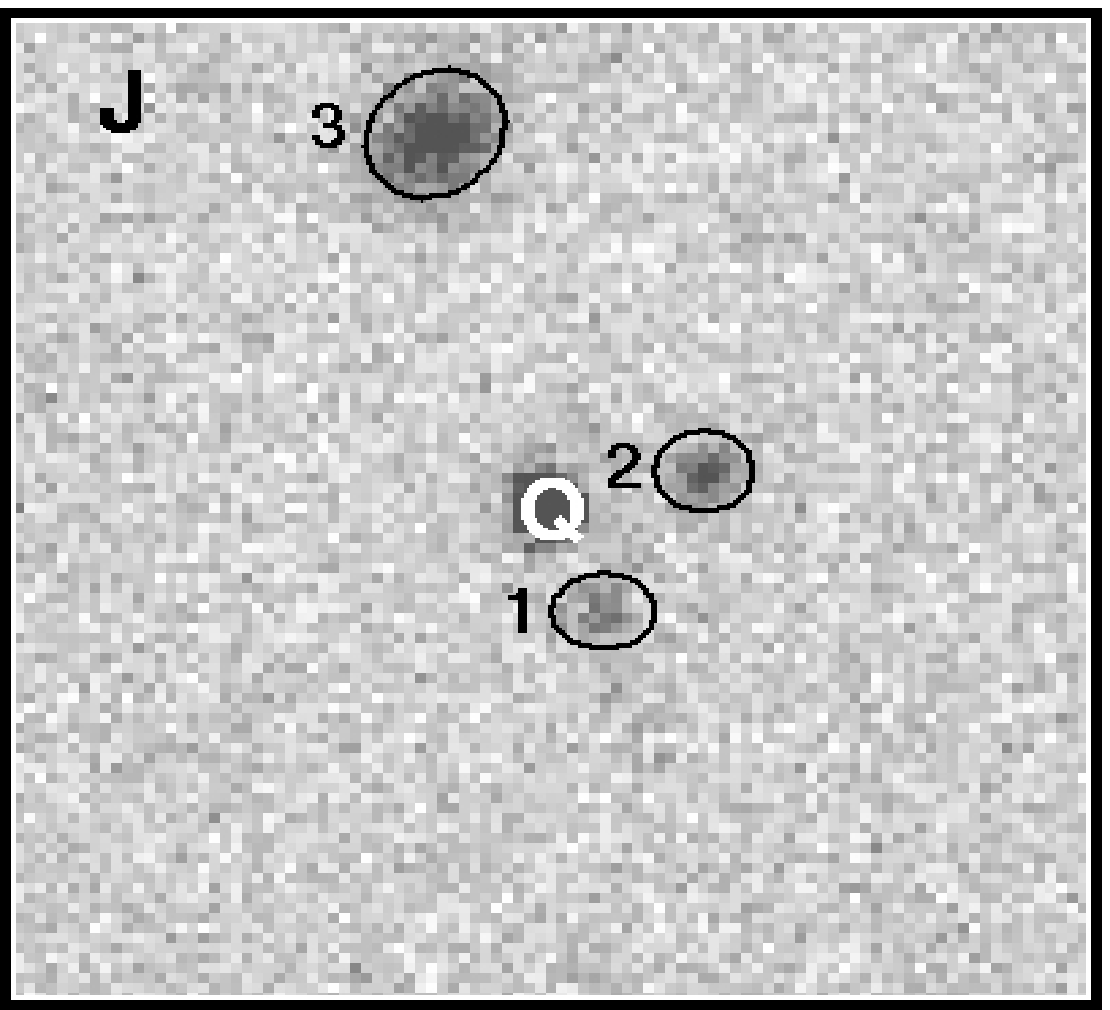}
\includegraphics[height=2.95in,height=2.95in]{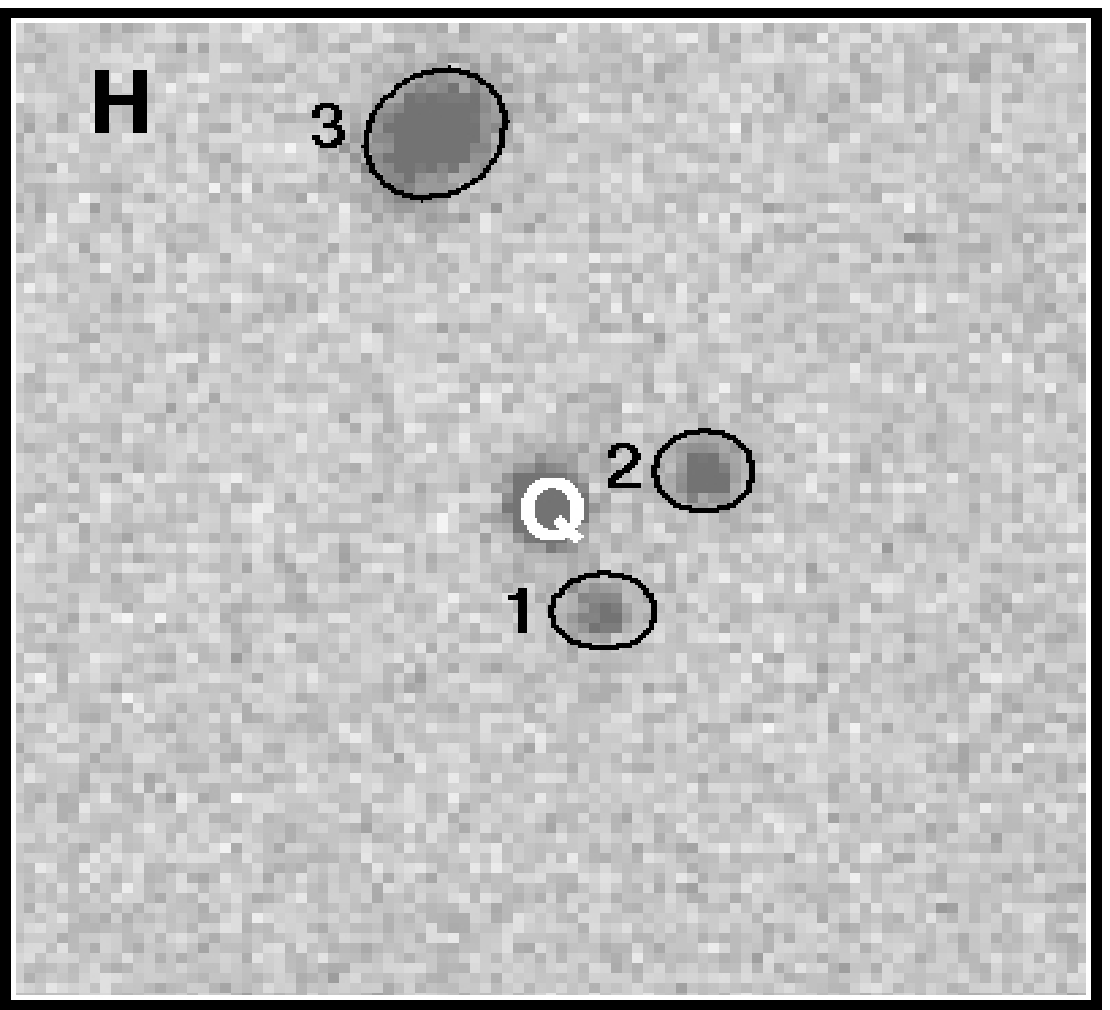}
\includegraphics[height=2.95in,height=2.95in]{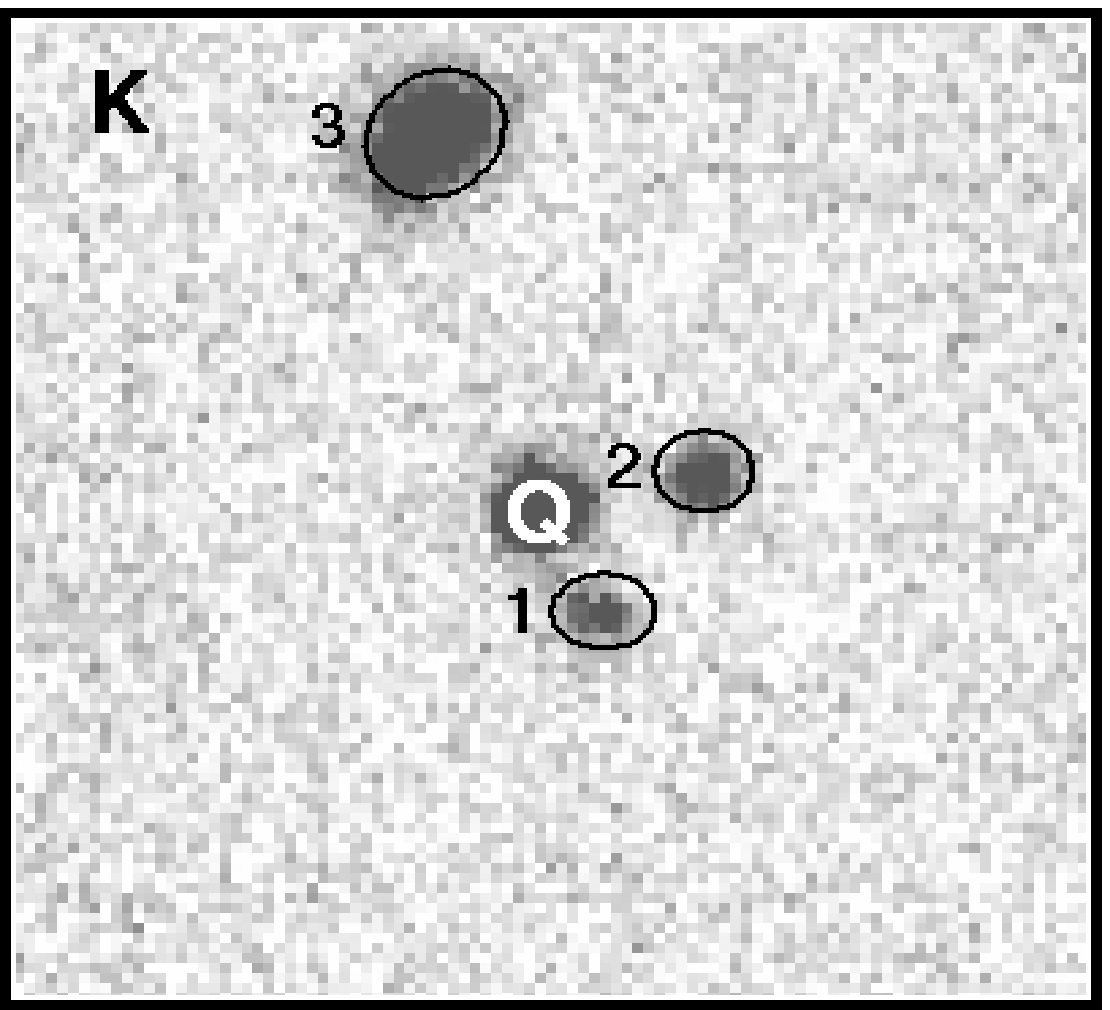}
\caption{30$\arcsec \times$ 30$\arcsec$ $J, H, K$ images of the field
1715+5747.  This field has a subDLA system at $z_{abs}$ = 0.5579. None of 
the three objects in this field is a 
galaxy at the absorption redshift, and so the absorber galaxy in this
 field remains unidentified (\S3.1.4). The
images shown above correspond to $\approx$ 220 $\times$ 220 kpc$^{2}$
at the absorber redshift. The quasar is marked by the letter ``Q''.
The quasar PSF could not be subtracted  as there were no suitable PSF
stars in the field. North is up and east is to the left.}
\label{1715IRImages}
\end{figure*}

This is a case where no galaxy is identified as the absorber.  The
sightline towards the quasar 1715+5747 (SDSS J171539.86+574722.2)
contains a subDLA system at $z_{abs}$ = 0.5579 with a column density
of $\log N_{HI} = 19.18^{+0.15}_{-0.18}$ $\rm cm^{-2}$ (RTN06). A
complete optical and  infrared dataset was obtained for this
field. Figures \ref{1715optImages} and \ref{1715IRImages} show that
only three objects are detected within 100 kpc of the quasar at
$z_{abs}$.  PSF subtractions were carried out on the optical images
and no objects were detected within the subtracted region. The quasar
PSF could not be subtracted on the infrared images as there were  no
suitable PSF stars in the field.  Object 1 is located 3.5$\arcsec$
from the quasar sightline which corresponds to 22.3 kpc at the
absorption redshift.  Objects 2 and 3 have $z_{phot} = 0.621\pm0.085$
and $0.398\pm0.053$, respectively, according  to the SDSS  database.

\begin{figure*}
\includegraphics[width=3.0in]{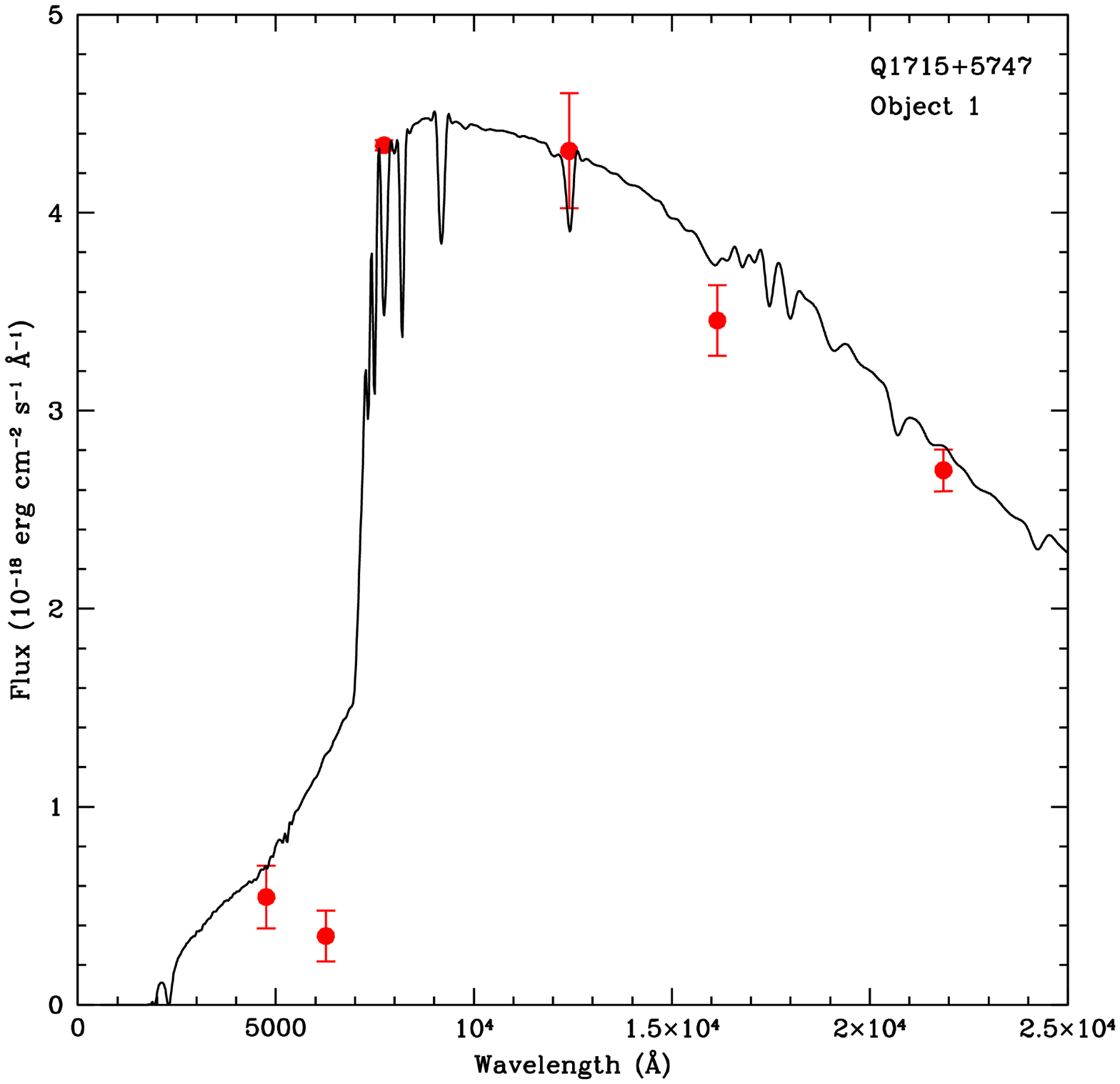}
\includegraphics[width=3.0in]{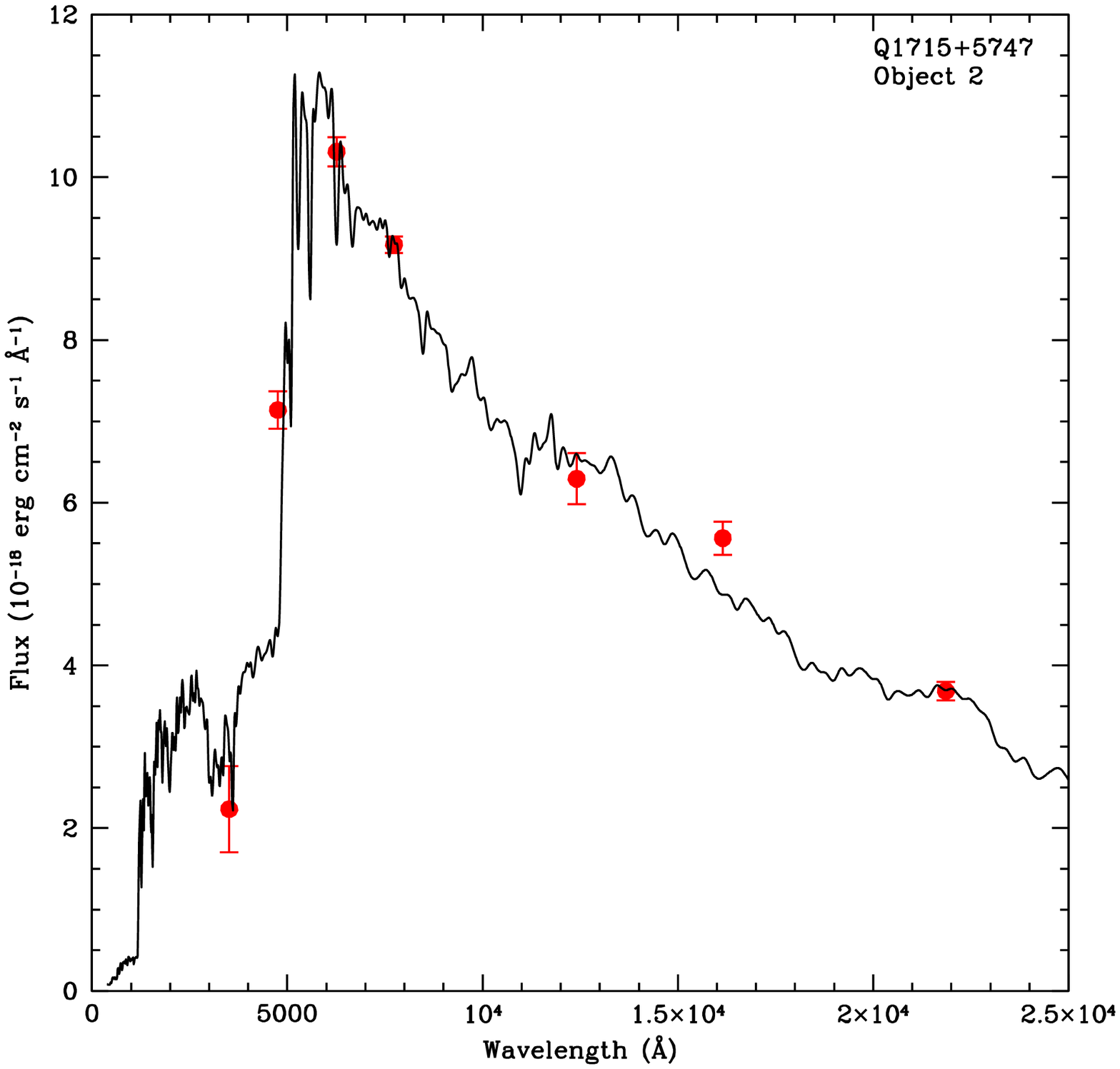}
\includegraphics[width=3.0in]{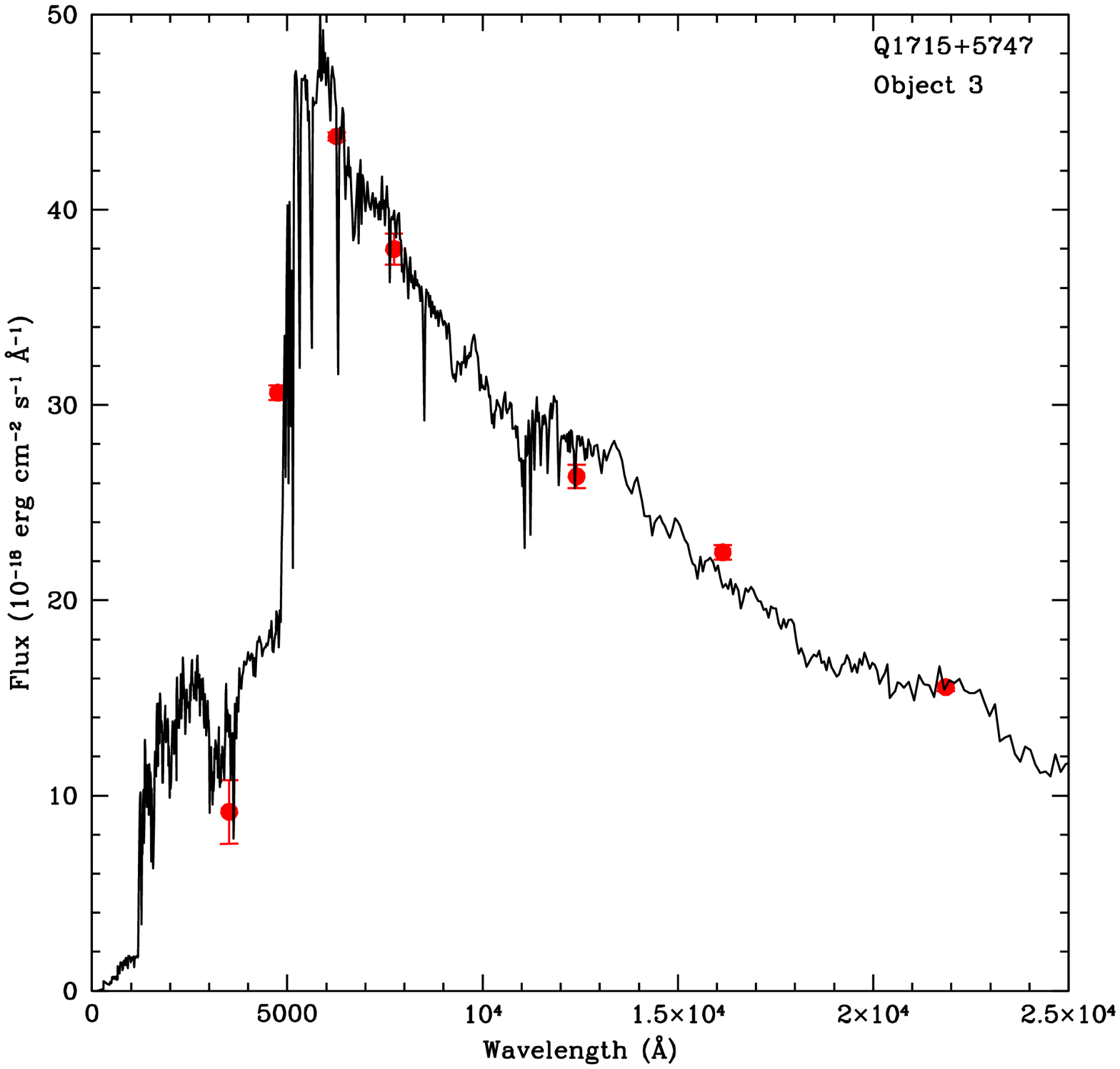}
\caption{The curves are the best-fit stellar population synthesis
models to our  photometry (solid circles) for Objects in the 1715+5747
field.  The best-fit model parameters are listed in Table
\ref{1715photozdata}. None of the three galaxies is at the absorption
redshift (\S3.1.4)}
\label{1715photoz}
\end{figure*}

The best-fit stellar population synthesis model to our photometry for
Objects 1, 2, and 3 are shown in Figure \ref{1715photoz}, and the
model  parameters are listed in Table \ref{1715photozdata}.  The
photometric redshift we derive for Object 1 does not match the
absorption redshift. While the SDSS photometric redshift for Object 2
is consistent with the redshift of the absorption system and our
optical photometric measurements agree well with those measured by the
SDSS, the addition of our IR data results in a very different
photometric redshift for Object 2. Again, as was the case for Object 2
in the 0735+178 field, the addition of IR data was crucial  for the
determination of the galaxy's redshift. With regards to Object 3, we
derive a photometric redshift that is  consistent with the one
obtained by the SDSS.  

Thus, none of the objects detected in this
field have photometric redshifts consistent with the absorption
redshift. Based on their proximity to the quasar sightline, one  might
expect either Object 1 or 2 to be the absorbing galaxy. However, until
spectroscopic data or better photometry are  available that might
prove our results to be incorrect, we consider  our data on this field
to be inconclusive, i.e., we do not have an absorbing  galaxy
identification. It may be one of the cases where the absorbing galaxy is
$<0.8\arcsec$ (5.2 kpc) from the quasar sightline, or at a larger 
impact parameter and fainter than the
brightness limit of our {\it K}-band data, $L_K$ = 0.09$L^{\star}_K$.

\begin{landscape}
\begin{table}
\scriptsize
  \caption{1715+5747: Optical Photometry}
  \begin{tabular}{c r r r c c c c c c c c}
\hline
Object & $\Delta\alpha$\tablenotemark{a} & $\Delta\delta$\tablenotemark{a} & $\theta$\tablenotemark{a} &
$u'\pm\sigma_{u'}$ & DS $(N_{pix})$\tablenotemark{b} & 
$g'\pm\sigma_{g'}$ & DS $(N_{pix})$\tablenotemark{b} & 
$r'\pm\sigma_{r'}$ & DS $(N_{pix})$\tablenotemark{b} &
$i'\pm\sigma_{i'}$ & DS $(N_{pix})$\tablenotemark{b} \\
 & $\arcsec$  & $\arcsec$  & $\arcsec$ & \multicolumn{8}{c}{} \\
\hline
 QSO& 0.0 & 0.0 & 0.0 & 19.69 $\pm$ 0.15 & 21.9 (496) & 18.21 $\pm$ 0.04 & 40.5 (758) & 18.44 $\pm$ 0.04 & 18.4
(770) & 18.55 $\pm$ 0.05 & 20.5 (424) \\ 
1  & $-$1.4 & $-$3.2 & 3.5 & \nodata & \nodata & 24.87 $\pm$ 0.41 & 2.1 (34) & 24.76 $\pm$ 0.53 &
2.0 (21) & 21.56 $\pm$ 0.08 & 3.5 (160) \\ 
2  & $-$4.3 & +0.9 & 4.4 & 23.99 $\pm$ 0.27 & 2.7 (78) & 22.07 $\pm$ 0.07 & 4.1 (224) & 21.07
$\pm$ 0.05 & 4.4 (282) & 20.74 $\pm$ 0.06 & 5.0 (230) \\ 
3  & +3.2 & +11.1 & 11.5 & 22.46 $\pm$ 0.19 & 3.3 (266) & 20.49 $\pm$ 0.05 & 8.4
(456) & 19.51 $\pm$ 0.04 & 8.4 (630) & 19.20 $\pm$ 0.05 & 8.8 (539) \\
\hline
\vspace{-0.5cm}
\tablenotetext{a}{Same as for Table \ref{0153mags}.}
\tablenotetext{b}{Same as for Table \ref{0153mags}.}
\label{1715optmags}
\end{tabular}
\end{table}

\begin{table}
\scriptsize
  \caption{1715+5747: Infrared Photometry}
  \begin{tabular}{c r r r c c c c c c}
\hline
Object & $\Delta\alpha$\tablenotemark{a} & $\Delta\delta$\tablenotemark{a} & $\theta$\tablenotemark{a} &
$J\pm\sigma_{J}$ & DS $(N_{pix})$\tablenotemark{b} & 
$H\pm\sigma_{H}$ & DS $(N_{pix})$\tablenotemark{b} & 
$K\pm\sigma_{K}$ & DS $(N_{pix})$\tablenotemark{b}\\
 & $\arcsec$  & $\arcsec$  & $\arcsec$ & \multicolumn{6}{c}{} \\
\hline
 QSO& 0.0 & 0.0 & 0.0 & 18.07 $\pm$ 0.01 & 11.8 (82) & 17.82 $\pm$ 0.01 & 12.2 (103) & 17.28 $\pm$ 0.01 & 16.0 (139) \\ 
1  & $-$1.4 & $-$3.2 & 3.5 & 20.54 $\pm$ 0.08 & 2.8 (35) & 20.21 $\pm$ 0.06 & 3.4 (41) & 19.82 $\pm$ 0.04 & 3.2 (68) \\ 
2  & $-$4.3 & +0.9 & 4.4 & 20.13 $\pm$ 0.06 & 3.2 (46) & 19.69 $\pm$ 0.04 & 3.7 (61) & 19.48 $\pm$ 0.03 & 4.0 (74) \\ 
3  & +3.2 & +11.1 & 11.5 & 18.57 $\pm$ 0.03 & 3.5 (174) & 18.17 $\pm$ 0.02 & 4.6 (196) & 17.92 $\pm$ 0.01 & 5.0 (249) \\ 
\hline
\vspace{-0.5cm}
\tablenotetext{a}{Same as for Table \ref{0153mags}.}
\tablenotetext{b}{Same as for Table \ref{0153mags}.}
\end{tabular}
\end{table}

\begin{table}
  \caption{1715+5747: Photometric Redshift Fits\tablenotemark{a}}
  \begin{tabular}{crrccccc}
\hline
\multicolumn{3}{c}{Galaxy} & \multicolumn{5}{|c}{Stellar Population Synthesis Model Parameters}\\
\hline
\# & $\theta$\tablenotemark{b} & $b$ & Age & $\tau$ & $E(B -V)$ & $Z$ & $z_{phot}\pm\sigma_{z_{phot}}$ \\
 & $\arcsec$ & kpc & Gyr & Gyr & & \\
\hline
1 & 3.5 & 22.3 & 1.00 & 0.10 & 0.50 & 0.0001 & 0.890$\pm$0.066 \\
2 & 4.4 & 28.6 & 0.50 & 0.10 & 0.20 & 0.0500 & 0.287$\pm$0.044 \\
3 & 11.5& 74.4 & 0.50 & 0.10 & 0.20 & 0.0500 & 0.341$\pm$0.107 \\
\hline
\vspace{-0.5cm}
\label{1715photozdata}
\tablenotetext{a}{$z_{abs}=0.5579$ }
\tablenotetext{b}{Relative to the quasar}
\end{tabular}
\end{table}
\end{landscape}

\section{Galaxy Properties}

\subsection{Earlier work}

Absorber galaxies that had been previously identified
are presented in Table \ref{IDprevious}.
The first results from our DLA imaging programme that were
presented in Turnshek et al. (2001) and Rao et al. (2003) are
included in this table. Additionally, our current sample
(Table 1) has five fields that were studied by other investigators, 
however, our new images did not alter the earlier conclusions. These
five are also included in Table \ref{IDprevious} with previous
studies referenced.

\mgii absorbers from RTN06 are tabulated in the first section of
Table \ref{IDprevious}, and those not in RTN06 are included in the
second section of the table. 
Column 1 is the quasar designation, column 2  gives the quasar
emission redshift, columns 3 and 4, the \mgii\ absorption
redshift and rest equivalent width, and column 5, the \hi 
column  density of the absorber. Column 6 gives the impact parameter
of the identified galaxy in kpc, and columns 7 and 8 give the galaxy's AB magnitude
and absolute luminosity with respect  to $L^*$ (see \S4.2). The relevant filter
is noted in parentheses.  The
reference for the galaxy's parameters is given in column 9 and the
method by which it was identified is given in column 10. ``Specz''
indicates that a spectroscopic redshift was used to identify the
galaxy, ``Photoz''  indicates that the galaxy's photometric redshift
matched the absorption  redshift, and ``Prox'' indicates that the
closest galaxy to the quasar  sightline was chosen as the absorbing
galaxy. Column 11 indicates the  confidence level, CL, assigned to the
identification (see \S3).  ``Photoz'' and ``Prox'' identifications are
assigned CL = 1 and CL = 2, respectively.  Specz is assigned CL = 1,
except for the $z_{abs}=0.656$ galaxy  in the 1622+239 field. Steidel
et al. (1997) obtained a spectroscopic redshift that matches $z_{abs}$
for the galaxy at $b=99.6$ kpc in this field, but commented that a
galaxy this  faint and this far away could not be the DLA
absorber. More recently,  Kacprzak et al. (2007) have identified this
galaxy as the absorber in their work, and we have adopted this new
interpretation as well. However, we have assigned this galaxy  an
identification confidence level of CL = 2, because we feel that we
cannot be as confident about the validity of this  identification as
we are about the rest of the spectroscopically-identified galaxy
candidates (also see \S6).

\begin{landscape}
\begin{table}
\caption{Absorbing Galaxy Identifications  ({\it Previously known})}
\begin{tabular}{ccccccccccc}
\hline
Quasar & $z_{em}$ & $z_{abs}$ & $W_0^{\lambda2796}$ & $\log N_{HI}$ & Impact par. & $m_{AB}$\tablenotemark{a} &
$L/L^{\star}$ & Ref.\tablenotemark{b} & \multicolumn{2}{c}{ID} \\[.2ex] 
 & & & (\AA) & (cm$^{-2}$) & $b$ (kpc) & & & & Method & CL \\
\hline
\multicolumn{11}{l}{\bf MgII Systems from RTN06:}\\
0002+051  & 1.899 & 0.8514 & 1.09  & 19.08 & 25.9 & 22.90($R$)\tablenotemark{c} & 0.92 & 1 & Specz & 1 \\
0058+019  & 1.959 & 0.6127 & 1.666\tablenotemark{d} & 20.04 &  7.3 & 23.7($R$) & 0.17 & 2,3 & Specz & 1 \\ 
0117+213  & 1.491 & 0.5764 & 0.91  & 19.15 &  7.3 & \nodata\tablenotemark{e} & 2.54 & 4 & Specz & 1 \\ 
0302$-$223& 1.409 & 1.0096 & 1.16  & 20.36 &  25.8\tablenotemark{f} & 24.56($R$) & 0.22 & 5,6   & Specz & 1 \\ 
0420$-$014& 0.915 & 0.6331 & 0.75\tablenotemark{d}  & 18.54 & 14.6 & \nodata\tablenotemark{e} & 0.34 & 4 & Specz & 1 \\ 
0454+039  & 1.343 & 0.8596 & 1.45  & 20.67 &  5.5 & 24.76($R$) & 0.14 & 4,5 & Specz & 1 \\
0738+313  & 0.630 & 0.2213 & 0.61  & 20.90 & 19.2 & 19.7($K$) & 0.10 & 7 & Specz & 1 \\
0827+243  & 0.941 & 0.5247 & 2.563\tablenotemark{g} & 20.30 & 32.8 & 18.97($K$) & 1.20 & 8,9 & Specz& 1 \\
0952+179  & 1.478 & 0.2377 & 1.087\tablenotemark{d}  & 21.32 &  4.2 & 22.10($K$) & 0.01 & 8 & Prox & 2 \\
1038+064  & 1.265 & 0.4416 & 0.66  & 18.30 & 56.0 & 21.26($R$)\tablenotemark{c} & 0.29  & 10 & Specz & 1 \\
1127$-$145& 1.187 & 0.3130 & 2.21  & 21.71 & 45.6 & 19.26($I$) & 0.59 & 1,8 & Specz  & 1 \\
1148+386  & 1.304 & 0.5533 & 0.482\tablenotemark{g}  & $<$18 & 20.3 & 21.50($R$)\tablenotemark{c} & 0.48   & 10 & Specz & 1 \\
1209+107  & 2.193 & 0.3930 & 1.00  & 19.46 & 34.9 & 22.22($R$) & 0.14 & 5 & Specz  & 1 \\
1229$-$021& 1.038 & 0.7571 & 0.384\tablenotemark{g}  & 18.36 & 10.5 & 25.66($R$) & 0.02 & 5,11& Prox  & 2 \\
1241+176  & 1.282 & 0.5505 & 0.570\tablenotemark{g}  & 18.90 & 21.4 & 21.96($R$)\tablenotemark{c} & 0.31   & 10 & Specz & 1 \\
1317+277  & 1.014 & 0.6601 & 0.34  & 18.57 & 103.2& 21.91($R$)\tablenotemark{c} & 0.61 & 10 & Specz & 1 \\
1622+239  & 0.927 & 0.6561 & 1.471\tablenotemark{g}  & 20.36 & 99.6 & 22.67($K$) & 0.05 & 10,12\tablenotemark{h}  & Specz   & 2 \\
\nodata & \nodata & 0.8913 & 1.622\tablenotemark{g}  & 19.23 & 21.4 & 21.43($K$) & 0.65 & 12 & Specz  & 1 \\ 
1623+269  & 2.521 & 0.8881 & 1.214\tablenotemark{g}  & 18.66 & 48.2 & 24.20($R$)\tablenotemark{c} & 0.21 & 10 & Specz & 1 \\
1629+120  & 1.792 & 0.5313 & 1.666\tablenotemark{g}  & 20.70 & 17.1 & 19.55($K$) & 1.04 & 8 & Photoz  & 1 \\
2128$-$123& 0.501 & 0.4297 & 0.41  & 19.18 & 48.8 & 20.98($R$)\tablenotemark{c} & 0.35 & 10 & Specz & 1 \\
\hline
\multicolumn{11}{l}{\bf Others:}\\
0051+0041 & 1.189 & 0.7397 & 2.4   & 20.4  & 24.1 & 22.45($I$) & 0.41 & 13 & Specz & 1 \\
0151+045  & 0.404 & 0.1602 & 1.55  & 19.84 & 17.7 & 19.31($R$) & 0.14 & 14 & Specz & 1 \\
0235+164  & 0.940 & 0.5243 & 2.42  & 21.70 & 13.1 & 20.2($I$) & 0.72 & 3 & Specz & 1 \\
0439$-$433& 0.593 & 0.1009 & 1.62  & 19.85 &  7.6 & 17.2($I$) & 0.98 & 3 & Specz & 1 \\
0809+483  & 0.871 & 0.4369 & 2.00  & 20.80 &  8.4 & 19.9($I$) & 0.59 & 3 & Specz & 1 \\
1122$-$1649&2.400 & 0.6850 & 1.83  & 20.45 & 25.2 & 22.4($I$) & 0.35 & 3,15 & Photoz & 1 \\
1137+3907 & 1.027 & 0.7195 & 3.0   & 21.1  & 10.8 & 19.8($K$) & 0.16 & 13 & Specz & 1 \\
1229$-$021& 1.038 & 0.3950 & 2.22  & 20.75 &  7.5 & 22.31($R$)& 0.08 & 5 & Prox & 2 \\
\hline
\vspace{-0.5cm}
\tablenotetext{a}{All quantities have been converted to the ``737'' cosmology.}
\tablenotetext{b}
{1. Kacprzak et al. 2010,
2. Pettini et al. 2000,
3. Chen et al. 2005,
4. Churchill et al. 1996,
5. Le Brun et al. 1997,
6. Peroux et al. 2010,
7. Turnshek et al. 2001,
8. Rao et al. 2003,
9. Steidel et al. 2003,
10. Kacprzak et al. 2007,
11. Steidel et al. 1994,
12. Steidel et al. 1997,
13. Lacy et al. 2003,
14. Guillemin \& Bergeron 1997,
15. Mshar et al. 2007}
\tablenotetext{c}{The $m(F702W)$ magnitudes provided by Kacprzak et al. (2007; 2010) has been 
converted to an $R(AB)$ magnitude.}
\tablenotetext{d}{Measurements of $W^{\lambda2796}_{0}$ have been changed from RTN06 values to 
reflect the more recent measurements of Mathes et al. (in preparation).}
\tablenotetext{e}{Magnitude not provided by Churchill et al. 1996.}
\tablenotetext{f}{We have taken Object \#4, as labeled in Le Brun et al. (1997) and Peroux et al. (2010), as the absorber.}
\tablenotetext{g}{Measurements of $W^{\lambda2796}_{0}$ have been changed from RTN06 values to 
reflect the more recent measurements of Quider et al. (2011).}
\tablenotetext{h}{Steidel et al. (1997) obtained a spectroscopic redshift that matches $z_{abs}$ 
for the galaxy at $b=99.6$ kpc, but commented that a galaxy this faint and this far away could not be the DLA absorber.
However, Kacprzak et al. (2007) have adopted this galaxy as the absorber in their work. We have adopted this
new interpretation as well, but have assigned it CL = 2 (see text).}
\label{IDprevious}
\end{tabular}
\end{table}
\end{landscape}

\subsection{Current work}

Table \ref{IDsummary} provides details on DLA candidate galaxies in
the 55 quasar fields that appear here for the first time. The first 
five columns are as described above for Table \ref{IDprevious}.
 We note here that column density values less than $10^{19}$
cm$^{-2}$ were obtained by fitting Voigt profiles with the same Doppler
broadening parameter as the stronger subDLA and DLA lines 
(Rao \& Turnshek 2000; RTN06). The column
densities should therefore be considered approximate, but 
less than $10^{19}$ cm$^{-2}$. Nevertheless, these are legitimate
\mgii\ absorption systems for which absorbing galaxies
have been identified. Column 6 gives the object in each field that was
identified as the absorbing galaxy candidate. As indicated in \S3, the
numbering is in order of increasing distance from the quasar
sightline. Columns 7 and 8 are the galaxy's impact parameter in
arcsec and kpc, respectively, and column 9 gives the galaxy's
photometric redshift and associated error,  which were determined 
if the field was observed through four or more filters.

Columns 10 and 11 give the galaxy's AB magnitude and absolute
luminosity with respect  to $L^*$. $K$-band AB magnitudes and
luminosities are provided unless the object was not observed (or
detected) in  $K$, in which case the non-$K$ filter is noted.
Our magnitude errors are typically 10 to 20\%. See, for example, the
photometry tables of individual fields in \S3.

\begin{table*}
\caption{Absorbing Galaxy Identifications (\it This work)}
\begin{tabular}{ccccccccccccc}
\hline \hline Quasar & $z_{em}$ & $z_{abs}$ & $W_0^{\lambda2796}$ &
$\log N_{HI}$ & Obj. &  \multicolumn{2}{c}{Impact parameter} &
$z_{phot}\pm\sigma$ & $m_{AB}$\tablenotemark{a} & $L/L^{\star}$ &
\multicolumn{2}{c}{ID} \\[.2ex]  & & & (\AA) & (cm$^{-2}$) & \# &
$\theta$ ($\arcsec$) & $b$ (kpc) & & & & Method & CL \\ \hline
0021+0043 & 1.245 & 0.5203 & 0.533 &19.54 & 1 & 10.8 & 67.3 & 0.549
$\pm$ 0.070 & 19.25 & 0.73 & Photoz  & 1  \\ \nodata &\nodata & 0.9420
& 1.777 & 19.38 & 2 & 10.8 & 85.2 & \nodata & 20.11 & 1.26 & Colour  &
2 \\ 0041$-$266 & 3.053 & 0.8626 & 0.67  &$<$18.00 & 1 & 11.6 & 89.2 &
\nodata & \nodata\tablenotemark{b} & \nodata & Prox  & 3 \\
0107$-$0019 & 0.738 & 0.5260& 0.784 & 18.48 & 1 & 2.6 & 16.3 & 0.564
$\pm$ 0.157 & 20.11 & 0.31 & Photoz  & 1  \\ 0116$-$0043 & 1.282 &
0.9127&  1.379& 19.95 & 1 & 8.1 & 63.4 & 0.717 $\pm$ 0.248 & 20.71 &
0.66 & Photoz  & 1 \\ 0123$-$0058 & 1.551 & 0.8686 & 0.757 &$<$18.62 &
1 & 1.3 & 10.0 & \nodata & 21.1: & 0.41 & Prox  & 3 \\ 0138$-$0005 &
1.340 & 0.7821 & 1.208 &19.81 & 2 & 6.5 & 48.4 & \nodata & 22.34 &
0.11 & Prox  &  3 \\ 0139$-$0023 & 1.384 & 0.6828 & 1.243 &20.60 & 2 &
5.7 & 40.3 & 0.661 $\pm$ 0.075 & 21.60 & 0.14 & Photoz  & 1 \\
0141+339 & 1.450 & 0.4709 & 0.78  & 18.88 & 1 & 5.3 & 31.3 & \nodata &
20.79 & 0.14 & Prox  & 3 \\ 0152+0023 & 0.589 & 0.4818 & 1.340 &19.78
& 2 & 5.3 & 31.7 & 0.518 $\pm$ 0.110 & 21.45($H$) & 0.10 & Photoz  & 1
\\ 0153+0009 & 0.837 & 0.7714 & 2.960 &19.70 & 1 & 4.9 & 36.3 & 0.745
$\pm$ 0.040 & 21.01 & 0.33 & Photoz  & 1 \\ 0253+0107 & 1.035 & 0.6317
& 2.571 &20.78 & 1 & 1.2 & 8.2 & 0.632\tablenotemark{c} & 21.5: & 0.14
& Photoz  & 2 \\ 0254$-$334 & 1.849 & 0.2125 & 2.23  & 19.41 & 1 & 5.5
& 19.0 & 0.030$\pm$0.220 & 22.86 & 0.005 & Photoz  & 2 \\ 0256+0110 &
1.349 & 0.7254 & 3.104 &20.70 & 1 & 2.4 & 17.4 & 0.815$\pm$0.080 &
19.55 & 1.17 & Photoz  & 2 \\ 0710+119 & 0.768 & 0.4629 & 0.62
&$<$18.30 & \nodata & \nodata & \nodata & \nodata & \nodata & \nodata
& \nodata  & \nodata \\ 0735+178 &$>$0.424 & 0.4240 & 1.32  &$<$19.00
& 4 & 12.7 & 70.7 & 0.423 $\pm$ 0.179 & 19.98 & 0.22 & Photoz  & 1 \\
0843+136 & 1.877 & 0.6064 & 0.938\tablenotemark{d}  & 19.56 & 8 & 10.1
& 68.0 & 0.443 $\pm$ 0.281 & 22.46 & 0.05 & Photoz  & 2 \\ 0953$-$0038
& 1.383 & 0.6381 & 1.668 &19.90 & 1 & 11.9& 81.8 & 0.644 $\pm$ 0.150 &
19.78 & 0.71 & Photoz  & 1 \\ 0957+003 & 0.907 & 0.6720 &
1.936\tablenotemark{d}  & 19.59 &  \nodata & \nodata &  \nodata &
\nodata & \nodata & \nodata  & \nodata& \nodata \\ 1009+0036 & 1.699 &
0.9714 & 1.093 &20.00 & 1 & 2.5 & 19.9 & \nodata &
\nodata\tablenotemark{b} & \nodata & Prox  & 3 \\ 1009$-$0026 & 1.244
& 0.8426 & 0.713 & 20.20 & 2 & 5.2 & 39.7 & \nodata & 21.68 & 0.23 &
Prox  & 3 \\ \nodata & \nodata & 0.8866 & 1.900 & 19.48 & \nodata &
\nodata & \nodata & \nodata & \nodata & \nodata & \nodata  &\nodata \\
1019+309 & 1.319 & 0.3461 & 0.70  & 18.18 & 3 & 9.2 & 45.1 & 0.244
$\pm$ 0.167 & 21.81 & 0.03 & Photoz  & 2 \\ 1028$-$0100 & 1.531 &
0.6322 & 1.579 & 19.95 & \nodata & \nodata & \nodata & \nodata &
\nodata & \nodata & \nodata  &  \nodata\\ \nodata & \nodata& 0.7087 &
1.210 & 20.04 & \nodata & \nodata & \nodata & \nodata & \nodata &
\nodata & \nodata  &  \nodata\\ 1047$-$0047 & 0.740 & 0.5727 & 1.063
&19.36 & 1 & 4.7 & 30.7 & \nodata & 21.10 & 0.17 & Prox & 2 \\
1048+0032 & 1.649 & 0.7203 & 1.878 &18.78 & 3 & 7.4 & 53.5 & 0.947
$\pm$ 0.300 & 20.54 & 0.46 & Photoz & 1 \\ 1107+0048 & 1.392 & 0.7404
& 2.952 &21.00 & 2 & 7.9 & 57.7 & \nodata & 23.83(r) & 0.17 & Colour &
3 \\
 1109+0051 & 0.957 & 0.4181 & 1.361 &19.08 & 1 & 1.3 & 7.2 & \nodata &
 22.24($J$) & 0.05 & Prox & 2 \\ \nodata & \nodata & 0.5520 & 1.417
 &19.60 & 3 & 10.0 & 64.2 & 0.645 $\pm$ 0.157 & 22.88 & 0.03 & Photoz
 & 2 \\ 1209+107  & 2.193 & 0.6295 & 2.619\tablenotemark{d}  &20.30 &
 1 & 1.7 & 11.6 & 0.644 $\pm$ 0.100 & 19.89 & 0.55 & Photoz & 1 \\
 1225+0035 & 1.226 & 0.7730 & 1.744 &21.38 & 1 & 8.2 & 60.8 & \nodata
 & \nodata\tablenotemark{b} & \nodata & Prox & 2 \\ 1226+105  & 2.305
 & 0.9376 & 1.646\tablenotemark{d}  &19.41 & 1 & 4.6 & 36.2 &
 0.947$\pm$0.060 & 20.44 & 0.77 & Photoz & 1 \\ 1323$-$0021 & 1.390 &
 0.7160 & 2.229 &20.54 & 1 & 1.4 & 10.1 & \nodata & 21.90($r$) & 0.97
 & Prox & 2 \\ 1342$-$0035 & 0.787 & 0.5380 & 2.256 &19.78 & \nodata &
 \nodata & \nodata & \nodata & \nodata & \nodata & \nodata  & \nodata
 \\ 1345$-$0023 & 1.095 & 0.6057 & 1.177 &18.85 & 2 & 7.6 & 51.0 &
 0.628 $\pm$ 0.040 & 22.24 & 0.06 & Photoz & 1 \\ 1354+258  & 2.006 &
 0.8585 & 1.176\tablenotemark{d}  &18.57 & 2 & 4.0 & 30.7 & \nodata &
 23.78($R$)& 0.48 & Prox & 3 \\ \nodata   &\nodata& 0.8856 &
 0.489\tablenotemark{d}  &18.76 & 3 &12.2 & 94.6 & \nodata &
 23.68($R$)& 0.49 & Prox & 3 \\ 1419$-$0036 & 0.969 & 0.6238 & 0.597
 &19.04 & 3 & 9.6 & 65.3 & 0.499 $\pm$ 0.177\tablenotemark{e} &
 22.89($r'$) & 0.17 & Photoz & 2\\ \nodata & \nodata & 0.8206 & 1.145
 &18.78 & \nodata & \nodata & \nodata & \nodata & \nodata & \nodata &
 \nodata & \nodata \\ 1426+0051 & 1.333 & 0.7352 & 0.857 & 18.85 & 1 &
 5.0 & 36.4 & \nodata & 22.96($r'$) & 0.44 & Prox & 2 \\ \nodata &
 \nodata & 0.8424 & 2.618 & 19.65 & \nodata & \nodata & \nodata &
 \nodata & \nodata & \nodata & \nodata  & \nodata \\ 1431$-$0050 &
 1.190 & 0.6085 & 1.886 & 19.18 & 1 & 2.6 & 17.5 & 0.737 $\pm$ 0.224 &
 21.2: & 0.17 & Photoz & 2 \\ \nodata & \nodata & 0.6868 & 0.613 &
 18.40 & 2 & 3.3 & 23.4 & \nodata & 23.2($r$) & 0.28 & Colour & 2 \\
 1436$-$0051 & 1.275 & 0.7377 & 1.142 & 20.08 &\nodata  & \nodata &
 \nodata & \nodata & \nodata & \nodata & \nodata &  \nodata\\ \nodata
 &\nodata & 0.9281 & 1.174 & $<$18.82 & \nodata & \nodata & \nodata &
 \nodata & \nodata & \nodata & \nodata & \nodata\\ 1437+624  & 1.090 &
 0.8723 & 0.71  &$<$18.00 & 2 & 9.5 & 73.3 & 0.694 $\pm$ 0.190 & 20.70
 & 0.60 & Photoz & 1 \\ 1521$-$0009 & 1.318 & 0.9590 & 1.848 &19.40 &
 1 & 8.5 & 67.4 & \nodata & 23.46($J$) & 0.10 & Colour & 2 \\
 1525+0026 & 0.801 & 0.5674 & 1.852 &19.78 & 1 & 4.8 & 31.3 & \nodata
 & 19.59 & 0.66 & Prox & 3 \\ 1704+608 & 0.371 & 0.2220 &
 0.562\tablenotemark{f} & 18.23 & 2 & 7.7 & 27.5 & 0.220 $\pm$ 0.050 &
 19.63 & 0.08 & Photoz & 1 \\ 1714+5757 & 1.252 & 0.7481 & 1.099
 &19.23 & 1 & 2.4 & 17.6 & \nodata & 24.33($r$) & 0.13 & Prox & 3 \\
 1715+5747 & 0.697 & 0.5579 & 1.001 &19.18 & \nodata & \nodata &
 \nodata & \nodata &\nodata  & \nodata & \nodata & \nodata \\
 1716+5654 & 0.937 & 0.5301 & 1.822 &19.98 & 1 & 1.1 & 6.9 &
 0.5301\tablenotemark{c} & 23.44($r$) & 0.07 & Photoz & 2 \\ 1722+5442
 & 1.215 & 0.6338 & 1.535 &19.00 & 2 & 6.6 & 45.2 & \nodata &
 24.10($r'$) & 0.09 & Colour & 3 \\ 1727+5302 & 1.444 & 0.9448 & 2.832
 &21.16 & 1 & 3.1 & 24.5 & 0.9448\tablenotemark{c} & 23.14 & 0.08 &
 Photoz & 2 \\ \nodata & \nodata & 1.0312 & 0.922 &21.41 & 2 & 3.6 &
 29.0 & \nodata & 20.94 & 0.71 & Prox & 2 \\  1729+5758 & 1.342 &
 0.5541 & 1.836 &18.60 & 2 & 5.6 & 36.0 & 0.503 $\pm$ 0.140 & 20.56 &
 0.26 & Photoz & 1 \\ 1733+5533 & 1.072 & 0.9981 & 2.173 &20.70 & 1 &
 8.4 & 67.2 & \nodata & 24.4($g$) & 0.47 & Colour & 2 \\ 1857+566 &
 1.578 & 0.7151 & 0.65  &18.56 & \nodata  & \nodata  & \nodata  &
 \nodata & \nodata  & \nodata  & \nodata  & \nodata \\ 2149+212  &
 1.538 & 0.9114 & 0.72  &20.70 & 1 & 1.7 & 13.3 & \nodata & 22.31($I$)
 & 0.31 & Prox & 2 \\ \nodata & \nodata & 1.0023 & 2.46  &19.30 & 3 &
 5.5 & 44.0 & \nodata & 22.37($I$) & 0.41 & Prox & 3 \\ \hline
\label{IDsummary}
\end{tabular}
\end{table*}

\begin{table*}
	\addtocounter{table}{-1}
\caption{Continued}
\begin{tabular}{ccccccccccccc}
\hline \hline Quasar & $z_{em}$ & $z_{abs}$ & $W_0^{\lambda2796}$ &
$\log N_{HI}$ & Obj. &  \multicolumn{2}{c}{Impact parameter} &
$z_{phot}\pm\sigma$ & $m_{AB}$\tablenotemark{a} & $L/L^{\star}$ &
\multicolumn{2}{c}{ID} \\[.2ex]  & & & (\AA) & (cm$^{-2}$) & \# &
$\theta$ ($\arcsec$) & $b$ (kpc) & & & & Method & CL \\ \hline
2212$-$299  & 2.706 & 0.6329 & 1.15\tablenotemark{f}  &19.75 & 1 & 2.3
& 16.0 & \nodata & 20.90 & 0.25 & Colour & 2 \\ 2223$-$052  & 1.404 &
0.8472 & 0.586\tablenotemark{f}  &18.48 & 1 & 6.9 & 52.8 & \nodata &
24.62($B$) & 0.32 & Prox & 2\\ 2328+0022 & 1.308 & 0.6519 & 1.896
&20.32 & 1 & 1.7 & 11.8 & 0.815 $\pm$ 0.242 &
22.61($r'$)\tablenotemark{g} & 0.33 & Photoz & 1 \\ 2334+0052 & 1.040
& 0.4713 & 1.226 &20.65 & 1 & 5.5 & 32.5 & 0.4713\tablenotemark{c} &
20.37 & 0.20 & Photoz & 2 \\ 2353$-$0028 & 0.765 & 0.6044 & 1.601
&21.54 & 1 & 4.9 & 32.9 & 0.844 $\pm$ 0.300 & 19.27 & 1.01 & Photoz &
1 \\ \hline
\end{tabular}
$^a$ Apparent $K$ magnitudes are provided, unless otherwise noted. The
symbol ``:'' indicates that the magnitude is uncertain because the
galaxy overlaps with the quasar PSF.\\ $^b$ Observations not
photometric.\\ $^c$ ``BestTemplate'' fit calculated by fixing the
redshift of the stellar population  template at the absorption
redshift, to illustrate that a galaxy template consistent with the
measured photometry exists.\\ $^d$ Measurements of
$W^{\lambda2796}_{0}$ have been changed from RTN06 values to  reflect
the more recent measurements of Quider et al. (2011).\\ $^e$ SDSS
photometric redshift, $AB$-converted SDSS magnitudes are provided.\\
$^f$ Measurements of $W^{\lambda2796}_{0}$ have been changed from
RTN06 values to  reflect the more recent measurements of Mathes et
al. (in preparation).\\ $^g$ SDSS magnitudes provided by S. Zibetti
(private communication).
\end{table*}

The following $M^*$ values were used to determine $L/L^*$ values for
the galaxies listed in Tables \ref{IDprevious} and \ref{IDsummary} in
the filters indicated:

\noindent
$M_B^* = -21.22$ ($0.1<z<0.5$, Dahlen et al. 2005)\\ $M_B^* = -21.46$
($0.5<z<1.0$, Dahlen et al. 2005)\\ $M_g^*= -21.47$ ($0.45<z<0.81$,
Gabasch et al. 2004)\\ $M_g^* = -21.72$ ($0.81<z<1.11$, Gabasch et
al. 2004)\\  $M_R^* = -22.38$ (Dahlen et al. 2005)\\ $M_r^* = -22.12$
($0.4<z<0.8$, Wolf et al. 2003)\\ $M_I^* = -23.4$ ($z=1$, Ilbert et
al. 2005)\\ $M_I^* = -22.17$ ($0.6<z<0.8$, Ilbert et al. 2005)\\
$M_J^* = -22.68$ ($0.1<z<0.5$, Dahlen et al. 2005)\\ $M_J^* = -23.09$
($0.75<z<1.0$, Dahlen et al. 2005)\\ $M_H^* = -22.93$ (Jones  et
al. 2006; assuming no evolution between $z=0.1$ and 0.5)\\ $M_K^* =
-22.86$ (Cirasuolo et al. 2006)\\
 
$K$-corrections for galaxies whose redshifts were obtained using
template fits to the photometry were determined from the template fits
themselves. For the rest, an Sb-type $K$-correction in the observed
filter at the redshift of the absorber was assumed.

Columns 12 and 13 give the method by which the DLA galaxy candidate
was identified, and CL, the confidence level of this identification
(\S3). Of the 66 absorbers in Table \ref{IDsummary}, 17 have
photometric redshifts that match the absorption redshift, and are
assigned CL $= 1$. Thirty-seven identifications were made either with
colours that were consistent with the galaxy  being at the absorption
redshift, the proximity criterion, or photometric-redshift matches
that were only marginally consistent with the absorption
redshift. These are labeled as having CL =  2 or 3, with 2 being the
more confident identification.  No galaxy identification was possible
for 12 absorbers.  Examples of some of these were given in \S3.

\begin{figure*}
\includegraphics[angle=0,width={0.9\textwidth}]{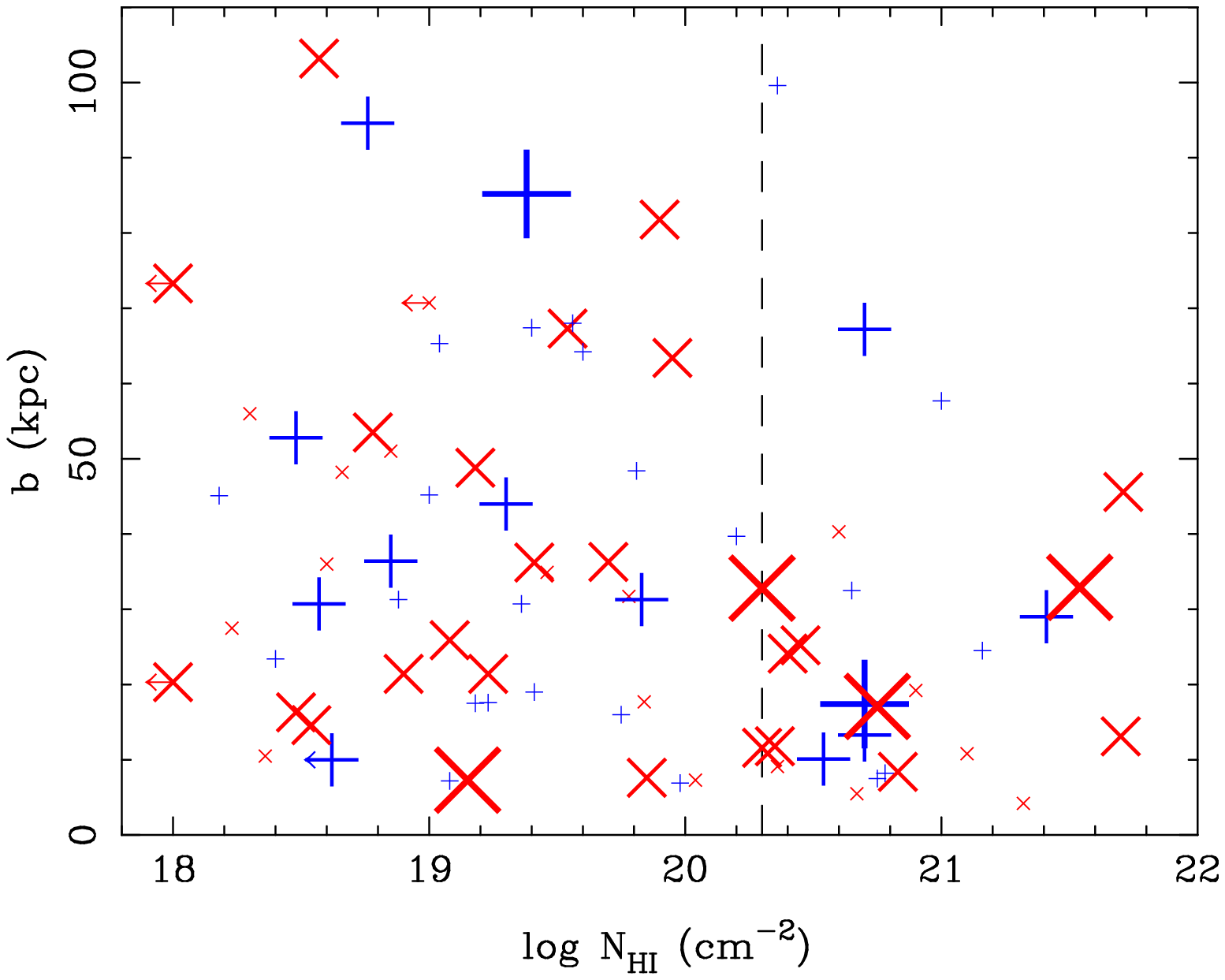}
\caption{Impact parameter, $b$, versus $\log N_{HI}$. Galaxy
luminosity is  represented by the size of the symbol. The smallest
symbols are galaxies with $L \le 0.3 L^*$, the  medium-sized symbols,
$0.3L^* < L \le L^*$, and the largest symbols represent $L > L^*$. Red
crosses represent galaxies with identification confidence level CL =
1, and blue `+' symbols are those that have CL = 2 or 3. The dashed
line is at the DLA threshold column density of $\log N_{HI}=20.3$.  }
\label{NbLgt18}
\end{figure*}

Figure \ref{NbLgt18} shows that the CL = 1 and CL = 2 or 3  samples
have very similar impact parameter and luminosity distributions. Here,
we have included galaxies  from Tables \ref{IDprevious} and
\ref{IDsummary}.  In Figure \ref{NbLgt18}, impact parameter,  $b$, is
plotted versus log \hi column density for the 80 galaxies that have
$b$ and $L$ measurements. Galaxy luminosity is  represented by the
size of the symbol (see caption). We will discuss the $b$-$\log
N_{HI}$ plane in more detail later, however, this plot clearly shows
that the CL values do not cluster with either parameter, or  with
luminosity. In addition, Kolmogorov-Smirnov (KS) tests show that the
two CL samples are drawn from the  same parent population: the KS test
probabilities are 0.08 for  the two luminosity distributions, 0.68 for
the two impact parameter distributions, and 0.94 for the two \hi
column density distributions.  Therefore, this is evidence that we are
{\it statistically} selecting similar candidate galaxies in all of
these samples.  Hereafter,  we will no longer separate  the sample by
CL value, and will explore the properties of all candidate  galaxies
irrespective of their indentification method.

\begin{figure*}
\includegraphics[angle=0,width={0.9\textwidth}]{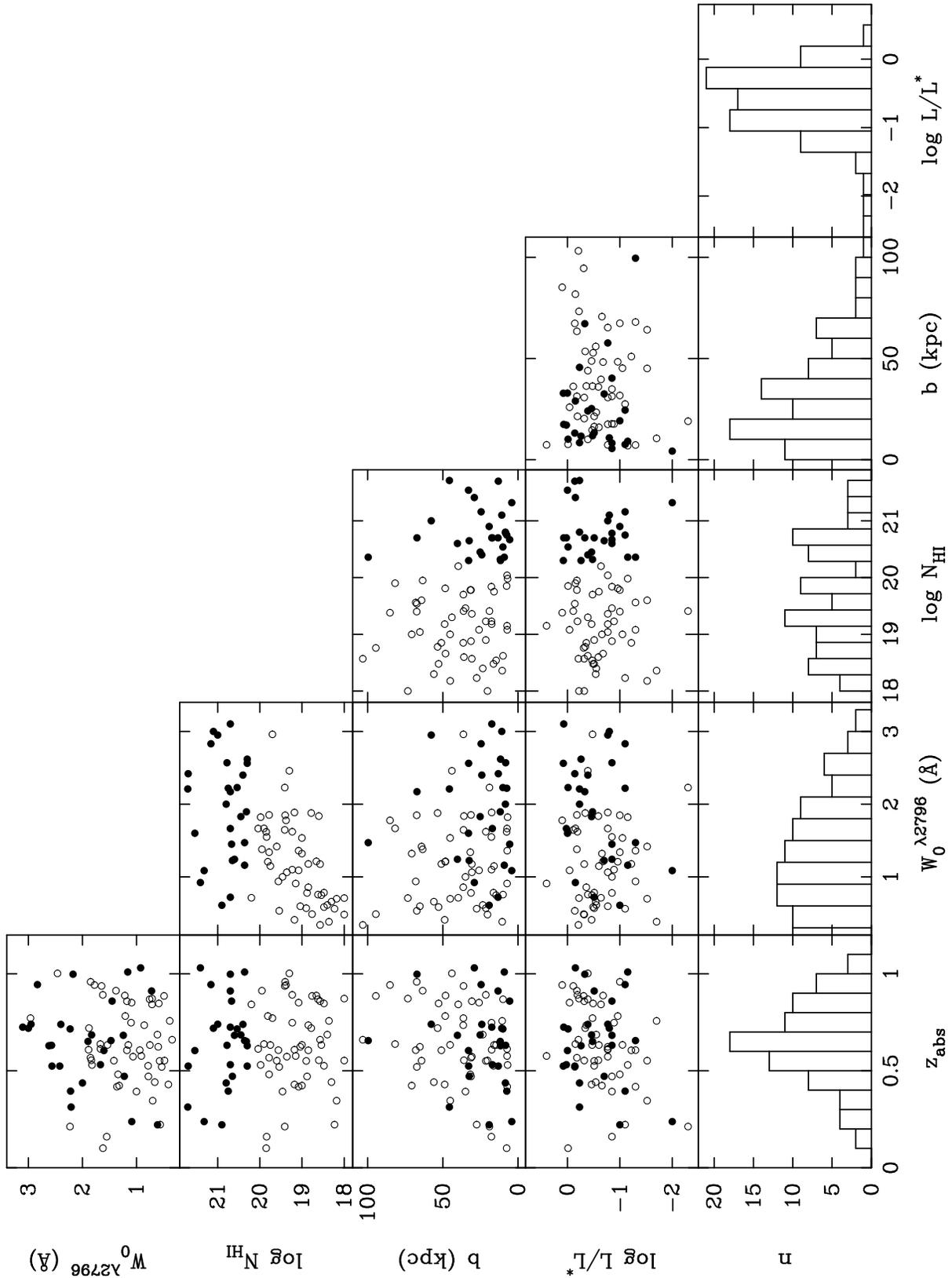}
\caption{Absorber properties at a glance. Filled circles are DLAs and
open circles  are systems with $\log N_{HI} < 20.3$. For clarity, the
upper limit arrows for  column density are not shown. See Tables
\ref{IDprevious} and \ref{IDsummary} for  the four systems with upper
limits for $\log N_{HI}$, but with measured values for  $b$ and $L$.}
\label{stairplot}
\end{figure*}

Our galaxy sample is essentially defined by five parameters:
absorption redshift, $z_{abs}$,  \mgii $\lambda$2796 rest equivalent
width, $W_0^{\lambda2796}$, \hi column density, $\log N_{HI}$, galaxy
impact parameter, $b$, and galaxy luminosity, which  we express as a
fraction of $L^*$, $L/L^*$. The three galaxy  identifications from
Table 15 that have measured $b$ values but no $L/L^*$ measurements are
not included. The sample we analyse includes 80 absorption systems and
their identified galaxies.  Figure \ref{stairplot} shows the
distributions of these properties at a glance.  Open circles are
systems with $\log N_{HI} < 20.3$ and solid circles are DLAs.  In
Table \ref{Spearman}, we provide results from the Spearman rank
correlation test in order to quantify possible correlations among the
various parameters\footnote{Note that while statistical correlations
are derived using $L/L^*$ values, we plot $\log L/L^*$ in figures for
clarity.}.  Column 1 gives the pair of parameters between which the
correlation test was  performed, and column 2 is the Spearman rank
coefficient, $r_S$. A value of $r_S=0$ indicates no correlation, while
$r_S=\pm1$ indicate a perfect correlation and a perfect
anticorrelation, respectively. Column 3 gives $P_S$, the significance
of the deviation of $r_S$ from 0. A small value of $P_S$ indicates
significant  correlation (positive $r_S$) or anticorrelation (negative
$r_S$).  Column 4 gives the number of standard deviations that the
given correlation deviates from the null hypothesis.

\begin{table}
\caption{Spearman Rank Correlation Test Results}
\begin{tabular}{lccc}
\hline  Parameters & $r_S$ & $P(r_S)$ & $N_{\sigma}$ \\ \hline
$z_{abs}$, $W_0^{\lambda2796}$ & 0.10 & 0.390 & 0.9 \\ $z_{abs}$,
$\log N_{HI}$ & $-$0.02 & 0.875 & 0.2 \\ $z_{abs}$,  $b$ & 0.18 &
0.103 & 1.6\\ $z_{abs}$, $L/L^*$ & 0.19 & 0.089 & 1.7 \\
$W_0^{\lambda2796}$, $\log N_{HI}$ & 0.53 & 5.1E-7& 4.7\\
$W_0^{\lambda2796}$, $b$ & $-0.21$ & 0.068 & 1.8\\
$W_0^{\lambda2796}$, $L/L^*$ & 0.09 & 0.419 & 0.8\\ $\log N_{HI}$,
$L/L^*$ & 0.07 & 0.526 & 0.7\\ $\log N_{HI}$, $b$ & $-0.34$ & 0.002 &
3.0\\ $b$, $L/L^*$ & 0.14 & 0.230 & 1.2\\ \hline
\label{Spearman}
\end{tabular}
\end{table}

In addition, Table \ref{bW_KS} lists KS test probabilities  that two
given samples are drawn from the same parent population. In the first
set, $z_{abs}$, $W_0^{\lambda2796}$, $b$  in kpc, and $L/L^*$ are
compared for three different samples of \hi column density: the DLA
($\log N_{HI} \ge 20.3$) and subDLA  ($19 < \log N_{HI} < 20.3$)
samples, and for the subDLA and Lyman limit  system (LLS, $\log N_{HI}
\le 19$) samples.  There are 27 galaxies in the  DLA sample, 30 in the
subDLA sample, and 23 in the LLS sample. Next, the sample is split by
the median value of $W_0^{\lambda2796}$\footnote{The
$W_0^{\lambda2796}$ median value applies to our observed sample and
not to the true median of the $W_0^{\lambda2796}$ distribution over
some minimum and maximum $W_0^{\lambda2796}$ range.},  and then by the
median value of $b$.

\begin{table}
\caption{KS Test Probabilities, $P_{KS}$, for Various Pairs of
Subsamples}
\begin{tabular}{lcc}
\hline \hline Parameter & DLA vs. SubDLA\tablenotemark{a} & SubDLA
vs. LLS\tablenotemark{a}\\ \hline $z_{abs}$ & 0.595 & 0.624\\
$W_0^{\lambda2796}$ & 0.003 & 0.0003\\ $b$ & 0.089 & 0.828 \\ $L/L^*$
& 0.976 & 0.087 \\ \hline & \multicolumn{2}{c}{$W_0^{\lambda2796}\le
1.35$\AA\ vs. $W_0^{\lambda2796} > 1.35$ \AA\tablenotemark{b}}\\
\hline $z_{abs}$ & \multicolumn{2}{c}{0.893}\\ $b$ &
\multicolumn{2}{c}{0.361}\\ $\log N_{HI}$ &
\multicolumn{2}{c}{0.0001}\\ $L/L^*$ & \multicolumn{2}{c}{0.139}\\
\hline & \multicolumn{2}{c}{$b\le 30.7$ kpc vs. $b > 30.7$
kpc\tablenotemark{c}} \\ \hline $z_{abs}$ &
\multicolumn{2}{c}{0.724}\\ $W_0^{\lambda2796}$ &
\multicolumn{2}{c}{0.531}\\ $\log N_{HI}$ &
\multicolumn{2}{c}{0.043}\\ $L/L^*$ & \multicolumn{2}{c}{0.983}\\
\hline
\vspace{-0.5cm} \tablenotetext{a}{The DLA sample: $\log N_{HI} \ge
  20.3$; the SubDLA sample: $19.0<\log N_{HI}< 20.3$; the LLS sample:
  $\log N_{HI}\le 19.0$} \tablenotetext{b}{The median rest equivalent
  width for the sample of 80 systems with identified galaxies is
  $W_0^{\lambda2796} = 1.35$ \AA.}  \tablenotetext{c}{The median
  impact parameter for the sample of 80 identified galaxies is $b =
  30.7 $ kpc.}
\label{bW_KS}
\end{tabular}
\end{table}

We now discuss these correlations in some detail. The  two most
significant correlations are between $\log N_{HI}$ and
$W_0^{\lambda2796}$, and between $\log N_{HI}$ and $b$.

\subsection{Trends with Redshift}

\begin{figure*}
\includegraphics[angle=0,width={0.9\columnwidth}]{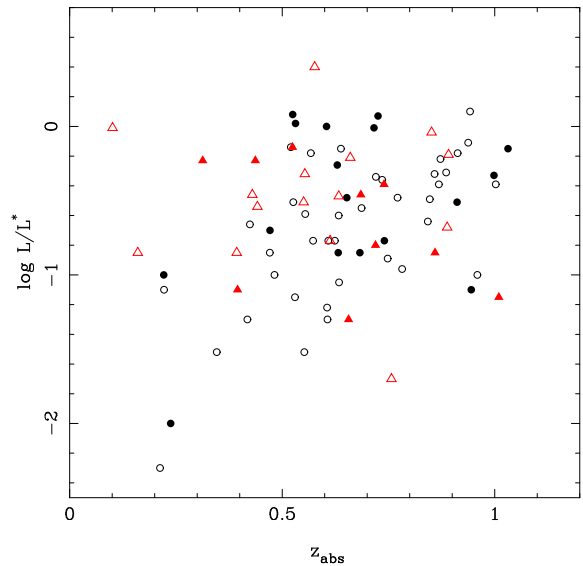}
\caption{ Galaxy luminosity vs. absorption redshift. The black circles
are galaxies that we have identified using our groundbased data.
Solid circles represent DLAs and open circles are systems with $\log
N_{HI} < 20.3$.  The red triangles are galaxy identifications from
other work (see Table \ref{IDprevious}).  Solid triangles are DLAs and
open triangles are systems with $\log N_{HI} < 20.3$.}
\label{Lvsz}
\end{figure*}

The first column in Figure \ref{stairplot} shows the redshift
distributions of the four primary properties of the sample of 80
galaxies: $W_0^{\lambda2796}$, $\log N_{HI}$, $b$, and $L/L^*$.  Table
\ref{Spearman} shows that the strongest correlations, albeit less than
$2\sigma$,   are between $z_{abs}$, and $L/L^*$ and $b$. The
luminosity - redshift correlation arises from the fact that the two
faintest galaxies, those with $\log L/L^* \le -2$, are among the
lowest redshift identifications ($z<0.3$). The two galaxies are the
$z_{abs}=0.2125$ absorber towards 0254$-$334 with  galaxy luminosity
$L=0.005L^*$ and the $z_{abs}=0.2377$ absorber towards 0952+179 with
galaxy luminosity $L=0.01L^*$.  A point of concern might be that these
faint galaxies were detected only because of their low redshifts, and
that similarly faint galaxies at higher redshifts, that are  the true
absorbers, are not being detected. However, Figure \ref{LKvsz} showed
that the $3\sigma$ detection threshold is as faint as $\sim 0.02L^*$
for galaxies (assuming a size of 10 kpc)  in a few fields between
redshifts 0.2 and 1.   While there is always the concern that imaging
studies miss very faint  galaxies, we do not believe that we are
severely limited by this bias in  comparison to other studies. In
fact, our images go deeper than most other similar studies, and most
of the galaxies fainter than $0.1 L^*$ have been identified in this
study.  To illustrate this, Figure \ref{Lvsz} shows the luminosity
distribution of galaxies with redshift. The black data points are from
our groundbased data, and the  red data points are galaxies identified
in other studies. A KS test shows that the two distributions are
similar, with $P_{KS}=0.873$.

The $b$ versus $z_{abs}$ panel in Figure \ref{stairplot} suggests that
there is an upper envelope to the  impact parameter distribution that
is a function of redshift. At the  lowest redshifts ($z_{abs}\la
0.3$), there are no absorbing galaxies  at $b \ga 30$ kpc.  We do not
believe that this is due to a bias, because we are unaware of
low-redshift systems with unidentified absorbing galaxies. For
example, such a bias may have arisen if large impact parameter
galaxies were overlooked assuming that such galaxies couldn't be  the
true absorbers. If this correlation is real, one speculation is that
by $z\sim 0.3$, mergers have resulted in fewer clumps of gas at large
distances  from the centers of galaxies. Also see Section 4.10.

Of course, the trends here are weak: the Spearman rank  test results
in only a $1.6\sigma$ correlation between  $b$ and $z_{abs}$ (Table
\ref{Spearman}), and the KS test between $z_{abs}$  samples split by
median $b$ (30.7 kpc) shows that the two distributions are similar
($P_{KS}=0.724$, Table \ref{bW_KS}).

Table \ref{bW_KS} also shows that the redshift distributions of
samples split by $\log N_{HI}$ or by  $W_0^{\lambda2796}$ are
statistically similar.

\subsection{Dependence of $\log N_{HI}$, $b$, and $L/L^*$ on $W_0^{\lambda2796}$}

The most significant correlation, one that has previously been
recognized (RTN06),  is seen between $W_0^{\lambda2796}$ and $\log
N_{HI}$, with $r_S = 0.53$, $P(r_S)=5.1\times 10^{-7}$, and
$N_{\sigma}=4.7$ (Table \ref{Spearman})\footnote{For the RTN06 sample,
which  included 195 \mgii systems with $0.1<z_{abs}<1.65$, the
$W_0^{\lambda2796}$ versus $\log N_{HI}$ correlation is significant at
the  8.4$\sigma$ level. For the 123 systems with $z_{abs}<1$, we find
the relation to be significant at the 6.1$\sigma$ level. The current
imaging  sample contains 80 systems with $z<1$.}.  The  \hi column
density clearly depends on $W_0^{\lambda2796}$, although the relation
is not tight (see Figure \ref{stairplot}, and Figures 2 and 3 of
RTN06), i.e., one cannot predict the value of  $\log N_{HI}$ given
$W_0^{\lambda2796}$.  As discussed in RTN06, the fraction of MgII
systems that are DLAs increases with $W_0^{\lambda2796}$, and there
are no DLAs with $W_0^{\lambda2796}<0.6$ \AA.  Since this imaging
sample is largely derived from the RTN06 \mgii sample (72 of the 80
systems are from RTN06), the same trend is seen here.

\begin{figure*}
\includegraphics[angle=0,width={0.9\textwidth}]{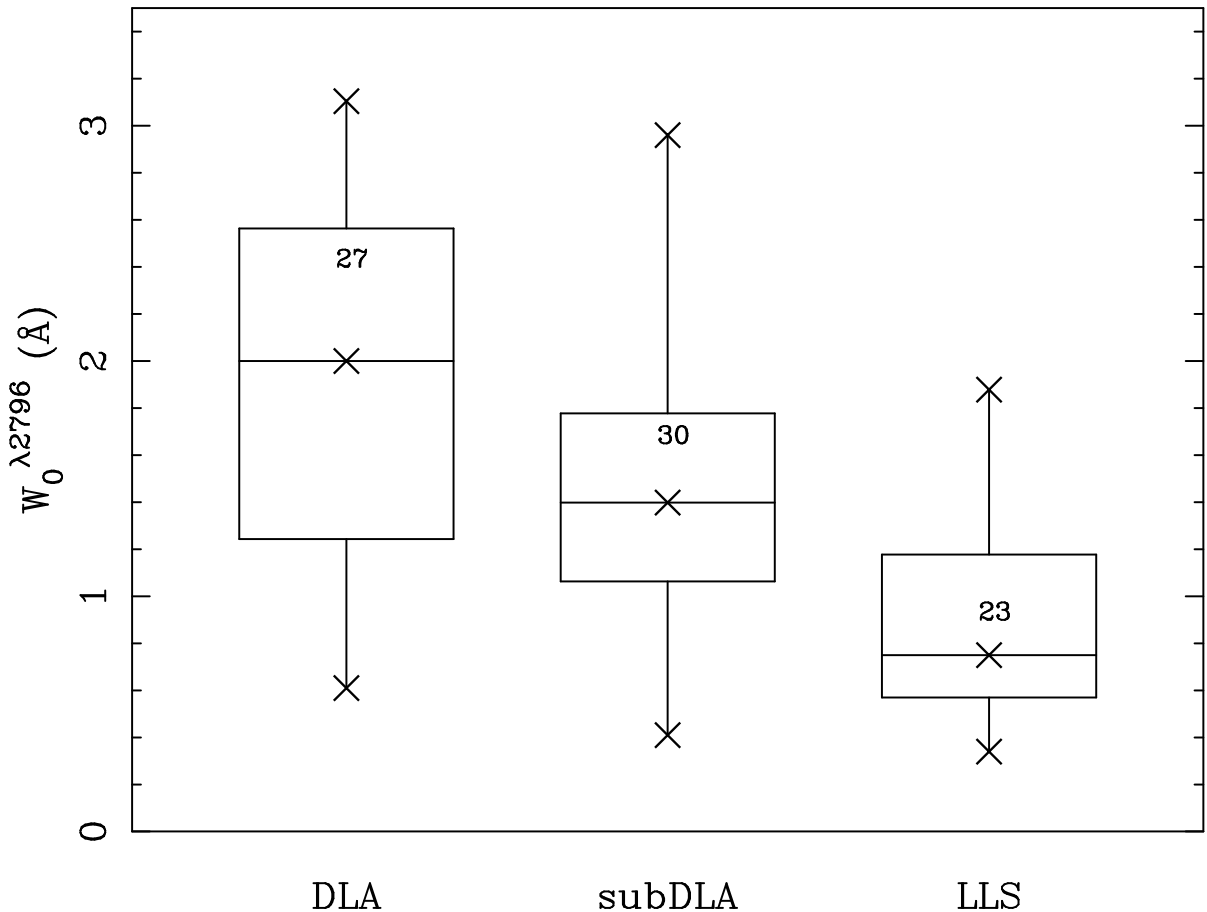}
\caption{A ``box and whisker'' plot showing the distribution of \mgii
$\lambda 2796$  rest equivalent widths for the indicated $N_{HI}$
subsamples. The DLA subsample includes  systems with $\log N_{HI} \ge
20.3$, the subDLA sample has $19.0<\log N_{HI} <  20.3$, and the LLS
sample has $\log N_{HI} \le 19.0$. The three crosses in each box and
whisker plot indicate the minimum, median, and maximum of the
distribution. The bottom and top of the box are the 25th and 75th
percentile  (the lower and upper quartile of the distribution)
respectively. The numbers in each box indicate the number of systems
in each subsample.}
\label{W_boxplot}
\end{figure*}

It is clear from Table \ref{bW_KS} that the $W_0^{\lambda2796}$
distributions of the DLA, subDLA,  and LLS samples are very
significantly different. They differ at the 99.7\% (or $3\sigma$)
confidence level for the DLAs and subDLAs, and at the 99.97\%
($4\sigma$) confidence level for the subDLA and LLS samples. This is
also seen in the KS test probability for the two $N_{HI}$ samples
divided by the median value of $W_0^{\lambda2796}$ ($P_{KS}=0.0001$).
The trends are better visualized in the ``box and whisker'' plots of
Figure \ref{W_boxplot}.  The crosses are the minimum, median, and
maximum values, and the bottom and top edges of the box are the first
and third quartile values of each subsample. The median values of
$W_0^{\lambda2796}$, which are 2.0 \AA\ for the DLA sample, 1.37 \AA\
for the  subDLA sample, and 0.78 \AA\ for the LLS sample, do show a
very significant trend with $\log N_{HI}$. (See also M\'enard \&
Chelouche 2009, who fit a power-law to the median values of $N_{HI}$
from RTN06 as a function of $W_0^{\lambda2796}$.)

Table \ref{Spearman} shows that $b$ and $L/L^*$ do not correlate
significantly with $W_0^{\lambda2796}$. The $W_0^{\lambda2796}$ versus
$b$ correlation is more significant, with $N_\sigma=1.8$. Moreover,
the correlation coefficient is negative reflecting the fact that there
are no large  $W_0^{\lambda2796}$, large $b$ absorbers. The
$W_0^{\lambda2796}$ versus $L/L^*$ correlation is significant only at
the  $0.8\sigma$ level.  Figure \ref{stairplot} shows that if the two
lowest luminosity galaxies are ignored, the $W_0^{\lambda2796}$ versus
$L/L^*$ distribution is essentially uniform with the vast majority of
galaxies having luminosities between 0.1 and 1$L^*$.  We note here
that Chen et al. (2010) fit a power-law model to $W_0^{\lambda2796}$
as a function of a luminosity-scaled impact parameter. Their sample
includes  $W_0^{\lambda2796}$  measurements as low as 0.1\AA\ and
upper limits as low as 0.02\AA, from which they derive a statistically
significant anti-correlation between $W_0^{\lambda2796}$ and the
luminosity-scaled $b$. However, in agreement with our results, their
data show no trend for $W_0^{\lambda2796} \ge  0.3$ \AA, which is the
lower limit of our sample.

Splitting the sample by the median value of $W_0^{\lambda2796}$
results in KS test probabilities  for the four other parameters as
shown in the second section of Table \ref{bW_KS}.  Statistically, the
two $z_{abs}$, $b$, and $L/L^*$ samples are drawn from the same parent
population. The median luminosity is $0.26 L^*$ for the
$W_0^{\lambda2796} < 1.35$ \AA\ sample and  $0.38 L^*$ for the
$W_0^{\lambda2796} >  1.35$ \AA\ sample. The two $N_{HI}$ samples
clearly separate out, with $P_{KS}=0.0001$.

\begin{figure*}
\includegraphics[angle=0,width={0.9\textwidth}]{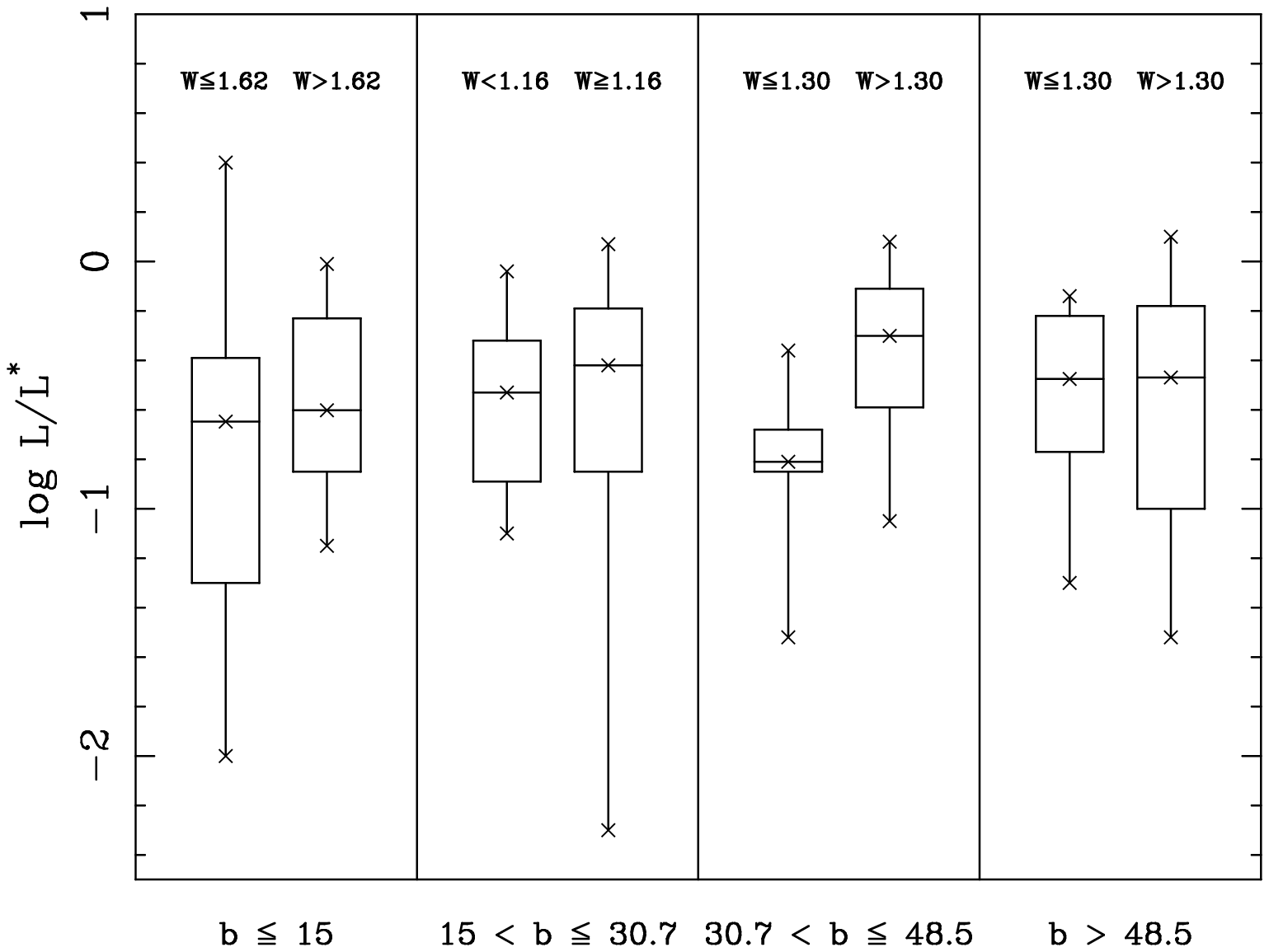}
\caption{Box and whisker diagram showing galaxy luminosity
distributions as a function of  $W_0^{\lambda2796}$ for four bins of
impact parameter. Each of the four $b$ samples,  labeled along the
x-axis in kpc, has 20 systems, and each $W_0^{\lambda2796}$ subsample
has 10 systems. The median value of $W_0^{\lambda2796}$ for each
impact parameter is indicated in units of \AA\ at the top of each
panel.}
\label{LWb_boxplot}
\end{figure*}

To investigate whether brighter galaxies cause stronger \mgii
absorption lines than do fainter ones at the same impact parameter, we
plot box and whisker diagrams of $\log L/L^*$ versus
$W_0^{\lambda2796}$ for four $b$ samples.  Figure \ref{LWb_boxplot}
shows the range of $\log L/L^*$ values for samples split by the median
value of $W_0^{\lambda2796}$ as a function of $b$. Each of the four
$b$ samples,  labeled along the x-axis in kpc, has 20 systems, and
each $W_0^{\lambda2796}$ subsample  has 10 systems. The median value
of $W_0^{\lambda2796}$ for each $b$ sample is indicated in units of
\AA\ at the top of each panel.  It is interesting, and perhaps
unexpected, that except perhaps in the third impact parameter bin,
galaxy luminosity and \mgii rest equivalent width are not correlated.

\subsection{Dependence of $b$ and $L/L^*$ on $\log N_{HI}$}

The third column of Figure \ref{stairplot} shows the $b$ and $L/L^*$
distributions as a function of $\log N_{HI}$. Statistically, we see a
strong,  $3.0\sigma$, correlation between $\log N_{HI}$ and $b$ (Table
\ref{Spearman}), and no significant correlation between  $\log N_{HI}$
and $L/L^*$. The third section of Table \ref{bW_KS} also shows that,
except for the $\log N_{HI}$ subsamples, the $z_{abs}$,
$W_0^{\lambda2796}$, and $L/L^*$  subsamples are statistically similar
when split by the median impact parameter of the sample,  $b=30.7$ kpc.

\begin{figure*}
\includegraphics[angle=0,width={0.45\textwidth}]{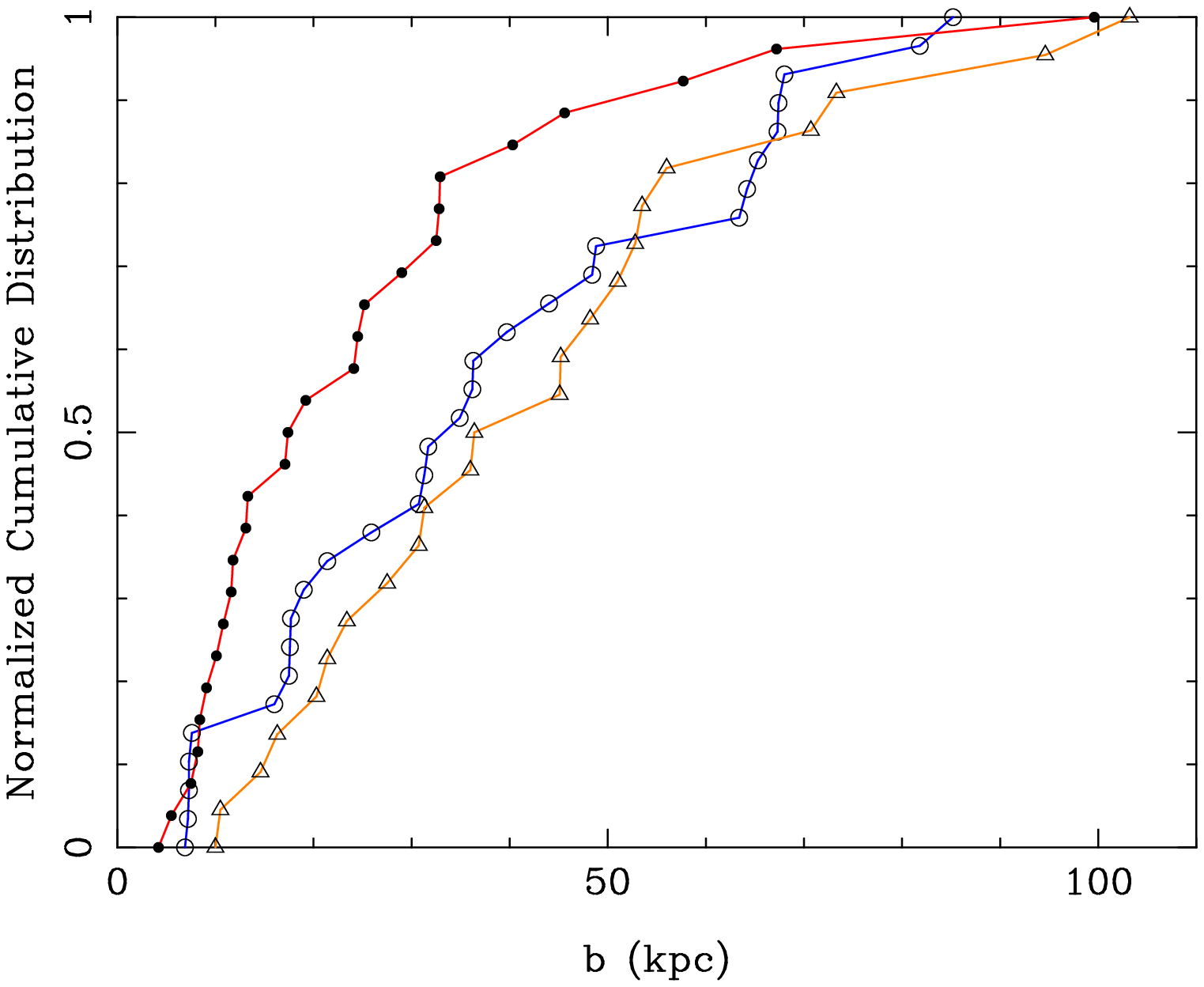} \hfil
\includegraphics[angle=0,width={0.45\textwidth}]{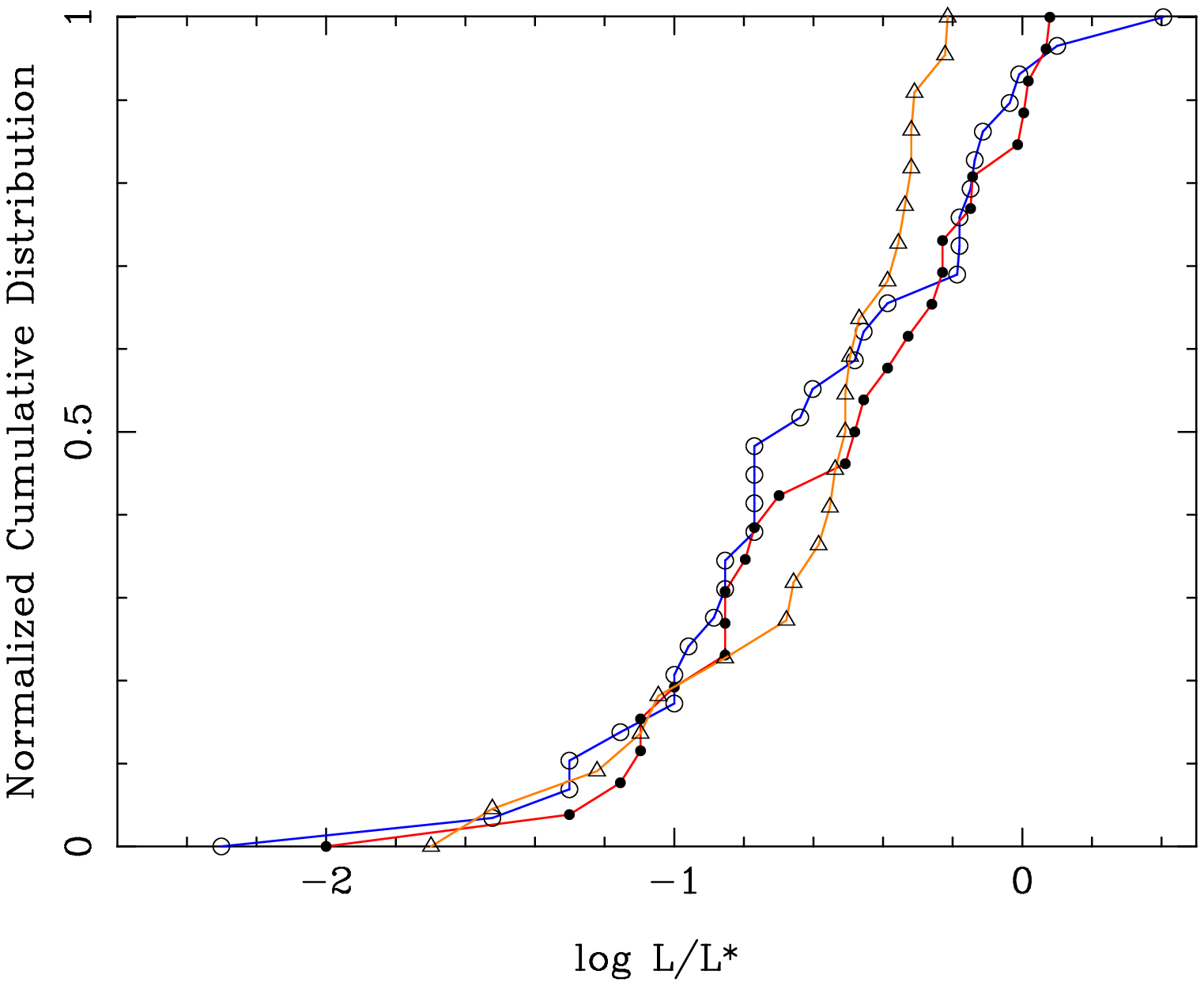}
\caption{Normalized cumulative distributions of impact parameter, $b$,
in the left panel, and luminosity, $\log L/L^*$, in the right
panel. Red solid circles are DLAs, blue open circles are subDLAs, and
orange open triangles are LLSs.}
\label{bL_KSplots}
\end{figure*}

Figure \ref{bL_KSplots} shows the normalized cumulative distributions
of $b$ and $\log L/L^*$ for three samples of \hi column density.  The
solid circles with red lines are $b$ and $\log L/L^*$ values for DLA
galaxies, the open circles with blue lines represent subDLA galaxies,
and the open triangles with  orange lines represent LLS galaxies. The
impact parameter distributions  for the subDLA and LLS galaxies are
very similar with KS test probability $P_{KS}=0.828$  (Table
\ref{bW_KS}). On the other hand the DLA galaxy $b$ distribution is
markedly different, with KS test probability  $P_{KS}=0.022$ that the
DLA and $\log N_{HI} < 20.3$  cm$^{-2}$ (subDLA plus LLS) sample's
galaxy impact parameters are drawn from the same population, i.e., the
null hypothesis is rejected at the $97.8$\% confidence level, or
better than $2\sigma$. Individually for the two non-DLA samples, we
find $P_{KS}(DLA, subLDA)=0.089$ and $P_{KS}(DLA, LLS) = 0.030$. The
two $\log N_{HI}$ subsamples split by median $b$ are also
statistically different at the $\sim 2\sigma$  level with
$P_{KS}=0.043$ (Table \ref{bW_KS}).

\begin{figure*}
\includegraphics[angle=0,width={0.9\textwidth}]{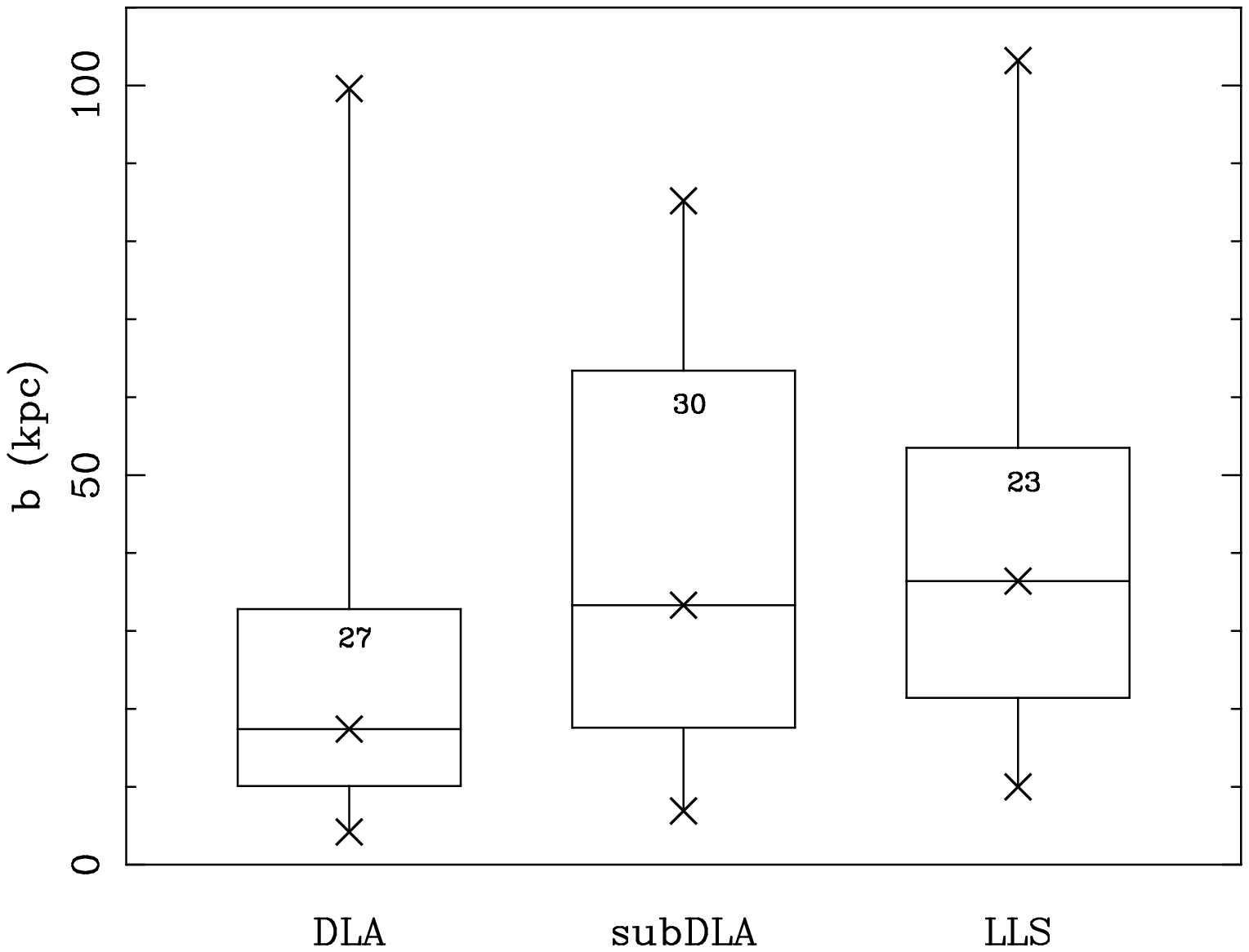}
\caption{Same as in Figure \ref{W_boxplot}, but for the impact
parameter distribution  of the sample.}
\label{b_boxplot}
\end{figure*}

Figure \ref{bL_KSplots}  shows that the median values of the DLA and
subDLA $b$ distributions are very different. In order to quantify the
dispersion in the data, we use the ``median absolute deviation
(MAD),'' which is defined as the median of the  absolute deviations
from the data's median for the sample under consideration.  It gives
an estimate of the spread in values about the median, and is a
statistic that is not sensitive to  extreme outliers in the data. We
find that the median  impact parameter for the DLA sample is $b=17.4$
kpc (with $b_{MAD}=9.0$ kpc), while the median value for the subDLA
sample is $b=33.3$ kpc (with $b_{MAD}=15.8$ kpc),  or twice as
large. For the LLS sample, the median is $b=36.4$ kpc (with
$b_{MAD}=16.1$ kpc).  The impact parameter distributions are shown
graphically in the box and whisker diagrams of Figure \ref{b_boxplot}.
As before, the crosses are the minimum, median, and maximum values,
and the bottom and top edges of the box are the first and third
quartile values of the sample. While there is overlap in the impact
parameter distributions of DLA and subDLA systems,  both the
cumulative (Figure \ref{bL_KSplots}) and the box and whisker plots
(Figure \ref{b_boxplot}) show that they are clearly
different\footnote{ In fact, the maximum value of the DLA $b$
distribution is 99.6 kpc. The identification of this galaxy is
controversial (see \S 6). The next highest value is $b=67.2$ kpc (see
Figure \ref{bL_KSplots} for example). So if the  $b=99.6$ kpc value is
ignored, there is less of an overlap in the $b$ values of the two
samples.}.  The subDLA (and LLS) galaxies tend to be farther away from
the quasar sightline.

\begin{figure*}
\includegraphics[angle=0,width={0.9\columnwidth}]{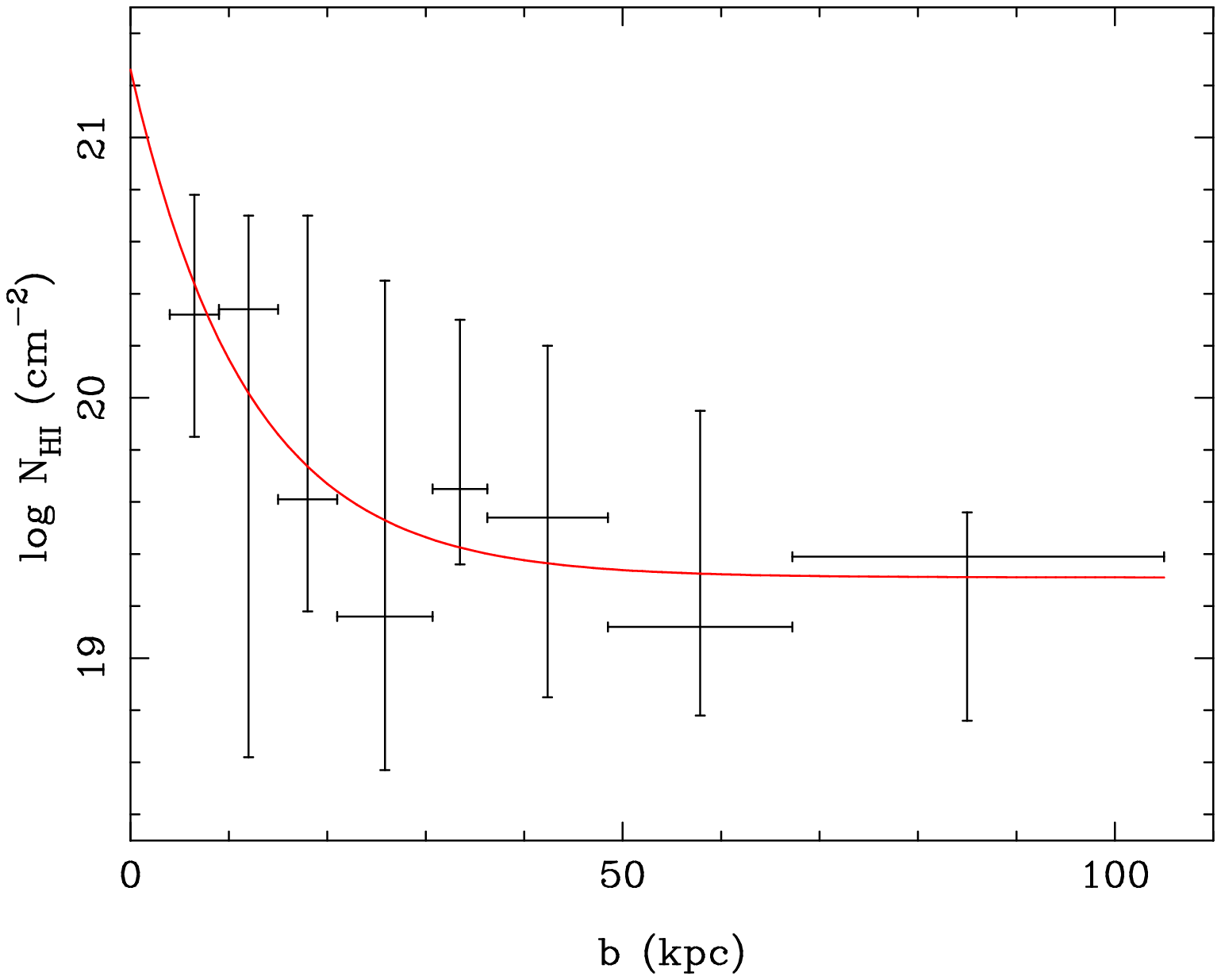}
\caption{Median values of $\log N_{HI}$ are plotted as a function of
$b$. The horizontal bars are bin sizes chosen to include an equal
number of systems in each bin. The vertical bars indicate lower and
upper quartiles of $\log N_{HI}$ values in each bin. Outliers are not
included in this plot. The red curve is an exponential fit to the
median values, and has an {\it e}-folding  length of 12 kpc.  Errors
are not included in the fit. }
\label{medianNHIvsb_fit}
\end{figure*}

In Figure \ref{medianNHIvsb_fit}, we plot the median value of $\log
N_{HI}$  as a function of $b$, with the binning chosen to include an
equal number  of systems in each bin. The horizontal bars indicate bin
size, and the vertical bars are the first and third quartile of the
range of $\log N_{HI}$ values in  each bin. Thus, outliers are not
represented in this plot. The red curve is an exponential fit to the
median values alone, and does not include errors.  It has an
$e$-folding length of 12 kpc, which occurs at $\log N_{HI} \approx
20.0$.  While there is a large spread in \hi column densities at any
given impact parameter,  this plot illustrates that the median value
of $\log N_{HI}$ declines roughly exponentially with impact parameter.

On the other hand, the galaxy luminosity distributions for the three
$N_{HI}$  samples are more similar (Table \ref{bW_KS}). The KS test
probability is $P_{KS}=0.976$ for the DLA and subDLA samples, 0.087
for the LLS and subDLA  samples, and 0.228 for the DLA and LLS samples
(not tabulated). The latter two are not considered to be statistically
significant. It can also be seen from Figure \ref{bL_KSplots} that the
median values of the three luminosity  distributions are very similar
($ 0.33L^*$, $0.20L^*$, and $0.31L^*$ for the DLA, subDLA, and LLS
samples, respectively).

Despite the fact that we have assembled the largest sample of DLA and
subDLA galaxies thus far, small number statistics still play a
relatively significant role. Consider for example, the case of the
galaxy identified as the $z=0.313$ absorber in the 1127$-$145
field. In Rao et al. (2003) we identified a dwarf galaxy, which is
perhaps associated with several brighter galaxies at the same
redshift, as the absorber because of its proximity to the quasar
sightline. Its luminosity is $0.01 L^*$, and it is at a projected
distance of $b=16.1$ kpc from the quasar sightline.  However, more
recently, Kacprzak et al. (2010) have shown that the kinematics of the
\mgii absorbing gas are more consistent with the gas being associated
with the more luminous $0.59 L^*$ galaxy that is 45.6 kpc from the
quasar sightline. We have adopted this new interpretation (see Table
\ref{IDprevious}) for this absorber. We believe that the
identification is  still debatable since no spectroscopic redshift
exists for the dwarf galaxy. Kacprzak et al. (2010) speculate that it
might be near the quasar redshift. Nevertheless, had we retained our
original  identification for this system, the $b$ and $L$
distributions would change slightly: the DLA and subDLA impact
parameter distributions would then be inconsistent with each other at
the 96\% confidence level ($P_{KS}=0.041$).  The DLA and subDLA
luminosity distributions would be nearly identical, with KS test
probability $P=0.9999$.

\begin{figure*}
\includegraphics[angle=0,width={0.9\textwidth}]{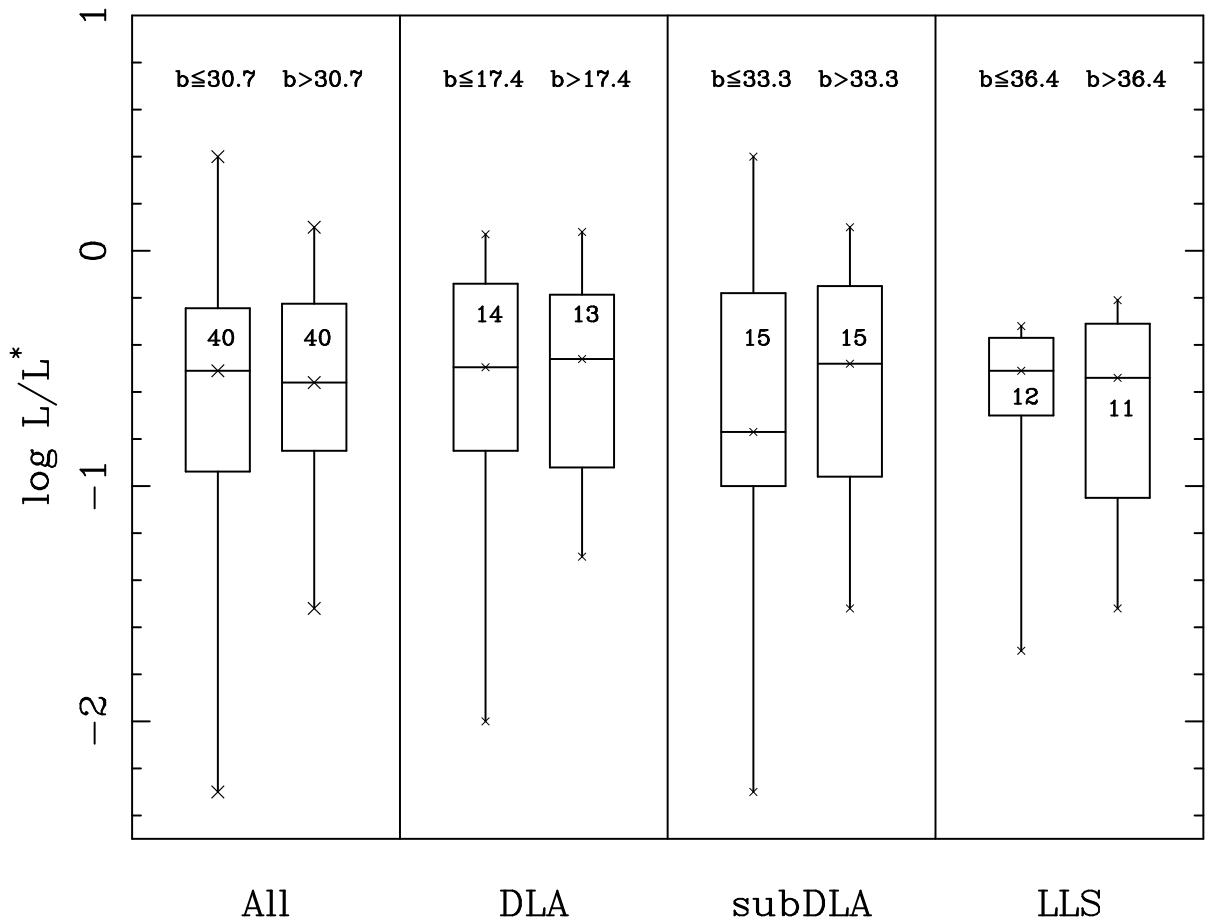}
\caption{Same as in Figure \ref{W_boxplot}, but for the luminosity
distribution of  the sample. The full sample is  shown in the first
panel. The samples have been split by impact parameter and the  median
impact parameter for each sample, in kpc, is indicated at the top of
each panel. }
\label{L_boxplot}
\end{figure*}

We now investigate whether brighter galaxies have larger impact
parameters within  each $N_{HI}$ sample. We have already seen that
brighter galaxies at the same impact parameter do not give rise to
higher values of $W_0^{\lambda2796}$ (\S 4.4), and so we do not expect
any trends here. The data are shown pictorially in Figure
\ref{L_boxplot} as box and whisker  plots.  There is significant
overlap within all three $N_{HI}$ subsamples, and there is no evidence
that galaxies that are farther away (larger $b$) are more luminous.

\subsection{Luminosity versus Impact Parameter}

This correlation is significant only at the $1.2\sigma$ level (Table
\ref{Spearman}).  The galaxy at $b = 99.6$ kpc is identified with the
$1622+239$ $z_{abs}=0.6561$ DLA system  (see the fourth column of
Figure \ref{stairplot}). It might seem that without this data point,
one would conclude that low luminosity absorber galaxies are not found
at high impact parameters. (The identification of this galaxy as the
DLA absorber is debatable. See \S 6.) However, the significance of the
correlation improves only to $1.5\sigma$ without this system.

\subsection{Principal Component Analysis}

In order to explore whether absorber galaxy properties occupy any
preferred direction in a multi-parameter space, we also performed a
principal component analysis  (PCA) using the four parameters that
characterize the absorber: $W_0^{\lambda2796}$, $\log N_{HI}$, $b$,
and $L/L^*$. The PCA did not reveal any useful  information. Each of
the four eigenvectors included multiple parameters with non-negligible
eigencoefficients.  In other words, four eigenvectors had to be used
to effectively explain the overall variance in the sample.  The four
eigenvectors are:
\begin{displaymath}
EV1 = 0.63W + 0.64N - 0.39b + 0.18L
\end{displaymath}
\begin{displaymath}
EV2 = 0.04W -0.05N + 0.40b + 0.91L
\end{displaymath}
\begin{displaymath}
EV3 = -0.35W -0.26N - 0.82b + 0.36L
\end{displaymath}
\begin{displaymath}
EV4 =  0.69W -0.72N - 0.08b - 0.03L
\end{displaymath}
with eigenvalues 1.75, 1.00, 0.84, and 0.42 respectively. In these
equations we have abbreviated $W_0^{\lambda2796}$ with $W$, $\log
N_{HI}$ with $N$, and $L/L^*$  with $L$.

\subsection{Galaxy Types}

The stellar population synthesis model fits used to calculate
photometric  redshifts also provide information on galaxy type. We
followed the classification  scheme of Budavari et al. (2003), whose
templates were also used in this study.  Details of our photometric
redshift fits are given in the Appendix.  Budavari et al. (2003) used
a spectral-type parameter that is essentially  derived from the
rest-frame colours of the best-fit spectral energy distribution (SED).
Using this same classification we find that of the 27 galaxies with
photometry for which we were able to fit templates, i.e., those  for
which photometry in four or more filters exists,  4 have SEDs that are
consistent with being ellipticals, 8 can be classified as  spiral type
Sbc, 9 as Scd, and 6 as irregular. Of the 27, eight are DLA galaxies
with template fits: one is an elliptical, two are of type Sbc, four
are Scd, and one is an irregular galaxy.  Similarly, of the 11 subDLA
and 8 LLS galaxies with template fits, the distribution is (2,2,5,2)
and (1,4,0,3), respectively,  for types (E, Sbc, Scd, Irr).

Thus, as has been known from previous studies, we can now confirm with
a larger sample  that DLA galaxies are predominantly late,
star-forming galaxies, but span the  entire range of galaxy spectral
types. We also find the same result for subDLA and LLS galaxies.
Since SED fits were made for a random subset of galaxies in our
sample, that depended mainly on observing parameters, we expect the
same distribution of types for the remainder of the sample, i.e., for
galaxies with insufficient photometry to carry out template
fits. Moreover, we also showed that the CL = 1 sample, i.e.,  the
sample for which spectral-types could be determined, has the same
properties as the CL = 2 or 3 sample (\S 4.2).

\subsection{Metallicity and Galaxy Properties}

\begin{figure*}
\includegraphics[angle=0,width={0.9\textwidth}]{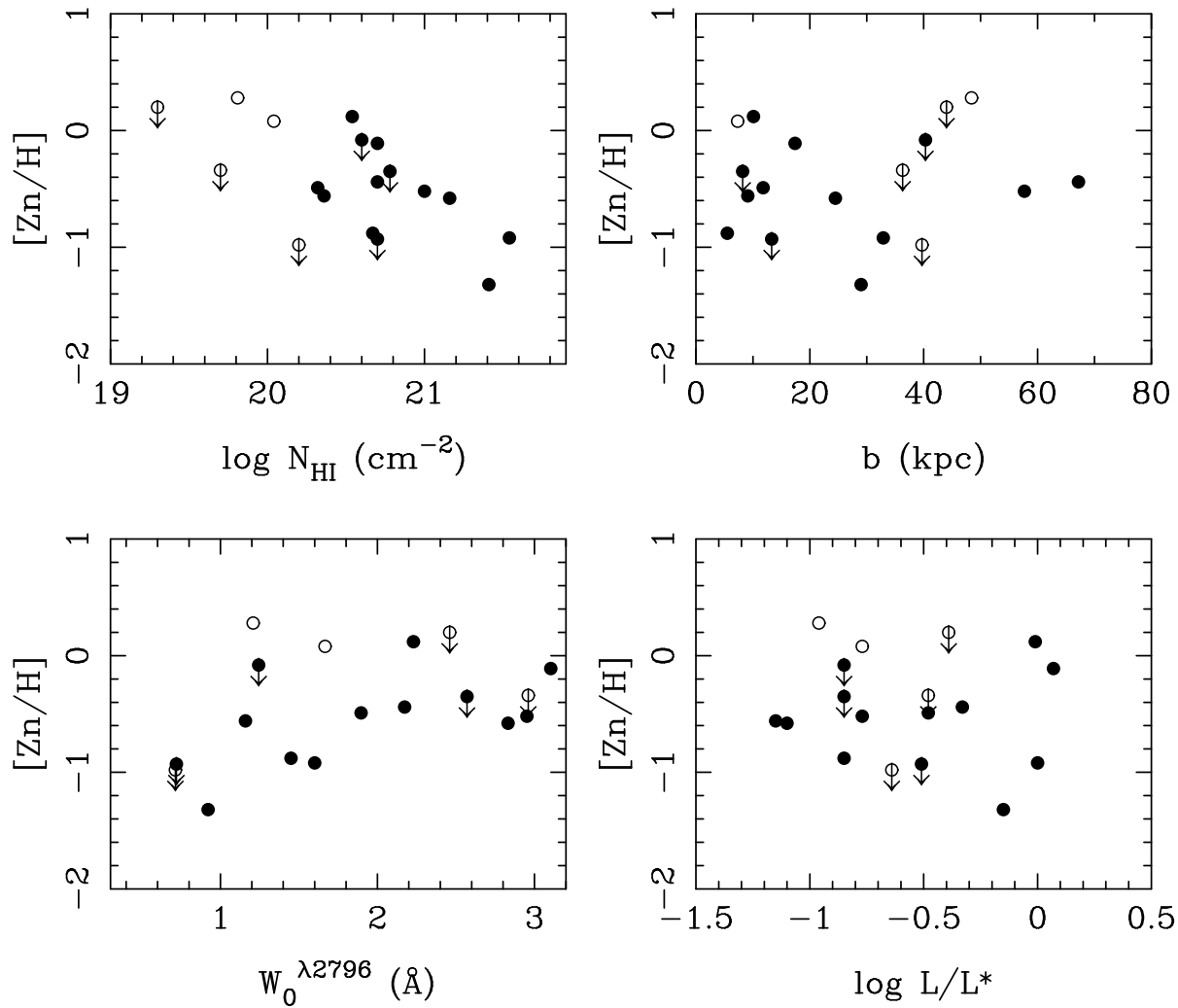}
\caption{Metallicity measurements from the literature versus
absorption line properties $\log N_{HI}$ and  $W_0^{\lambda2796}$ on
the left, and galaxy properties $b$ and  $\log L/L^*$ on the right. }
\label{Zn_NWbL}
\end{figure*}

Figure \ref{Zn_NWbL} shows plots of metallicity measurements,
specifically [Zn/H], from the literature as a function of absorption
line parameters $W_0^{\lambda2796}$ and $\log N_{HI}$, and galaxy
parameters $b$ and $L/L^*$. Even with this small sample, it can be
seen that higher $W_0^{\lambda2796}$ systems tend to have higher
metallicities (e.g., Nestor et al. 2003, Turnshek et al. 2005,
Kulkarni et al. 2010).  The tendency for subDLAs to have higher
metallicities than  DLAs is also apparent (e.g., Meiring et
al. 2009). However, no trends with galaxy properties can be deduced
from the two panels on the right in Figure \ref{Zn_NWbL}. One might
expect  that more luminous, and therefore more massive, galaxies would
tend to give rise to higher $W_0^{\lambda2796}$ systems (since rest
equivalent width is an indicator of the velocity spread of the gas),
but this is not seen in our sample (see Table \ref{Spearman}). One
might also expect some of the more luminous galaxies to be more
evolved, and thus exhibit higher metallicities, but  this is also not
seen in our sample. There also does not seem to be any correlation
between metallicity and galaxy impact parameter. Of course, the sample
is small. More conclusive results must await more measurements of
metallicities at $z<1$.

\subsection{Comparison with the $z=0$ Galaxy Distribution}

\begin{figure*}
\includegraphics[angle=0,width={0.9\textwidth}]{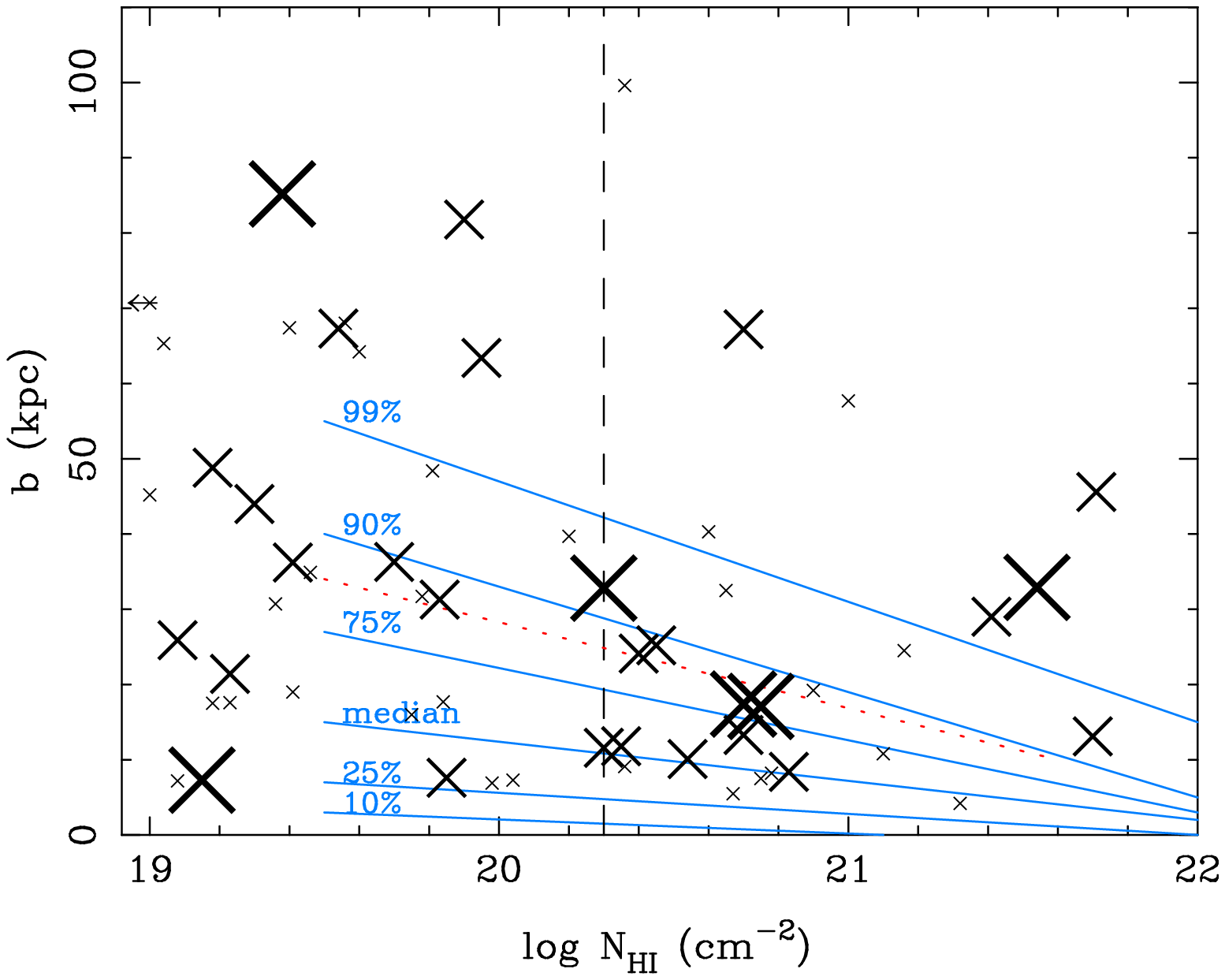}
\caption{Impact parameter, $b$, versus $\log N_{HI}$. The smallest
crosses are galaxies in our sample with $L\le 0.3L^*$, the medium
sized crosses are those with  $0.3<L\le L^*$, and the large crosses
represent galaxies with $L> L^*$. The subDLAs and the DLAs are
separated by the vertical dashed line. The blue solid lines, adapted
from Zwaan et al. (2005), approximate the curves drawn in their Figure
15, which extend only to $\log N_{HI} = 19.5$. The line  labeled
``median'' indicates that there is a 50\% probability that the
absorber galaxy will  lie below this line. The other lines show the
10th, 25th, 75th, 90th, and 99th percentiles  for galaxies at
$z=0$. The red dotted line has been adapted from Figure 18 of Zwaan et
al. (2005), and indicates that above this line, more than 50\% of the
galaxies are more luminous than $L^*$ at $z=0$.}
\label{Zwaan}
\end{figure*}

Figure \ref{Zwaan} is a plot of $b$ versus $\log N_{HI}$.  Our data
for $\log N_{HI} > 19$ are shown as crosses, with the size of  the
crosses indicating three different bins in luminosity. The smallest
crosses are galaxies with $L<0.3L^*$, the medium sized crosses are
those with  $0.3<L<1L^*$, and the large crosses represent galaxies
with $L>1L^*$. The subDLAs and the DLAs are separated by the vertical
dashed line. The blue solid lines, adapted from Zwaan et al. (2005),
approximate the curves drawn in their Figure 15 (their curves only
extend to $\log N_{HI} = 19.5$ at the low \hi  column density end).
They represent the conditional probability of impact parameter  as a
function of \hi column density for local galaxies. Using  \hi 21-cm
line maps of nearby galaxies taken with the Westerbork Synthesis Radio
Telescope, they present absorber and galaxy properties in a form that
can be used to compare with the properties of higher redshift \hi
absorbers.  Specifically, from their analysis of essentially $z\approx
0$ galaxies, they derive relevant probabilities within this $b-\log
N_{HI}$ plane. The line  labeled ``median'' indicates that there is a
50\% probability that the absorber galaxy will  lie below this
line. The other lines show the 10th, 25th, 75th, 90th, and 99th
percentiles  for galaxies at $z \approx 0$. Based on the low-redshift
DLA galaxy data available at  that time, they concluded that the
distributions were similar, although there were somewhat fewer low $b$
systems than expected. Now, with our much larger sample we see that
the distributions are markedly different. For $\log N_{HI} > 19.5$,
only eight out of 42 galaxies lie below the median line, and we have
no galaxies below the 25th percentile line. As we had discussed in \S
3, we might have misidentified about four galaxies in our sample due
to our inability to probe close to the quasar  sightline. Therefore,
at most, we would have 4/42, or 9.5\%  of the sample below the 25th
percentile line and 12/42 (28.6\%) below the median line. At the high
end of the $b$ distribution, we see that 11/42 (26\%) of the galaxies
are beyond the 99th percentile line. If four of these were the
misidentifications 7/42 (17\%) would be above this line.

Zwaan et al. (2005) also find that at $z\approx 0$ the most luminous
galaxies are most likely  to be associated with high $N_{HI}$ DLAs at
large impact parameters. The red dotted line in Figure \ref{Zwaan} has
been adapted from their Figure 18, and indicates that above this line,
more than 50\% of the galaxies are more luminous than $L^*$ at $z=0$.
We find that out of the 24 galaxies in our sample that have $\log
N_{HI} > 19.5$ and  are above this line, only  two are more luminous
than $L^*$.

Clearly, the properties of low-redshift DLA and subDLA galaxies differ
considerably in comparison to the local population, and this might be
indicative of evolution in the neutral-gas environments of galaxies.
Although the statistics at the low redshift end of our sample are
small, we find that the impact parameter distribution (Figure
\ref{stairplot} and Section 4.3) has an upper envelope that declines
between redshifts $z\sim 0.5$ and $z\sim 0$. Taken in combination with
the Zwaan et al. results, this is perhaps suggestive of the process of
galaxy assembly over the last 5 Gyr.  The factor of two decline in
$\Omega_{DLA}$ over the same redshift  interval (RTN06) might also be
related to the same phenomenon.

\subsection{HST versus Groundbased Identifications}

Here we address the question of whether the galaxies identified in HST
data have different properties in comparison to galaxies identified in
groundbased  data. For example, since HST images can probe closer in
to the quasar sightline, the impact parameters of galaxies identified
in HST images might be systematically smaller.

Of the 29 galaxies that were known prior to this work (Table
\ref{IDprevious}),  17 have been identified in HST images and 12 were
first identified in groundbased studies.  The 1127$-$145 absorber  was
first identified by Bergeron \& Boiss\'e (1991) in groundbased data.
An HST image of this field was later discussed by Kacprzak et
al. (2010). In  addition, the 0827+243 field which was in our initial
groundbased sample (Rao et al. 2003), was imaged with HST, but the
absorber identification remained the same (Steidel  et al. 2003). For
the purpose of the comparison being made here,  these two fields will
be considered HST fields.
 
A KS test gives a probability of $P_{KS}=0.637$ that the groundbased
and HST impact parameter samples are drawn from the same parent
population. While it may appear that the HST-identified galaxies have
smaller impact parameters (the median $b$ value for the HST
identifications is 21.4 kpc versus 31.3 kpc for the groundbased
identifications),  the distributions of HST and groundbased
identifications are statistically similar. Their luminosity
distributions are also similar with $P_{KS}=0.909$.

\section{Conclusions}

We have presented the results of an optical/IR imaging programme aimed
at studying the properties of $0.1 \la z \la 1$ galaxies giving rise
to quasar absorption systems with available neutral hydrogen column
densities.  Results from 55 quasar fields with 66  absorbers are
presented here for the first time. We were able to identify  the
absorbing galaxy for 54 of these. By combining galaxy identifications
from  previous studies, we have analysed the properties of a sample of
27 DLA ($N_{HI} \ge 2\times 10^{20}$ atoms cm$^{-2}$), 30 subDLA
($10^{19}< N_{HI} < 2\times 10^{20}$ atoms cm$^{-2}$), and 23 LLS
(with $N_{HI}$ not large enough to be DLAs or subDLAs -- $3\times
10^{17} < N_{HI} \le 10^{19}$ atoms cm$^{-2}$) galaxies. All of these
absorbers were \mgii-selected. While our sampling of
$W_0^{\lambda2796}$ values is sufficient to include all DLAs, it is
unlikely that this holds for subDLAs and LLSs. But we have no {\it a
priori} reason to believe that this significantly affects our results.

Here we summarize the main results from \S3 and \S4.

\begin{itemize}

\item[1.] An analysis of our uniform $K$-band dataset shows that the
surface density of galaxies falls off exponentially with increasing
impact parameter, $b$, from the quasar sightline relative to a
constant background of galaxies, with an {\it e}-folding length of
$\approx 46$ kpc (Figure 6).  Galaxies with $b \ga 100$ kpc calculated
at the absorption redshift are statistically consistent with being
unrelated to the absorption system, and are either background or
foreground galaxies. See \S3 and conclusion 4 (below).

\item[2.] The correlation between the log of the neutral hydrogen
column density, $\log N_{HI}$, and \mgii rest equivalent width,
$W_0^{\lambda2796}$, first reported by RTN06 but subject to the noted
caveats, is also present in this imaging sample (\S 4.4 and Figure
21). Since most of the sample was selected from RTN06, a correlation
is expected. The $\log N_{HI}$ and $W_0^{\lambda2796}$ parameters are
positively correlated at the $4.7\sigma$ level of significance (see
footnote \#8).  The median values of $W_0^{\lambda2796}$ in the
imaging sample  are 2.0 \AA\ for the DLAs, 1.37 \AA\ for the subDLAs,
and 0.78 \AA\ for the LLSs.

\item[3.] The galaxy luminosity relative to $L^*$, $L/L^*$, is not
correlated with $W_0^{\lambda2796}$, and the $b$ value is only weakly
correlated with $W_0^{\lambda2796}$ (\S 4.4). There is an inverse
correlation between $b$ and $W_0^{\lambda2796}$ at only the
1.8$\sigma$ level of significance (Figure 19).  Also, galaxies which
give rise to higher $W_0^{\lambda2796}$ are not significantly
systematically more luminous, even if the comparison is made as a
function of $b$ (Figure 22).

\item[4.] The $\log N_{HI}$ value is inversely correlated with $b$ at
the 3.0$\sigma$ level of significance in the sense that DLA galaxies
are found systematically closer to the quasar sightline, by a factor
of two, than are galaxies which give rise to subDLAs or LLSs (\S 4.5
and Figure 24). The median impact parameter is 17.4 kpc for the DLA
galaxy sample, 33.3 kpc for the subDLA sample, and 36.4 kpc for the
LLS sample.  This is not unexpected, since higher column density gas
tends to exist at smaller galactic radii, but this is the first time
it has been definitively demonstrated among galaxies identified as DLA
or subDLA absorbers.  We also find that the decline in the median
value of $\log N_{HI}$ with $b$ can be roughly described by an
exponential with an {\it e}-folding length of 12 kpc that occurs at
$\log N_{HI} = 20.0$ (Figure 25).

\item[5.] $\log N_{HI}$ is not correlated with galaxy luminosity (\S
4.5 and Figure 26). The median values of luminosity are $0.33L^*$,
$0.20L^*$, and $0.31L^*$ for the DLA, subDLA, and LLS samples,
respectively. There is also no evidence that, within each sample,
galaxies with large impact parameters are more luminous (Figure 26).

\item[6.] The $b$ value is not significantly correlated with $L/L^*$,
as a positive correlation is present at only a 1.2$\sigma$ level of
significance (\S 4.6 and Figure 19).

\item[7.] A PCA did not reveal any useful information, i.e., the
absorbers do not occupy a preferred direction in the multiparameter
space defined by $W_0^{\lambda2796}$, $\log N_{HI}$, $b$, and $L/L^*$
(\S 4.7).

\item[8.] DLA, subDLA, and LLS galaxies comprise a mix of spectral
types, but are inferred to be predominantly late type galaxies based
on their spectral energy distributions (\S 4.8).

\item[9.] Using measurements of metallicity from the literature, we
find no trends between metallicity and $b$ or $L/L^*$ (\S 4.9 and
Figure 27). This is somewhat surprising, but we caution that the
samples are small.

\item[10.] We find that the properties of low-redshift DLAs and
subDLAs are very different in comparison to the properties of gas-rich
galaxies at the present epoch (\S 4.10 and Figure 28). A significantly
higher fraction of low-redshift absorbers have large $b$ values, and a
significantly higher fraction of the large $b$ value galaxies have
luminosities $L<L^*$.

\end{itemize}

\section{Discussion}

We have presented results from a \mgii -based quasar absorption line
search for galaxies that give rise to DLA, subDLA, and LLS absorption
at redshifts $0.1 \la z \la 1$ in the spectra of background
quasars. The sample we studied was formed from a larger sample of
strong \mgii absorbers ($W_0^{\lambda2796} \ge 0.3$ \AA) whose \hi
column densities were determined by measuring the Ly$\alpha$ line in
HST UV spectra.

Analysis of our data revealed two main correlations. First, by
considering the three different $N_{HI}$ column density regimes (i.e.,
DLAs, subDLAs, and LLSs), we find that $\log N_{HI}$ is positively
correlated with $W_0^{\lambda2796}$ at the 4.7$\sigma$ significance
level (i.e., \S 5 conclusion 2 and Figure 21).  It is important to
realize that because the \mgii absorption lines are generally
saturated, $W_0^{\lambda2796}$ is a proxy for the sightline velocity
spread of the absorbing gas associated with the galaxy. Therefore, one
can statistically infer that sightlines that intercept larger \hi
columns of gas generally encounter larger gas velocity spreads.
However, this is not a tight correlation. The $N_{HI}$ value cannot be
used to predict $W_0^{\lambda2796}$, nor can $W_0^{\lambda2796}$ be
used to predict $N_{HI}$.  One interpretation is that the gas that is
primarily responsible for a DLA is one of many clouds along the line
of sight. The  larger the \mgii rest equivalent width, the more clouds
along the  sightline, and the higher the probability of one of them
being the  cloud that produces a DLA. This explains why, for example,
a strong \mgii system is not always a DLA, and why weaker  \mgii
systems can occassionally be DLAs. The threshold
$W_0^{\lambda2796}=0.6$ \AA,  below which no DLAs are found, is  then
representative of the minimum  velocity spread of a region containing
DLA clouds.

Second, we found that the median impact parameter of a sample of DLA
galaxies is approximately half that found for samples of subDLA and
LLS galaxies (i.e., \S5 conclusion 4 and Figure 24).  SubDLA and LLS
galaxies have similar impact parameter distributions.  This second
correlation has a 3.0$\sigma$ level of significance. Again, this is
not a tight correlation. To emphasize this, we note that three of our
DLA galaxy identifications have impact parameters $b> 50$ kpc, while
six of our subDLA and LLS galaxy identifications have $b\le10$ kpc. It
does, however, seem unreasonable to expect a tight correlation for
either of these two main correlations.  This is because the observed
impact parameter is set by the chance separation between the absorbing
galaxy and the quasar sightline. That is, the observed impact
parameter for an absorbing galaxy only indicates that the radial
extent of the gas surrounding the galaxy extends at least as far out
as the impact parameter (see discussion of equation 1, below).  The
median impact parameter for DLA galaxies is $\approx 17$ kpc, whereas
it is $\approx 35$ kpc for subDLA and LLS galaxies. But there is wide
overlap in the distributions of impact parameters among the three
identified galaxy samples, as expected.  In combination these two
findings suggest that systems with lower $b$ values should generally
have larger $W_0^{\lambda2796}$ values and vice versa, and indeed a
weak inverse correlation at a 1.8$\sigma$ level of significance is
seen (i.e., \S5 conclusion 3 and Figure 19).

Taken together, the observed trends, although weak, indicate that DLAs
generally have  higher values of \mgii $W_0^{\lambda2796}$, smaller
impact parameters, lower metallicities, and  similar luminosities in
comparison to subDLAs. That subDLA and DLA galaxies have similar
luminosities implies that the subDLAs are not more massive, which was
a suggestion made by Kulkarni et al. (2010) to explain their higher
metallicities. The mass-metallicity relation does not appear to play a
role here.  That DLA sightlines  have higher velocity spreads but
lower impact parameters  is an indication that the gas is not
rotationally supported. As suggested by several studies,  superwinds
and tidal gas from mergers are likely to be involved (see below).

The absence of tight correlations may also be due to the patchiness of
\hi absorbing gas and misidentifications of ``true'' absorbing
galaxies.  For example, with regard to the \hi gas being patchy,
Monier et al. (2009b) found that $N_{HI}$ changed from the DLA to the
subDLA regime over sightline changes as small as  $\approx 5$ kpc at
$z\approx 1.5$.  Cooke et al. (2010) studied a $z\approx1.63$ DLA with
$N_{HI} \approx 5\times10^{20}$ atoms cm$^{-2}$, but along an adjacent
sightline separated by 2.7 kpc found that the \hi column density
dropped to $N_{HI} < 1.3\times10^{18}$ atoms cm$^{-2}$. This suggests
that a slight change in sightline separation, which is much smaller
than the observed $b$ value, influences classification of the
absorbing galaxy as a DLA, subDLA, or LLS galaxy. This would clearly
increase the intrinsic spread in correlations with $N_{HI}$.  Since we
did not form a control sample of galaxies that {\it do not} give rise
to absorption lines in the spectra of background quasars, our current
work offers no conclusions on the covering factor of
$W_0^{\lambda2796} \ge 0.3$ \AA\ \mgii absorbers. However, we note
that with their absorbing and non-absorbing galaxy samples, Kacprzak
et al. (2008) obtain a mean covering factor of $\approx50$\% for
$W_0^{\lambda2796} \ge 0.3$ \AA\ \mgii gas. Chen et al. (2010) derive
a covering factor of $\approx70$\% for $W_0^{\lambda2796} \ge 0.3$
\AA\ absorbers and $\approx80$\% for $W_0^{\lambda2796} \ge 1.0$ \AA\
absorbers. Even though information on \hi column density is
unavailable in these other analyses, these results also provide
evidence for patchiness which would increase the intrinsic spread in
correlations.

With regard to the possible misidentification of absorbing galaxies,
consider our identifications for two DLA absorbing galaxies (\S3) with
impact parameters of $b\approx67$ kpc (for a 0.7$L^*$ galaxy) and
$b\approx100$ kpc (for a 0.05$L^*$ galaxy). These may both be
outliers, but certainly the identification of the second one seems
far-fetched. However, there is another possibility which may put such
identifications in context. Recently, Kacprzak et al. (2010) published
cosmological simulations to aide in the interpretation of their
observations of \mgii-absorption-selected galaxies at
intermediate redshift. Their simulations indicate that, relative to a
central galaxy, \mgii absorption selects metal-enriched halo
gas, tidal streams, filaments, and small satellite galaxies. In
particular, they find that \hi column densities in the DLA regime can
arise in low-mass satellite galaxies at impact parameters as large at
$\approx 100$ kpc.  Our large impact parameter DLA galaxies may be
examples of such cases, and in that sense they may be
misidentifications since the small satellite galaxies would not be
identified because of the glare of the background quasar.  It is
therefore useful to use our results to consider the overall radial
gaseous extent of our identified absorbing galaxies as a function of
galaxy of luminosity $L$, i.e., $R(L)$. However, it seems appropriate
to interpret results on $R(L)$ in the context of the Kacprzak et
al. (2010) simulations in which ``halo'' gas can have a variety of
origins, including the possibility that the observed absorption arises
in a neutral-gas-rich satellite galaxy which resides in the
environment of the galaxy we identify as the absorbing galaxy.

Another result worth mentioning is the $z_{abs}=0.006$, $\log
N_{HI}=19.3$,  subDLA system towards PG 1216+019 (Tripp et al. 2005),
which  is an example of an absorbing galaxy that was detected in \hi
emission alone.  Briggs \& Barnes (2006) report on the detection of a
21 cm line emitter  that has $M_{HI}$ between 5 and 15 $\times 10^6
M_\odot$, and is within $\sim 4$ kpc of the quasar sightline. No
optical counterpart to this \hi emitter has been detected thus far
(Chen et al. 2001. Tripp et al. 2005), implying that any optical emission from this
galaxy might be hidden under the glare of either the quasar PSF or
that of a nearby (10\arcsec\ away) star.  Other \hi emitting objects
within 100 km s$^{-1}$ of the absorption redshift are a $0.25L^*$
galaxy at a distance of 92 kpc from the quasar  sightline and an
optically-undetected, $M_{HI}\sim 3 \times 10^8 M_\odot$, object 120
kpc  from the quasar sightline. This suggests that the absorber is
likely to be tidal debris or diffuse gas in the halo of one or both of
the  two more massive galaxies (Briggs \& Barnes 2006).  Thus, here is
a case where the $b=92$ kpc galaxy would have been identified as the
absorber in the absence of the 21 cm data, when in fact, a much
smaller $b$, low-mass dwarf galaxy is the true site of absorption.

\begin{figure*}
\includegraphics[angle=0,width={0.45\textwidth}]{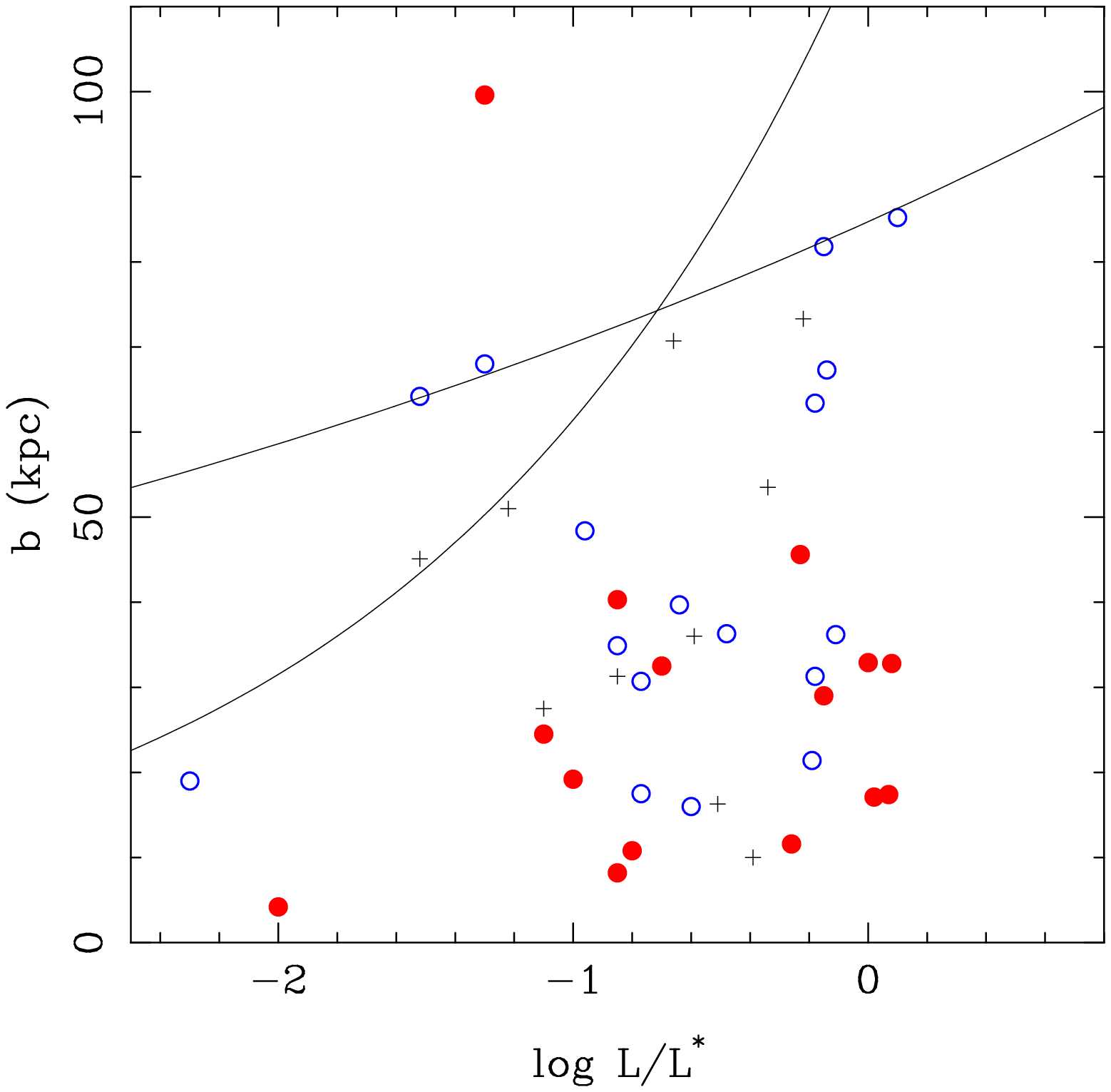}
 \hfil
\includegraphics[angle=0,width={0.45\textwidth}]{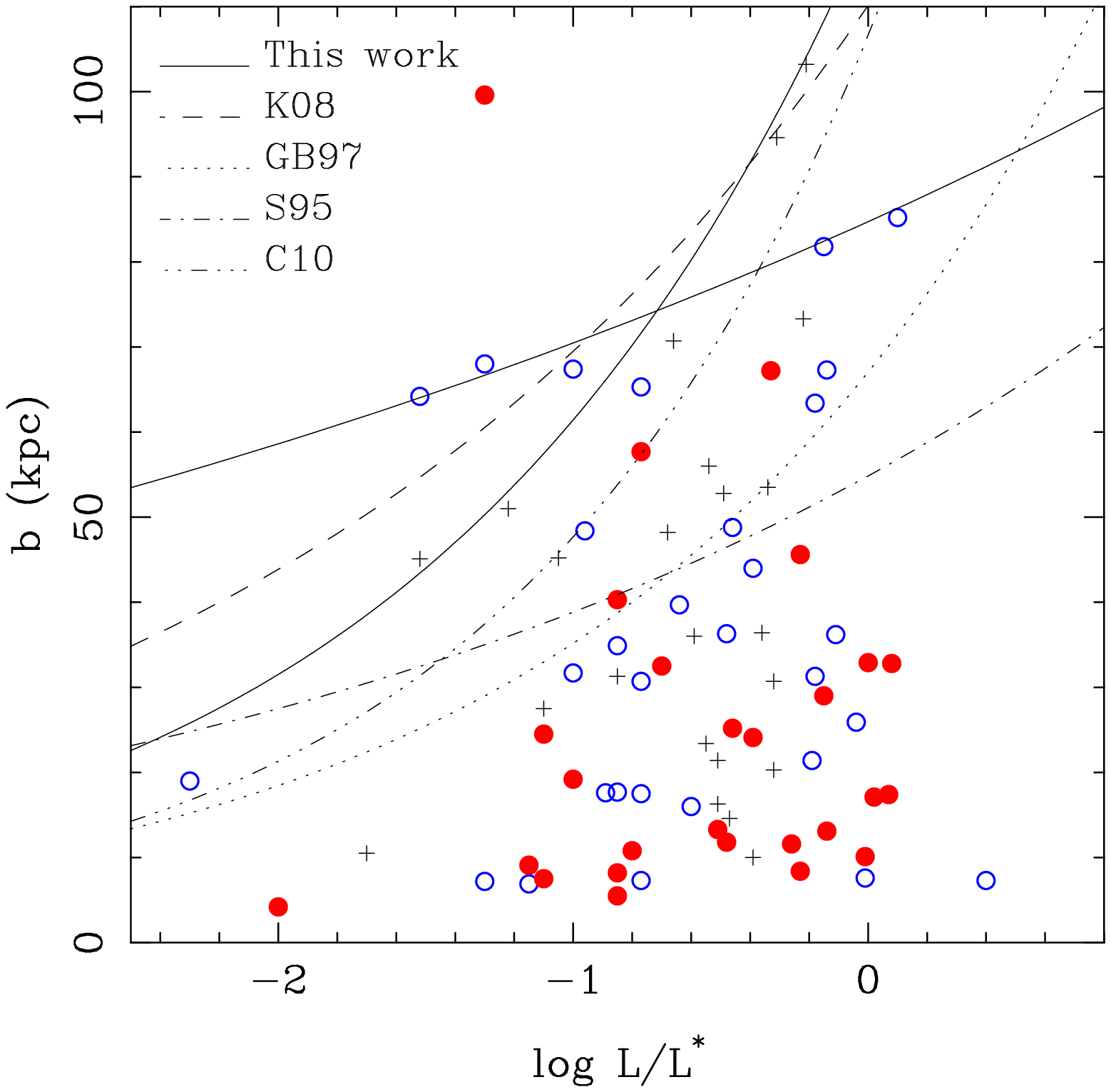}
\caption{Impact parameter, $b$, versus luminosity, $\log L/L^*$, for 
$K$-band data (left panel), and for the entire sample (right panel). 
Red solid circles are DLAs, blue open circles are subDLAs, and plus symbols
are LLSs. 
The solid lines are the same in both panels: they are power laws of 
the form $b=89 (L/L^*)^{0.08}$ kpc (the shallower one) and $b=120 (L/L^*)^{0.29}$ kpc,
 and are drawn to encompass the range of possible upper envelopes to the data. 
The other power laws are from previously published work:
K08: Kacprzak et al. (2008), GB97: Guillemin \& Bergeron (1997),
S95: Steidel (1995), and C10: Chen et al. (2010). See text.
}
\label{bLpowerlaw}
\end{figure*}

Past attempts have been made to derive a radius-luminosity scaling
relationship for absorbers in order to infer the extent of gaseous halos 
(e.g., Steidel 1995, Guillemin \& Bergeron 1997, Kacprzak et al. 2008, 
Chen et al. 2010). The scaling relation is a power law of the form
\begin{equation}
R(L) = R_* (L/L^*)^\beta,
\label{equationRL}
\end{equation}
where $R$ denotes the radial gaseous extent of a galaxy of luminosity
$L$.  For our data this is illustrated by the upper envelope of a plot
of $b$ versus $L/L^*$.  In the left panel of Figure \ref{bLpowerlaw}
we plot  $b$ versus $\log L/L^*$  for galaxies in our sample with $K$-band data. In
the right panel we plot this for our entire sample using a compilation
of measurements in several wavebands (see Tables \ref{IDprevious} and
\ref{IDsummary}). We show the $K$-band data separately because
measurements in the literature generally use a single waveband,
however a comparison between the left and right panels shows that the
two distributions of data points are similar, and that the conclusions
do not change when we include our entire sample.  The two solid curves
in the two panels are power laws of the form shown in Equation
\ref{equationRL}, and encompass the range of possible upper envelopes
to the data points. The shallower power law has $R_*=89$ kpc and
$\beta = 0.08$, and the steeper power law has $R_*=120$ kpc and
$\beta=0.29$. Both power laws do not include the outlier at
$b\approx100$ kpc and $L\approx 0.05L^*$. This is the galaxy  that is
identified as the $z_{abs}=0.656$ DLA absorber towards 1622+239
(Kacprzak et al.  2007). It is also inconsistent with the other models
from the literature, and is therefore, almost certainly a
misidentification.  
The two data points at $b\approx65$ kpc and $\log L/L^* \approx -1.3$ 
are identifications with confidence levels CL $=2$, where the 
identifications were not straightforward. It
could well be that these galaxies were misidentified. On the other hand,
of the two LLSs with impact parameters $b>90$ kpc, one is a
spectroscopic identification, while the other has CL $=3$. Therefore, the
steeper power law, i.e., the solid curve with $\beta = 0.29$, may be
a more adequate representation of the upper envelope of gaseous halos 
of strong \mgii absorbers, although the shallower power law cannot be 
ruled out because of the small number of galaxies that define the 
upper envelope. In any case, these data indicate that the characteristic
size of the gaseous halo (environment) of an $L^*$ galaxy is likely to
be as large as 120 kpc, larger than any of the previously derived values. 

The K08 (Kacprzak et al. 2008) and C10 (Chen et al. 2010) models 
shown in Figure \ref{bLpowerlaw} are roughly consistent with the data, although
the C10 model  excludes more data points. The S95 (Steidel et al. 1995) 
and GB97 (Guillemin and Bergeron 1997) models are
clearly ruled out by our data.  

It is now clear from the more recent
studies  that gas extends much farther out from the centers of
galaxies  than previously thought. For example, the conclusion from
initial studies that  there is always a bright, $L^*$, galaxy
associated with strong \mgii absorbers,  and that the galaxies
have gaseous disks that are $\sim 40$ kpc in radius, is  no longer
supported. Only nine out of 80 galaxies in our sample are $\sim L^*$ 
or brighter.  It is also clear that the gas distribution within
galaxies is patchy, much like what is seen in high resolution 21 cm
maps of local galaxies (e.g., Zwaan et al. 2005; Braun et
al. 2009). However, at larger galactocentric distances processes such
as star-formation induced outflows (Weiner et al. 2009), radiatively
driven winds (Murray et al. 2010), infalling gas in filaments, tidal
streams and satellite galaxies (Kacprzak et al. 2010), and superwinds
and tidal gas from mergers and interactions (Zwaan et al. 2008) appear
to be playing  an important role. Illustration of our results at
$z\approx0.1-1.0$ in Figure 28  apparently demonstrates the importance
of some of these processes at low-to-moderate redshift. In Figure 28
we show our results in conjunction with local ($z\approx 0$) results on \hi 
gas in galaxies adapted from Zwaan et al. (2005). Our results at
low-to-moderate redshift are inconsistent with the $z\approx 0$ results,
suggesting that we are detecting evolution in the neutral-gas
environments of galaxies.

\section*{Acknowledgments}

SMR, DAT, and MB-M acknowledge support from NSF grant AST 03-07743. 
AMQ acknowledges support REU support from NSF grant AST 03-07743. 
We are grateful for the help and support provided by the NOAO, MDM, and 
NASA IRTF staff. EMM acknowledges the allocation of telescope 
time at the MDM observatory. We are grateful to S. Zibetti for 
providing us with SDSS photometry for some of our galaxies and to 
T. Budavari for his photometric redshift code. We thank J. Busche for
his help in setting up the Web page. We also greatly appreciate the 
anonymous referee's comments and encouraging remarks.

Funding for the SDSS and SDSS-II has been provided by the Alfred 
P. Sloan Foundation, the Participating Institutions, the National 
Science Foundation, the U.S. Department of Energy, the National 
Aeronautics and Space Administration, the Japanese Monbukagakusho, 
the Max Planck Society, and the Higher Education Funding Council 
for England. The SDSS Web Site is http://www.sdss.org/.

The SDSS is managed by the Astrophysical Research Consortium for 
the Participating Institutions. The Participating Institutions are the 
American Museum of Natural History, Astrophysical Institute Potsdam, 
University of Basel, University of Cambridge, Case Western Reserve 
University, University of Chicago, Drexel University, Fermilab, 
the Institute for Advanced Study, the Japan Participation Group, 
Johns Hopkins University, the Joint Institute for Nuclear Astrophysics, 
the Kavli Institute for Particle Astrophysics and Cosmology, the Korean 
Scientist Group, the Chinese Academy of Sciences (LAMOST), Los Alamos 
National Laboratory, the Max-Planck-Institute for Astronomy (MPIA), 
the Max-Planck-Institute for Astrophysics (MPA), New Mexico State University, 
Ohio State University, University of Pittsburgh, University of Portsmouth, 
Princeton University, the United States Naval Observatory, and the 
University of Washington.

This research made use of the NASA/IPAC Extragalactic Database,
which is operated by Jet Propulsion Laboratory, California
Institute of Technology, under contract with the National 
Aeronautics and Space Administration.

\appendix

\section{Photometric Redshift Determinations}

Photometric redshifts are determined using the template
fitting method through a $\chi^2$ minimization procedure.
Conti et al. (2003) describe the details of this procedure, and we
summarize it here.  This approach compares  the expected colours of a
galaxy derived from template spectral energy distributions (SEDs)
with those observed for an individual  galaxy.  Each SED template is
redshifted, convolved with the photometric filter response curves, and
compared with the observed fluxes  through each filter.  A redshift
dependent $\chi^{2}(z)$ is defined as
\begin{equation}
\chi^2(z,T) = \sum_{i=1}^{N_{filters}}\left[\frac{F_{obs,i}- b_j
\times F_{i,j}(z,T)}{\sigma_{i}}\right]^2
\end{equation}
where $F_{obs,i}$ is the flux of the galaxy observed through the $i$th
filter, $F_{i,j}$ is the flux of the $j$th template observed  through
the $i$th filter at redshift $z$, $\sigma_{i}$ is the error in the
observed flux in the $i$th filter, and $b_j$ is a  normalization
constant. The sum is carried out over all available filters,
$N_{filters}$.  The resulting $\chi^2$ is minimized as a  function of
$z$ and template, $T$, giving an estimate of the galaxy redshift with
its variance and spectral type parameters.  This  algorithm is
courtesy of T. Budav\'ari (2003, private communication).

Conti et al. (2003) used the Bruzual \& Charlot (2003) population
synthesis models to generate SED templates that are used to fit the
photometry of the detected galaxies.  Each SED template is the result
of modeling the detailed physical processes affecting star  formation
efficiency and gas properties. These SEDs are used for the current
analysis.  The parameters selected  to generate the templates are
chosen to sample a wide range of physical characteristics, i.e., age,
star formation rate (SFR),  obscuration, and metallicity.  A Salpeter
initial mass function (IMF) with low- and high-mass cutoffs equal to
0.1 $M_{\odot}$ and 125.0  $M_{\odot}$ is assumed for all of the SEDs.
The stellar populations sample 10 ages, ranging from extremely young
(0.001, 0.01, 0.1, 0.5  Gyr), middle (1.0, 3.0, 5.0 Gyr) to old (9.0,
12.0, 15.0 Gyr).  An exponential SFR with an $e$-folding time $\tau$
of the form,  $\Psi(t) = \Psi_0\:e^{-t/\tau}$ is applied.  This form
describes an instantaneous burst when $\tau \rightarrow 0$ and a
constant rate of  star formation when $\tau \rightarrow \infty$.  The
$e$-folding times used in generating the SEDs are $\tau$ = 0.1, 1.0,
3.0, 5.0, 9.0,  and 12.0.  The metallicity is allowed to be
$\frac{1}{200}$ to 2.5 times solar: $Z$ = 0.0001, 0.0004, 0.004, 0.008,
0.02, 0.05, where $Z$  is the metal mass fraction with $Z_{\odot}$ =
0.02. A Calzetti et al. (2000) extinction law of the form  $k(\lambda)
= A(\lambda)/E_{B-V}$ is applied where the following values of
magnitudes of extinction are allowed: 0.0, 0.1, 0.2, 0.3,  0.5, 0.9.
This results in a total of 2160 templates (Conti et al. 2003)
  
Photometric redshifts are calculated for objects detected in 
fields for which data in four or more filters are available.  SDSS optical
photometry, when available, is used to supplement our photometry of
fields for which we have only infrared data.  The photometric
redshift, $z_{phot}$, is deemed consistent with the absorption
redshift, $z_{abs}$, when $z_{abs}$ is included in the range spanned
by  the 1$\sigma$ $z_{phot}$ errors.

\label{lastpage}
\end{document}